\documentclass{jfm}
\usepackage{amsmath}
\usepackage{amssymb}
\usepackage{graphicx}
\usepackage{epstopdf, epsfig}
\usepackage{psfrag}
\usepackage{subcaption}
\usepackage{cleveref}
\usepackage{xcolor}
\usepackage{mathtools}
\usepackage{mathrsfs}

\newtheorem{lemma}{Lemma}
\newtheorem{remark}{Remark}

\newcommand{\bs}[1]{\boldsymbol{#1}}
\newcommand*{\QEDB}{\hfill\ensuremath{\square}}

\shorttitle{Optimal bounds in Taylor--Couette flow}

\shortauthor{A. Kumar}

\title{Optimal bounds in Taylor--Couette flow}

\author{Anuj Kumar\aff{1}
  \corresp{\email{akumar43@ucsc.edu}}}

\affiliation{\aff{1}Department of Applied Mathematics, University of California, Santa Cruz, CA 95064, USA}

\begin{document}

\maketitle

\abstract
This paper is concerned with the optimal upper bound on mean quantities (torque, dissipation and the Nusselt number) obtained in the framework of the background method for the Taylor--Couette flow with a stationary outer cylinder. Along the way, we perform the energy stability analysis of the laminar flow, and demonstrate that below radius ratio $0.0556$, the marginally stable perturbations are not the axisymmetric Taylor vortices but rather a fully three-dimensional flow. The main result of the paper is an analytical expression of the optimal bound as a function of the radius ratio. To obtain this bound, we begin by deriving a suboptimal analytical bound using analysis techniques. We use a definition of the background flow with two boundary layers, whose relative thicknesses are optimized to obtain the bound. In the limit of high Reynolds number, the dependence of this suboptimal bound on the radius ratio (the geometrical scaling) turns out to be the same as that of numerically computed optimal bounds in three different cases: (1) where the perturbed flow only satisfies the homogeneous boundary conditions but need not be incompressible, (2) the perturbed flow is three dimensional and incompressible, (3) the perturbed flow is two dimensional and incompressible. We compare the geometrical scaling with the observations from the turbulent Taylor--Couette flow, and find that the analytical result indeed agrees well with the available DNS data. In this paper, we also dismiss the applicability of the background method to certain flow problems and therefore establish the limitation of this method.

\section{Introduction}
\label{Introduction}
An important problem in the study of turbulent flows is to estimate the functional dependence of global properties (such as energy dissipation, drag force, heat and mass transport, and mixing efficiency) on input parameters. The lack of analytical solutions of the Navier--Stokes equations in the fully turbulent regime has forced the scientific community to adopt a multi-faceted approach to this problem, in which simple physical theories and reduced models are proposed, and then corroborated by direct numerical simulations (DNS) and/or results from laboratory experiments.  However, the inability to perform simulations and experiments in the extreme parameter regimes that often concern atmospheric, oceanic and astrophysical flows and engineering applications leaves these theories unsubstantiated.


In these extreme parameter regimes, an alternative approach that can provide meaningful information is to obtain rigorous bounds on the aforementioned global properties. The first method to obtain bounds was developed by \citet{howard1963heat} and \citet{busse1969howard}, but it was not until the 1990s that bounding techniques gained general popularity, with the introduction of the so-called ``Background Method" by Doering and Constantin \citep{PhysRevLett.69.1648, PhysRevE.49.4087, PhysRevE.51.3192, PhysRevE.53.5957}. The background method is based on ideas from Hopf to produce \textit{a priori} estimates for the solutions of the Navier--Stokes equations with inhomogeneous boundary conditions \citep{hopf1955lecture}. It has so far been applied to many different fluid mechanics problems \citep{PhysRevLett.69.1648, PhysRevE.51.3192, PhysRevE.53.5957, caulfield2001maximal, tang2004bounds, whitehead2011ultimate, goluskin2016bounds, fantuzzi2018boundsA, fantuzzi2018boundsB, kumar2020bound, arslan2021bounds, fan2021three, arslan2021bounds2, kumar2021IH}. See \citet{fantuzzi2021background} for a recent review.

In the background method, we write the total flow field as a sum of two flow fields: the background flow and the perturbed flow. To obtain a bound on the desired bulk quantity requires choosing a background field that satisfies a certain integral constraint (extracted from the governing equations of the perturbed flow). Generally, one takes one of the two following routes. The first route is to specify a functional form of the background flow and then use standard inequalities. This route leads to an analytical but suboptimal bound on the bulk quantity as a function of system parameters. The second route is to find the \textit{best} possible bound (optimal bound) through a variational formulation of the background method in which one solves the corresponding Euler--Lagrange equations usually numerically. Numerous studies pertaining to the background flow have concentrated on the scaling of optimal bounds as a function of the principal flow parameter, such as the Reynolds number and the Rayleigh number. However, only a handful of them studied the variation of these bounds with the shape of the domain. One such study is by \citet{wen2013computational}, where the authors were interested in determining the dependence on aspect ratio of the optimal bound on heat transfer in porous medium convection. 

In this paper, we are concerned with the question of whether it is possible to obtain the analytical expression for the dependence of optimal bounds on the geometrical parameters of the system. Indeed, while the numerically obtained optimal bounds usually follow an easily-identifiable simple power-law in the principal flow parameter, the variation of the optimal bounds with geometrical parameters, however, is not so readily apparent. Furthermore, we also aim to determine whether this analytical form bears any resemblance to the actual dependence of the corresponding bulk quantity on system geometry in fully turbulent flows. This question is motivated by engineering applications where the geometry plays an important role.


In a recent study, we attempted to provide bounds on the friction factor in the context of pressure driven helical pipe flows \citep{kumar2020pressure}. We focussed in particular on the dependence of this bound on the geometrical parameters: the curvature and torsion of the pipe.   We took the first route described above, and used standard functional inequalities to find a suboptimal bound on the friction factor. In order to account for the geometry, we constructed a background flow in which we allowed for a boundary layer thickness that varies along the circumference of the pipe, and optimized the shape of that boundary layer to find the best possible bound for any curvature and torsion. Without giving any further evidence, we hypothesized that the suboptimal bound thus produced might have the same geometrical dependence as the optimal bound.


This paper demonstrates that this hypothesis holds true for Taylor--Couette flow; i.e., the analytical geometrical dependence of the suboptimal bound obtained using traditional functional inequalities (but with a definition of the background flow with optimized boundary layer thickness) is the same as for the optimal bounds obtained using the variational approach.

There are several reasons why we choose to work with the Taylor--Couette flow to test this hypothesis. The Taylor--Couette flow is one of the most extensively investigated problems in fluid mechanics, going back to the seminal paper of Taylor \citep{taylor1923viii} and laboratory experiments of Wendt \citep{wendt1933turbulente}, which are one of the early major contributions to the field. It is known that the Taylor--Couette system exhibits  rich flow structures and complex fluid dynamical phenomena and has served as a testing ground for the theories of turbulent flows. The simplicity of the Taylor--Couette setup makes it amenable to conduct direct numerical simulations and experiments with high precision at high Reynolds numbers. As a result, starting with the work of \citet{PhysRevA.46.6390, PhysRevLett.68.1515}, the last two decades have seen a tremendous activity in the study of high Reynolds number Taylor--Couette flow  from the computational and experimental point of view (see a review by \citet{grossmann2016high}).

Concurrently, progress has also been made on obtaining rigorous bounds in Taylor--Couette flow. \citet{nickerson1969upper} was the first to derive an upper bound on the torque in Taylor--Couette flow using the technique developed by \citet{howard1963heat} and \citet{busse1969howard}. \citet{constantin1994geometric} later revisited the problem using the background method of Doering and Constantin, and also derived an analytical upper bound on the torque. More recently, \citet{ding2019upper} computed the corresponding optimal bounds numerically for systems where the ratio of the inner to outer cylinder radii, called the radius ratio hereafter, is $0.5$, $0.714$ and $0.909$. Note that these three studies concentrated on the dependence of the bounds on the Reynolds number. 

The primary goal of this paper is to obtain the correct functional dependence of the optimal bounds on the torque with respect to the radius ratio. To do so, we shall begin by obtaining an analytical bound using standard inequalities, with the aim of optimizing this bound simultaneously for all values of the radius ratio. Subsequently, we obtain numerical optimal bounds for several values of the radius ratio considering  three different scenarios for the perturbations, which are the following:
\begin{itemize}
    \item[case 1:] The perturbations that satisfy the homogeneous boundary conditions but are not necessarily incompressible;
    \item[case 2:] Additionally, the perturbations are three-dimensional and incompressible;
    \item[case 3:] The perturbations, along with satisfying the boundary conditions and being incompressible, are only two-dimensional (invariant in the axial direction).
\end{itemize}
These scenarios impose increasingly stringent constraints on the type of admissible perturbations and allow us to systematically test the hypothesis described above. We shall demonstrate that the optimal bounds computed in each case  not only have the same dependence in the radius ratio in all scenarios as the Reynolds number tends to infinity, but also that this dependence is the same as the one obtained from the suboptimal analytical bound.


The arrangement of the paper is as follows. We begin by describing the problem configuration, the definitions of the relevant mean quantities and the relations between those quantities in \S \ref{Problem setup}. In \S \ref{Energy stability analysis}, we perform the energy stability analysis of the laminar flow. In \S \ref{An analytical bound}, we obtain analytical bounds on the mean quantities. \S \ref{Optimal bounds: Numerical scheme} presents optimal bounds obtained in the three cases listed above and compares the results with the analytical bounds from \S \ref{An analytical bound}. In \S \ref{A note on the applicability of the background method}, we show that the background method cannot be applied to certain flow problems past certain Reynolds numbers. Finally, \S \ref{Discussion and conclusion}  presents a discussion, comparison with DNS results, the broad applicability of the present study and open problems.

\section{Problem setup}
\label{Problem setup}
Consider the flow of an incompressible Newtonian fluid of density $\rho$ and kinematic viscosity $\nu$ between two coaxial circular cylinders, where the inner cylinder rotates with a constant angular velocity $\Omega$ and the outer cylinder is stationary. The radius of the inner cylinder is $R_i$ and the radius of the outer cylinder is $R_o$. The quantity $\eta = R_i/R_o$ is referred to as the radius ratio hereafter, and $d = R_o - R_i$ is the gap width. We non-dimensionalize the variables as follows:
\begin{eqnarray}
\boldsymbol{x} = \frac{\boldsymbol{x}^\ast}{d}, \quad \boldsymbol{u} = \frac{\boldsymbol{u}^\ast}{\Omega R_i}, \quad t = \frac{t^\ast}{(d / \Omega R_i)}, \quad p = \frac{p^\ast - p_0}{\rho \Omega^2 R_i^2},
\label{Problem setup: Non-dimensionalization}
\end{eqnarray} 
where $p_0$ is the reference pressure and $\boldsymbol{x}$, $\boldsymbol{u}$, $t$ and $p$ denote the non-dimensional position, velocity, time and pressure, respectively. The starred variables are the corresponding dimensional quantities. In non-dimensional form, the governing equations are
\begin{eqnarray}
\bnabla \bcdot \boldsymbol{u} = 0, 
\label{Problem setup: div-free condition}
\\
\frac{\partial \boldsymbol{u}}{\partial t} + \boldsymbol{u} \bcdot \bnabla \boldsymbol{u} = - \bnabla p + \frac{1}{\Rey}\nabla^2 \boldsymbol{u},
\label{Problem setup: momentum equation}
\end{eqnarray}
where 
\begin{eqnarray}
Re = \frac{\Omega R_i d}{\nu}
\label{Problem setup: The Reynolds number}
\end{eqnarray}
is the Reynolds numbers which, along with the radius ratio $\eta$, fully characterizes the flow field. Note that instead of the Reynolds number, one can also use the Taylor number
\begin{eqnarray}
Ta = \frac{(1+\eta)^4}{64 \eta^2} \frac{d^2 (R_i + R_o)^2 \Omega^2}{\nu^2} = \frac{(1+\eta)^6}{64 \eta^4} \Rey^2,
\label{Problem setup: The Taylor number}
\end{eqnarray}
to characterize the flow field.
We use a cylindrical coordinate system $(r, \theta, z)$. The boundary conditions are
\begin{eqnarray}
(u_r, u_{\theta}, u_z) = (0, 1, 0) \quad \text{at} \quad r = r_i, 
\label{Problem setup: The boundary condition inner}
\\
(u_r, u_{\theta}, u_z) = (0, 0, 0) \quad \text{at} \quad r = r_o,
\label{Problem setup: The boundary condition outer}
\end{eqnarray}
where $r_i$ and $r_o$ are the non-dimensional inner and outer cylinder radii. In this paper, we will assume that the flow is periodic in the spanwise direction $z$ with non-dimensional length $L$. The domain of interest, denoted by $V$, is therefore given by
\begin{eqnarray}
V = \{(r, \theta, z)| r_i \leq r \leq r_o, 0 \leq \theta < 2 \pi, 0 \leq z < L \}.
\label{Problem setup: The domain of interest}
\end{eqnarray}

At sufficiently small Reynolds numbers, or equivalently, at small Taylor numbers, the flow is laminar and can be expressed as
\begin{eqnarray}
\boldsymbol{u}_{lam} = \frac{1}{1-\eta^2} \left(\frac{r_i}{r} - \frac{r r_i}{r_o^2}\right) \boldsymbol{e}_{\theta}.
\label{Problem setup: The laminar flow-field}
\end{eqnarray}
Before proceeding further, it is useful to introduce a few convenient notations. We use angle brackets for the volume integration and overbar for the long-time average of a quantity:
\begin{eqnarray}
\left\langle [\; \cdot \;] \right\rangle = \int_{V} [\; \cdot \;] \; d \boldsymbol{x}, \quad \overline{[\; \cdot \;]} = \lim_{T \to \infty} \frac{1}{T}\int_{t = 0}^{T}  [\; \cdot \;] \; dt.
\label{Problem setup: Averaged quantity}
\end{eqnarray}
The $L^2$-norm of a quantity is henceforth denoted as
\begin{eqnarray}
\left\|[\;\cdot\;]\right\|_2 = \left\langle  [\; \cdot \;]^2\right\rangle^{\frac{1}{2}}.
\label{Problem setup: L^2 norm}
\end{eqnarray}

In what follows, the three quantities that we are interested in bounding are the energy dissipation rate, the torque and the equivalent of a Nusselt number (defined based on the transverse current of azimuthal velocity). These quantities are not independent, as we now demonstrate. We start by writing the dimensional expression of the time-averaged torque required to rotate the inner cylinder:
\begin{eqnarray}
G^\ast = - R_i \times \int_{0}^{L^\ast} \int_{0}^{2 \pi} \left. \overline{\tau_{r \theta}^\ast} \; \right|_{r^\ast = R_i} \; R_i d\theta^\ast dz^\ast, 
\label{Bulk quantities of interest: dimensional torque}
\end{eqnarray}
where $\tau_{r \theta}^\ast$ denotes the shear-stress. In non-dimensional form the torque is given by
\begin{eqnarray}
G = \frac{G^\ast}{\rho \nu^2 L^\ast} = - \frac{\Rey r_i^2}{L} \int_{0}^{L} \int_{0}^{2 \pi} \left[ \; \overline{\frac{1}{r} \frac{\partial u_r}{\partial \theta} + \frac{\partial u_\theta}{\partial r} - \frac{u_\theta}{r}} \; \right]_{r = r_i} \; d\theta dz.
\label{Bulk quantities of interest: non-dimensional torque}
\end{eqnarray}
In a statistically stationary state, the work done by the torque to rotate the inner cylinder eventually dissipates in the fluid, i.e.,
\begin{eqnarray}
G^\ast \Omega = \varepsilon^\ast,
\label{Bulk quantities of interest: work balance identity}
\end{eqnarray}
where $\varepsilon^\ast$ is the time-averaged total dissipation given by
\begin{eqnarray}
\varepsilon^\ast = 2 \rho \nu \int_{V^\ast} \overline{ \bnabla^\ast \boldsymbol{u}^\ast \boldsymbol{:} \bnabla^\ast  \boldsymbol{u}^\ast}_{sym} \; \; d \boldsymbol{x}^\ast, 
\label{Bulk quantities of interest: dimensional energy dissipation}
\end{eqnarray}
where
\begin{eqnarray}
\bnabla^\ast \boldsymbol{u}^\ast_{sym} = \frac{\bnabla^\ast  \boldsymbol{u}^\ast + \bnabla^\ast \boldsymbol{u}^{\ast T}}{2}.
\label{Bulk quantities of interest: def. strain rate tensor}
\end{eqnarray}
The total kinetic energy of the fluid can be shown to be uniformly bounded in time within the framework of the background method \citep[see][for example]{PhysRevLett.69.1648}. The identity (\ref{Bulk quantities of interest: work balance identity}) can therefore be obtained by taking the long-time average of the evolution equation of the total kinetic energy.
The dissipation per unit mass non-dimensionalized by $\Omega^3 R_i^3 / d$ is given by
\begin{eqnarray}
\varepsilon = \frac{\varepsilon^\ast}{\Omega^3 R_i^3 / d} = \frac{2}{(\pi r_o^2 - \pi r_i^2) L \; \Rey} \overline{ \left\langle \bnabla \boldsymbol{u} \boldsymbol{:} \bnabla \boldsymbol{u}_{sym} \right \rangle}.
\label{Bulk quantities of interest: non-dimensional energy dissipation}
\end{eqnarray}
From the divergence-free condition (\ref{Problem setup: div-free condition}), the boundary conditions (\ref{Problem setup: The boundary condition inner}) and (\ref{Problem setup: The boundary condition outer}) along with the use of the divergence theorem, one finds that
\begin{eqnarray}
\left\langle \bnabla \boldsymbol{u} \boldsymbol{:} \bnabla \boldsymbol{u}^T \right\rangle = \left\langle \bnabla \bcdot \bnabla \bcdot (\boldsymbol{u} \otimes \boldsymbol{u})  \right\rangle = 2 \upi L.
\label{Bulk quantities of interest: inter identity}
\end{eqnarray}
As a result, the non-dimensional dissipation can also be written as
\begin{eqnarray}
\varepsilon = \frac{1}{(\pi r_o^2 - \pi r_i^2) \Rey} \left[\frac{1}{L} \overline{\|\bnabla \boldsymbol{u}\|_2^2} + 2 \upi\right].
\label{Bulk quantities of interest: non-dimensional energy dissipation 2}
\end{eqnarray}
Using (\ref{Bulk quantities of interest: non-dimensional torque}), (\ref{Bulk quantities of interest: work balance identity}) and (\ref{Bulk quantities of interest: non-dimensional energy dissipation}), we finally obtain a relation between the non-dimensional torque and the non-dimensional dissipation as
\begin{eqnarray}
G = \pi (r_i + r_o) r_i \Rey^2 \varepsilon.
\label{Bulk quantities of interest: G epsilon relation}
\end{eqnarray}
which is the non-dimensional version of (\ref{Bulk quantities of interest: work balance identity}).

Another quantity of interest is the transverse current of azimuthal velocity, defined as
\begin{eqnarray}
J^{\omega \ast} = \frac{1}{2 \pi L^\ast}\int_{0}^{L^\ast} \int_{0}^{2 \pi} r^{\ast 3}\left[ \overline{u_r^\ast \omega^\ast} - \nu \partial_{r^\ast} \overline{\omega^\ast}\right] \; r^\ast d\theta^\ast dz^\ast,
\label{Bulk quantities of interest: J omeaga}
\end{eqnarray}
where $\omega^\ast = u_\theta^\ast / r^\ast$ is the local angular velocity. As shown by \citet{eckhardt2007torque}, $J^{\omega \ast}$ is independent of the radial direction. In an analogy with Rayleigh--B\'enard convection, one defines the Nusselt number as the ratio of the transverse current of azimuthal velocity to its corresponding value in the laminar regime, i.e.
\begin{eqnarray}
Nu = \frac{J^{\omega \ast}}{J^{\omega \ast}_{lam}}.
\label{Bulk quantities of interest: Nusselt number}
\end{eqnarray}
Substituting $r^\ast = R_i$ in the right-hand-side of (\ref{Bulk quantities of interest: J omeaga}), one obtains the following relation between the torque and the transverse current of azimuthal velocity:
\begin{eqnarray}
J^{\omega \ast} = \frac{G^\ast}{2 \pi L^\ast \rho},
\end{eqnarray}
implying that the Nusselt number can also be written as
\begin{eqnarray}
Nu = \frac{G}{G_{lam}} = \frac{\varepsilon}{\varepsilon_{lam}},
\label{Bulk quantities of interest: G epsilon Nu relation}
\end{eqnarray}
where $G_{lam}$ and $\varepsilon_{lam}$ are the values of the non-dimensional torque and dissipation in the laminar regime, respectively.

\section{Energy stability analysis}
\label{Energy stability analysis}
We begin by discussing the energy stability of the laminar flow $\boldsymbol{u}_{lam}$. The importance of energy stability analysis in the context of bounding theories comes from the fact that bounds on mean quantities introduced in the last section are by definition saturated by the laminar state below the energy stability threshold. The energy stability of the laminar Taylor--Couette flow has been studied before both theoretically and numerically, by e.g. \citet{Serrin1959stability} and \citet{ddjoseph1976}. In these studies, the general conclusion was that at the energy stability threshold, the least stable perturbations are axisymmetric Taylor vortices. However, as we shall demonstrate in this section, this commonly accepted result does not hold below a certain radius ratio ($\eta < 0.0556$). Instead, we find that the least stable perturbations at the energy stability threshold in that case are fully three-dimensional.

We begin by defining the functional
\begin{eqnarray}
\mathcal{H} (\boldsymbol{\tilde{v}})  = \left[\frac{1}{2 \Rey} \| \bnabla \boldsymbol{\tilde{v}}\|_2^2 +  \int_{V} \boldsymbol{\tilde{v}} \bcdot (\bnabla \boldsymbol{u}_{lam})_{sym} \bcdot \boldsymbol{\tilde{v}} \; d \boldsymbol{x}\right],
\end{eqnarray}
where $\boldsymbol{\tilde{v}}$ is a perturbation over the laminar flow which satisfies the homogeneous boundary conditions at the inner and outer cylinders . From the governing equations, one can show that the laminar flow $\boldsymbol{u}_{lam}$ is energy stable when $\mathcal{H} (\boldsymbol{\tilde{v}})$ is nonnegative. We shall consider three types of constraints on the perturbations $\boldsymbol{\tilde{v}}$: no constraints, other than  the homogeneous boundary conditions (case 1), 3D incompressible perturbations (case 2) and 2D ($z$-invariant) incompressible perturbations (case 3). We perform an energy stability analysis for each of these cases, and present the results as a function of the radius ratio.

The critical Taylor number $Ta_c$ defining the energy stability threshold is the largest Taylor number for which the functional $\mathcal{H} (\boldsymbol{\tilde{v}})$ is nonnegative. For clarity, we add superscripts and use the notation $Ta_c^{nc}$, $Ta_c^{3D}$ and $Ta_c^{2D}$ when referring to case 1, case 2 and case 3, respectively. The statement of the nonnegativity of the functional $\mathcal{H} (\boldsymbol{\tilde{v}})$ can be posed as a convex optimization problem, where we require that the minimum value of $\mathcal{H}$ to be nonnegative. Then , it can be shown using the corresponding Euler--Lagrange equations that the nonnegativity of the functional $\mathcal{H} (\boldsymbol{\tilde{v}})$ is equivalent to the nonnegativity of the smallest eigenvalue in the eigenvalue problem
\begin{subequations}
\begin{eqnarray}
\bnabla \bcdot \boldsymbol{\tilde{v}} = 0, \\
2 \lambda \boldsymbol{\tilde{v}} = \frac{1}{\Rey}  \nabla^2 \boldsymbol{\tilde{v}} - 2 \boldsymbol{\tilde{v}} \bcdot \bnabla (\boldsymbol{u}_{lam})_{sym} - \bnabla \tilde{p}.
\end{eqnarray}
\label{Energy stability analysis: eigenvalue problem}
\end{subequations}
Note that for case 1, the eigenvalue problem corresponds just to equation (\ref{Energy stability analysis: eigenvalue problem}b) without the pressure term.   


\begin{figure}
\centering
\psfrag{A}[][][0.9]{$0$}
\psfrag{B}[][][0.9]{$0.1$}
\psfrag{C}[][][0.9]{$0.2$}
\psfrag{D}[][][0.9]{$0.3$}
\psfrag{E}[][][0.9]{$0.4$}
\psfrag{F}[][][0.9]{$0.5$}
\psfrag{G}[][][0.9]{$0.6$}
\psfrag{H}[][][0.9]{$0.7$}
\psfrag{I}[][][0.9]{$0.8$}
\psfrag{J}[][][0.9]{$0.9$}
\psfrag{K}[][][0.9]{$1$}
\psfrag{L}[][][0.9]{$0.02$}
\psfrag{O}[][][0.9]{$0.04$}
\psfrag{P}[][][0.9]{$0.06$}
\psfrag{7}[][][0.9]{$0.01$}
\psfrag{8}[][][0.9]{$0.03$}
\psfrag{9}[][][0.9]{$0.05$}
\psfrag{!}[][][0.9]{$0.07$}
\psfrag{1}[][][0.9]{$10^2$}
\psfrag{R}[][][0.9]{$10^3$}
\psfrag{S}[][][0.9]{$10^4$}
\psfrag{T}[][][0.9]{$10^5$}
\psfrag{U}[][][0.9]{$10^6$}
\psfrag{V}[][][0.9]{$10^7$}
\psfrag{W}[][][0.9]{$10^8$}
\psfrag{X}[][][0.9]{$10^9$}
\psfrag{Y}[][][0.9]{$10^{10}$}
\psfrag{Z}[][][0.9]{$10^{11}$}
\psfrag{Q}[][][0.9]{$10^{12}$}
\psfrag{M}[][][0.9]{$\eta$}
\psfrag{N}[][][0.9][180]{$Ta_c$}
\psfrag{2}[][][0.9]{$ \substack{0.0556 \\ \uparrow} $}
\psfrag{\$}[][][0.9]{$ \substack{0.0188 \\ \uparrow} $}
\psfrag{#}[l][][0.7]{$\to 1.86 \times 10^{10}$}
\psfrag{3}[l][][0.7]{$\to 5.8 \times 10^7$}
\psfrag{4}[l][][0.7]{$\to 31641$}
\psfrag{5}[l][][0.7]{$\to 6831$}
\psfrag{6}[l][][0.7]{$\to 389.6$}
\psfrag{a}[][][0.9]{$0$}
\psfrag{b}[][][0.9]{$10$}
\psfrag{c}[][][0.9]{$20$}
\psfrag{d}[][][0.9]{$30$}
\psfrag{e}[][][0.9]{$40$}
\psfrag{f}[][][0.9]{$50$}
\psfrag{g}[][][0.9]{$60$}
\psfrag{h}[][][0.9]{$70$}
\psfrag{i}[][][0.9]{$80$}
\psfrag{j}[][][0.9]{$90$}
\psfrag{k}[][][0.9]{$10^{-4}$}
\psfrag{l}[][][0.9]{$10^{-3}$}
\psfrag{m}[][][0.9]{$10^{-2}$}
\psfrag{o}[][][0.9]{$10^{-1}$}
\psfrag{p}[][][0.7]{$0.33$}
\psfrag{n}[][][0.9][180]{$Ta_c^{2D}/Ta_c^{nc}$}
\psfrag{q}[][][0.9][180]{$Ta_c^{2D}/Ta_c^{nc}-1$}
\psfrag{r}[][][0.9]{$2$}
\psfrag{s}[][][0.9]{$4$}
\psfrag{t}[][][0.9]{$6$}
\psfrag{u}[][][0.9]{$8$}
\psfrag{v}[][][0.9]{$12$}
\psfrag{w}[][][0.9]{$14$}
\psfrag{x}[][][0.9]{$16$}
\psfrag{y}[][][0.9]{$18$}
\psfrag{z}[][][0.9][180]{$Ta_c^{3D}/Ta_c^{nc}$}
\psfrag{*}[][][0.7]{$0.29$}
\psfrag{?}[][][0.9][180]{$Ta_c^{3D}/Ta_c^{nc}-1$}
\begin{tabular}{lc}
\begin{subfigure}{0.5\textwidth}
\centering
 \includegraphics[scale = 0.9]{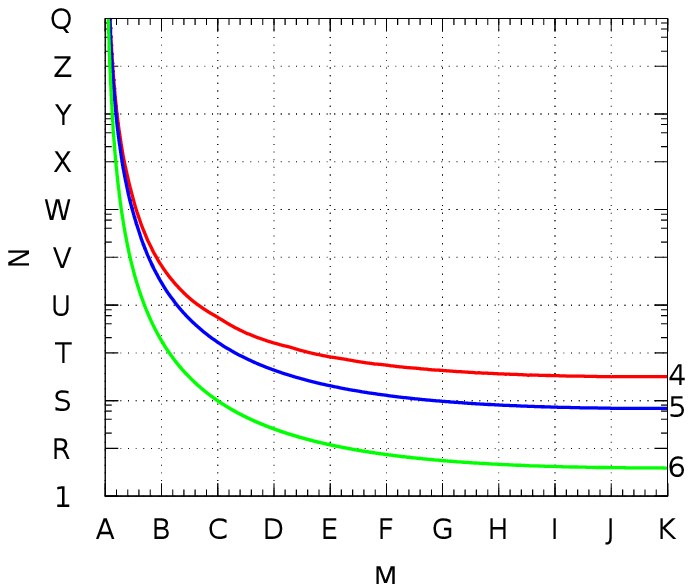}
\caption{}
\end{subfigure} &
\begin{subfigure}{0.5\textwidth}
\centering
 \includegraphics[scale = 0.9]{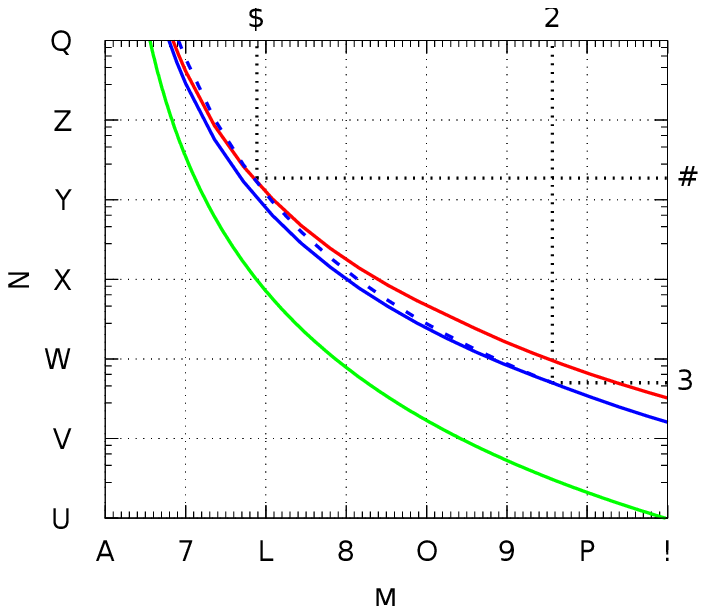}
\caption{}
\end{subfigure}
\end{tabular}
\begin{tabular}{lc}
\begin{subfigure}{0.5\textwidth}
\centering
 \includegraphics[scale = 0.9]{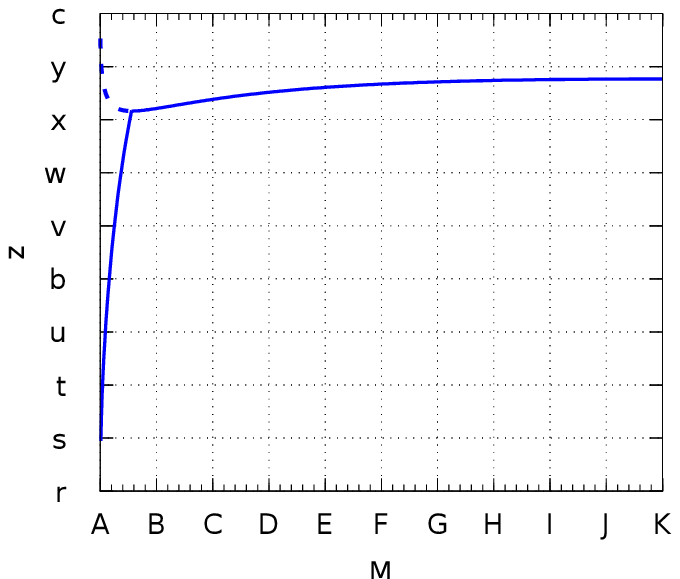}
\caption{}
\end{subfigure} &
\begin{subfigure}{0.5\textwidth}
\centering
 \includegraphics[scale = 0.9]{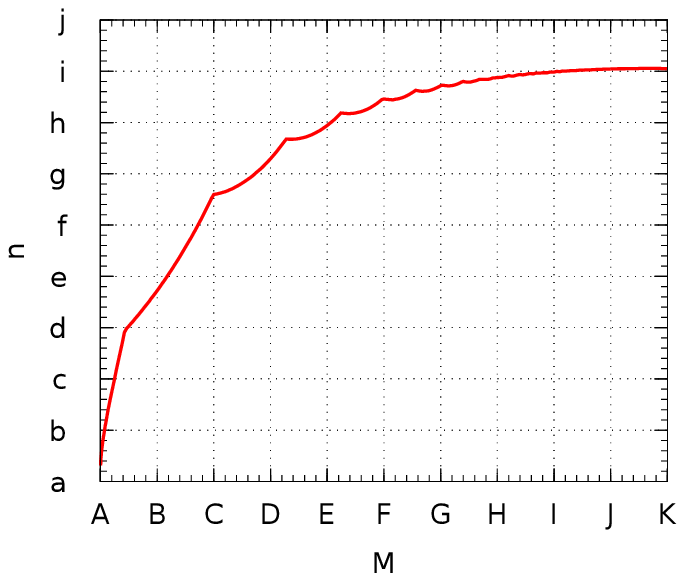}
\caption{}
\end{subfigure}
\end{tabular}
 \caption{Panel (a) shows the critical Taylor number $Ta^{nc}_c$ (green line), $Ta^{3D}_c$ (blue line) and $Ta^{2D}_c$ (red line) as a function of the radius ratio $\eta$ and (b) shows a close-up view of the same plot for small $\eta$. The dashed blue line corresponds to the marginally stable axisymmetric Taylor vortices, while $Ta^{3D}_c$  is continued to be shown with the solid blue line. Panels (c) and (d) shows the critical Taylor number $Ta^{3D}_c$ and $Ta^{2D}_c$ normalized by $Ta^{nc}_c$ as a function of $\eta$. }
 \label{Energy stability analysis: energy stability analysis main figure}
\end{figure}

We can actually obtain the critical Taylor number analytically for case 1. Indeed, in this case, we first simplify the eigenvalue problem using two pieces of information. From lemma \ref{lemma of only radial functions} (see appendix \ref{Analytical solution case 1 moma}), we note that the least stable perturbed flow (which optimizes $\mathcal{H}$) is a function of  the radial direction only.  Furthermore, the laminar flow $\bs{u}_{lam}$ satisfies the required condition in lemma  \ref{lemma of only radial functions}, therefore, the least stable perturbation also satisfies $\tilde{v}_r = \tilde{v}_\theta$. Using these two facts, we find that the marginally stable solution of (\ref{Energy stability analysis: eigenvalue problem}b) which satisfies the homogeneous boundary condition at $r = r_i$ is given by
\begin{eqnarray}
\tilde{v}_r = \tilde{v}_\theta = c \sin \left(\xi \log \frac{r}{r_i}\right), \qquad \xi = \sqrt{\frac{ \eta}{(1+\eta)(1-\eta)^2} \Rey + 1}.
\label{Energy stability analysis: Marginally stable flow nc}
\end{eqnarray}
The critical Reynolds number for energy stability is the smallest value of $\Rey$ for which this solution also satisfies the homogeneous boundary condition at $r = r_o$. We then obtain the critical Taylor number using (\ref{Problem setup: The Taylor number}), which leads to
\begin{eqnarray}
Ta_c^{nc} = \frac{(1 + \eta)^8 (1-\eta)^4}{64 \eta^6} \left(1 + \frac{\upi^2}{\log^2 \eta}\right)^2.
\label{Energy stability analysis: Critical Taylor number nc}
\end{eqnarray}
In case 2 (3D incompressible $\tilde{\bs{v}}$) and case 3 (2D ($z$-invariant) incompressible $\tilde{\bs{v}}$), we must turn to numerical computations to calculate the critical Taylor number. To find the eigenvalues of the equations (\ref{Energy stability analysis: eigenvalue problem}), we first transform the equations into a generalized eigenvalue problem using the spatial discretization described in \S \ref{Optimal bounds: Numerical scheme} and then use the DGGEV routine by Lapack for the computation. Let's call the critical wavenumbers of the least stable perturbation at the energy stability threshold $2 \upi/ L_c$ (where $L_c$ would then be known as the critical aspect ratio) in the $z$-direction and $m_c$ in the $\theta$-direction. We use the bisection algorithm in the Taylor number and the ternary search algorithm in aspect ratio or azimuthal wavenumber (depending on the case at hand) to accurately determine $Ta_c$, $L_c$  and $m_c$.

\begin{figure}
\centering
\psfrag{a}[][][0.9]{$1.5$}
\psfrag{0}[][][0.9]{$2$}
\psfrag{1}[][][0.9]{$2.5$}
\psfrag{2}[][][0.9]{$3$}
\psfrag{3}[][][0.9]{$3.5$}
\psfrag{4}[][][0.9]{$4$}
\psfrag{5}[][][0.9]{$4.5$}
\psfrag{6}[][][0.9]{$5$}
\psfrag{R}[][][0.9]{$0$}
\psfrag{S}[][][0.9]{$5$}
\psfrag{T}[][][0.9]{$10$}
\psfrag{U}[][][0.9]{$15$}
\psfrag{V}[][][0.9]{$20$}
\psfrag{W}[][][0.9]{$25$}
\psfrag{X}[][][0.9]{$30$}
\psfrag{Y}[][][0.9]{$35$}
\psfrag{Z}[][][0.9]{$40$}
\psfrag{A}[][][0.9]{$0$}
\psfrag{B}[][][0.9]{$0.1$}
\psfrag{C}[][][0.9]{$0.2$}
\psfrag{D}[][][0.9]{$0.3$}
\psfrag{E}[][][0.9]{$0.4$}
\psfrag{F}[][][0.9]{$0.5$}
\psfrag{G}[][][0.9]{$0.6$}
\psfrag{H}[][][0.9]{$0.7$}
\psfrag{I}[][][0.9]{$0.8$}
\psfrag{J}[][][0.9]{$0.9$}
\psfrag{K}[][][0.9]{$1$}
\psfrag{b}[][][0.9]{$m_c = 1$}
\psfrag{c}[][][0.9]{$m_c = 0$}
\psfrag{M}[][][0.9]{$\eta$}
\psfrag{N}[][][0.9][180]{$2 \upi / L_c$}
\psfrag{O}[][][0.9][180]{$m_c$}
\begin{tabular}{lc}
\begin{subfigure}{0.5\textwidth}
\centering
 \includegraphics[scale = 0.9]{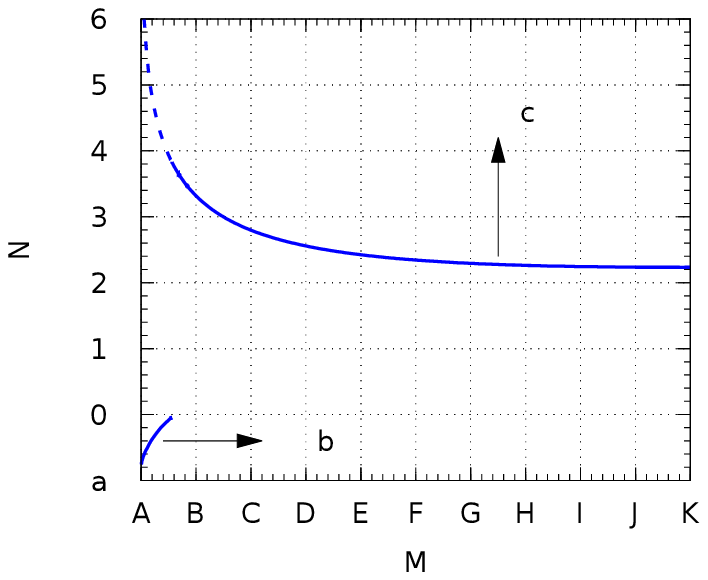}
\caption{}
\end{subfigure} &
\begin{subfigure}{0.5\textwidth}
\centering
 \includegraphics[scale = 0.9]{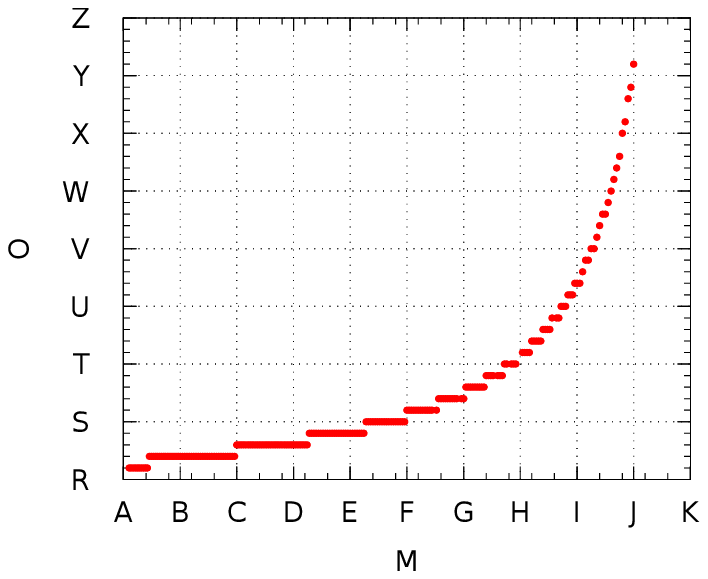}
\caption{}
\end{subfigure}
\end{tabular}
\caption{Variation of the critical axial wavenumber $2 \upi / L_c$ and critical azimuthal wavenumber $m_c$ with radius ratio $\eta$ for (a) case 2 and (b) case 3. In panel (a), the critical azimuthal wavenumber changes from $m_c = 0$ above $\eta = \eta_s = 0.0556$ to $m_c = 1$ below $\eta_s$, as discussed in the main text. }
 \label{Energy stability analysis: critical wavenumbers}
\end{figure}

The dependence of  the critical Taylor number for energy stability on the radius-ratio $\eta$ is shown in figure \ref{Energy stability analysis: energy stability analysis main figure} for all 3 cases. The critical axial wavenumber ($2 \upi/ L_c$) and the critical azimuthal wavenumber ($m_c$) of the corresponding perturbations in case 2 and case 3 are shown in figure \ref{Energy stability analysis: critical wavenumbers}. From figure \ref{Energy stability analysis: energy stability analysis main figure}a, we see that the critical Taylor number increases as we go from case 1 (green line) to case 3 (red line), which is not surprising since we correspondingly  increase the number of constraints on the perturbations. In all three cases, the critical Taylor number monotonically increases with decreasing $\eta$ and tends to infinity as $\eta \to 0$. By contrast, the critical Taylor number tends to a constant in the small gap width limit ($\eta \to 1$): in case 1 $Ta_c^{nc} \to 4 \upi^4 \approx 389.6364$, whereas, in case 2 and case 3, $Ta_c^{3D} \to 6831$ and $Ta_c^{2D} \to 31641$, which are, respectively, $17.5$ and $81.2$ times larger than in case 1. In this limit, the marginally stable perturbation in case 2 recovers the well-known axisymmetric Taylor vortices \citep{Serrin1959stability, ddjoseph1976}. In case 3, the marginally stable perturbation is composed of vortices whose axis is parallel to the cylinder axis \citep{harrison1921stability}.

Figure \ref{Energy stability analysis: energy stability analysis main figure}b shows a zoomed-in version of figure \ref{Energy stability analysis: energy stability analysis main figure}a for small values of $\eta$. We also show, for case 2 (blue line), a separate curve that assumes that perturbations are axially symmetric (dashed blue line). For large radius ratio, the two are identical, confirming that the axisymmetric Taylor vortices are indeed the least stable perturbations. However, we note that below radius ratio $\eta_{s} = 0.0556$, the marginally stable perturbation switches from the axisymmetric Taylor vortices to being fully three-dimensional.

\begin{figure}
\centering
\begin{tabular}{lc}
\begin{subfigure}{0.5\textwidth}
\centering
 \includegraphics[scale = 0.08]{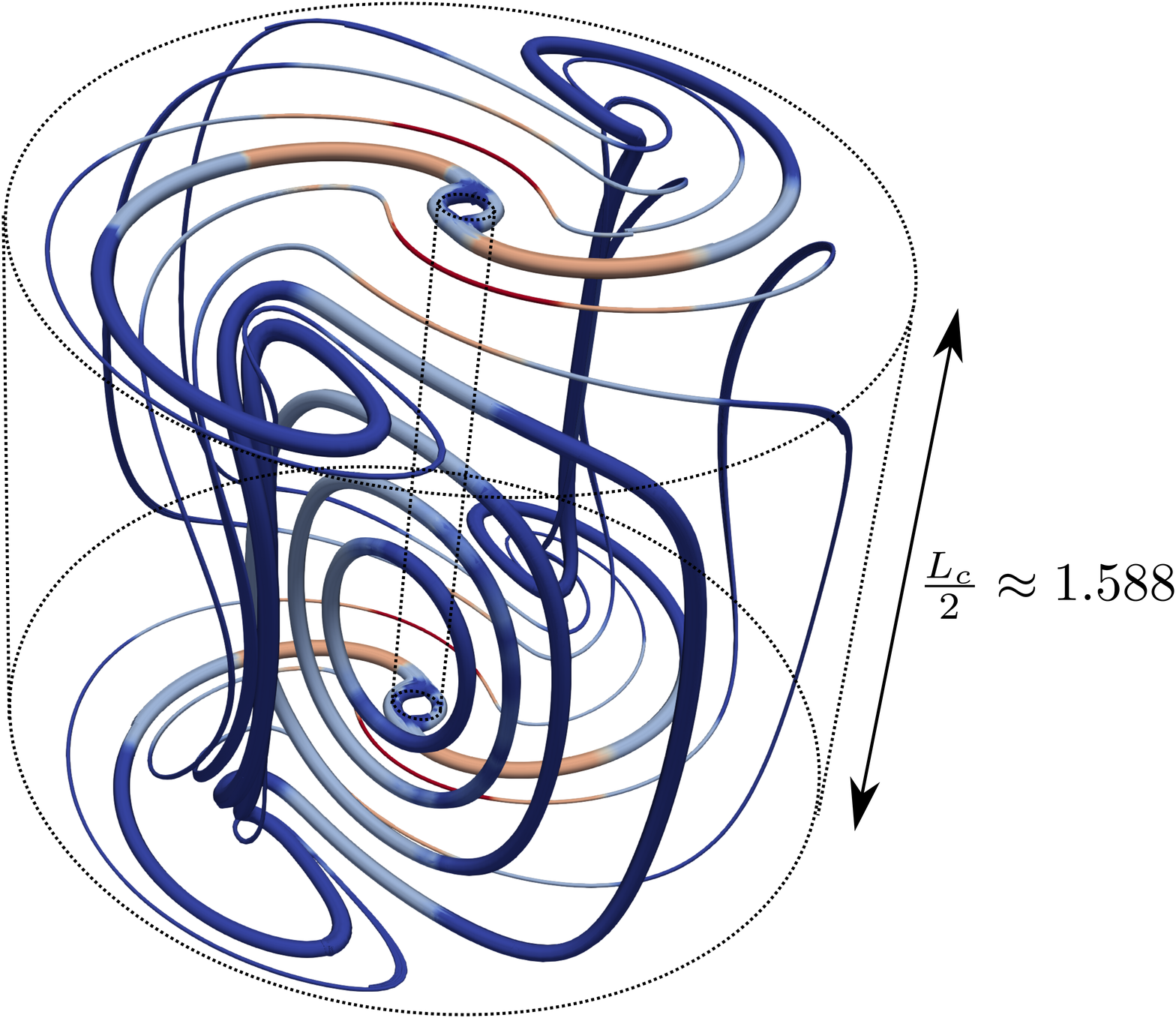}
\caption{}
\end{subfigure} &
\begin{subfigure}{0.5\textwidth}
\centering
 \includegraphics[scale = 0.07]{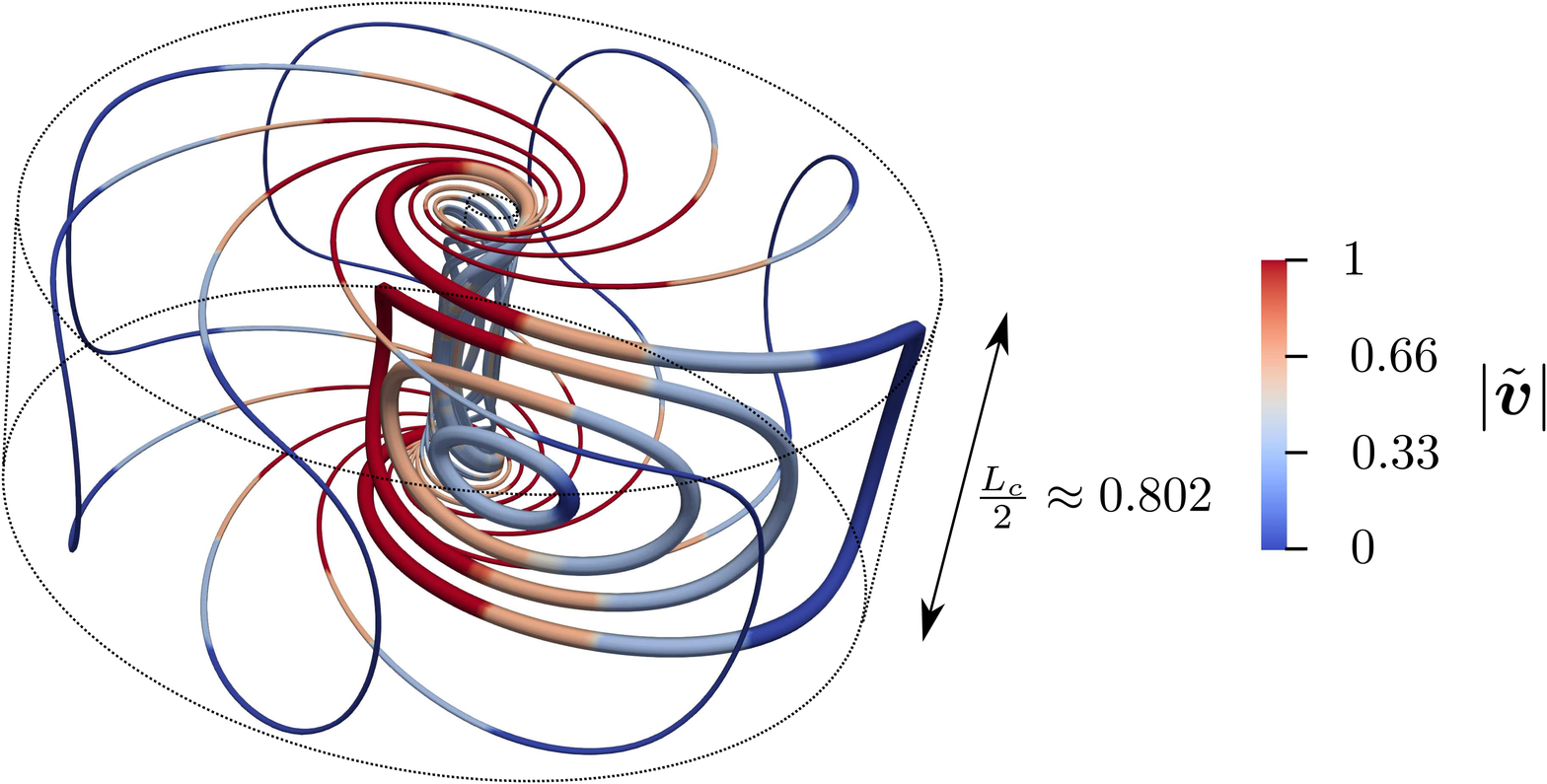}
\caption{}
\end{subfigure}
\end{tabular}
 \caption{Panel (a) shows selected streamlines of the marginally stable 3D flow and panel (b) shows selected streamlines of the marginally stable axisymmetric Taylor vortices, in both cases at a radius ratio ($\eta_s = 0.0556$). The corresponding critical Taylor numbers in both  cases are equal. The streamlines are colored according to the magnitude of the velocity. In both the cases the velocity field has been scaled such that the maximum magnitude is $1$. A typical vortex is shown using relatively thicker lines in both cases. Note that only half the vortex is shown in the axial direction. }
 \label{Energy stability analysis: marginally stable flow}
\end{figure}

Figure \ref{Energy stability analysis: marginally stable flow} shows the marginally stable 3D flow and Taylor vortices at $\eta = \eta_s$. A distinctive feature of the marginally stable 3D flow, compared to marginally stable axisymmetric Taylor vortices, is that one end of a typical vortex lies near the outer cylinder but the other end lies at one of the two lines that are offset from the inner cylinder. Also, the critical aspect ratio corresponding to marginally stable 3D flow is larger than the one corresponding to the Taylor vortices. In fact, with further decrease in the radius-ratio, the axisymmetric critical aspect-ratio corresponding to the marginally stable 3D flow grows, whereas the one corresponding to Taylor vortices shrinks, as can been seen from figure \ref{Energy stability analysis: critical wavenumbers}a. The decrease of the aspect ratio of the critical perturbations implies that the term $\|\bnabla \boldsymbol{\tilde{v}}\|_2^2$ increases rapidly as $\eta \to 0$, which causes the corresponding critical Taylor number for axisymmetric flows to do the same. This explains why the axisymmetric perturbations are no longer preferred for very low $\eta$.  At $\eta = 0.0188$, the critical Taylor number for the marginally stable Taylor vortices becomes even larger than the one corresponding to the two-dimensional flow ($Ta_c^{2D}$).  

Given that we were able to compute the critical Taylor number in case 1 analytically as a function of $\eta$, it is worth investigating whether the dependence of $Ta_c$ on $\eta$ in cases 2 and 3 is similar to that of case 1. To do so, we look at the figures \ref{Energy stability analysis: energy stability analysis main figure}c and \ref{Energy stability analysis: energy stability analysis main figure}d, which show the ratios  $Ta_c^{3D}/Ta_c^{nc}$  and $Ta_c^{2D}/Ta_c^{nc}$ respectively. One striking observation is that $Ta_c^{3D}/Ta_c^{nc}$ remains within $3.6\%$ of $16.92$ for a fairly large range of radius ratio $0.0556 \leq \eta \leq 1$. So, for this range of $\eta$
\begin{eqnarray}
Ta_c^{3D} \approx \frac{16.92 (1 + \eta)^8 (1-\eta)^4}{64 \eta^6} \left(1 + \frac{\upi^2}{\log^2 \eta}\right)^2.
\label{Energy stability analysis: Critical Taylor number 3D above 0.22}
\end{eqnarray}
However, the same is not valid for case 3, where $Ta_c^{2D}/Ta_c^{nc}$ varies substantially with $\eta$. The spikes in figure \ref{Energy stability analysis: energy stability analysis main figure}d, which are not visible in figure \ref{Energy stability analysis: energy stability analysis main figure}a, correspond to the discrete change in critical azimuthal wavenumber when $\eta$ varies, shown in figure \ref{Energy stability analysis: critical wavenumbers}b.


For small radius ratio it is possible to predict the asymptotic behavior of $Ta_c^{3D}$ and $Ta_c^{2D}$. We find that both $Ta_c^{3D}/Ta_c^{nc}$ and $Ta_c^{2D}/Ta_c^{nc}$ decrease as $\eta \to 0$ as can be seen in \ref{Energy stability analysis: energy stability analysis main figure}c and \ref{Energy stability analysis: energy stability analysis main figure}d . By construction, the asymptotic value of the ratios have to be larger than $1$. Therefore, in the small radius ratio limit, we can obtain the asymptotic behavior of $Ta_c^{3D}$ and $Ta_c^{2D}$ as
\begin{eqnarray}
Ta_c^{3D} = C_0^{3D} \lim_{\eta \to 0} Ta_c^{nc} =  \frac{C_0^{3D} \upi^4}{\eta^6 \log^4 \eta}, \; Ta_c^{2D} =  C_0^{2D} \lim_{ \eta \to 0} Ta_c^{nc} =  \frac{C_0^{2D} \upi^4}{\eta^6 \log^4 \eta} \; \text{as } \eta \to 0,
\end{eqnarray}
where $1 \leq C_0^{3D}, C_0^{2D} < \infty$ are two constants. 




\section{An analytical bound}
\label{An analytical bound}
In this section, we obtain a simple, suboptimal, analytical bound on the torque, the rate of energy dissipation and the Nusselt number defined in \S \ref{Problem setup}. We use the well-known background method \citep{PhysRevLett.69.1648, PhysRevE.49.4087} whose exact formulation in the context of the present problem is given in appendix \ref{The background method formulation}. As usual, we define $\boldsymbol{U}$ to be the background flow and $\bs{v}$ to be the perturbed flow such that the total flow is $\bs{u} = \bs{U} + \bs{v}$. The background flow $\boldsymbol{U}$ is divergence-free and satisfies the same boundary conditions as $\bs{u}$, so the perturbed flow $\bs{v}$ satisfies the homogeneous version of the boundary conditions. For mathematical convenience (see appendix \ref{The background method formulation}) we further define the so-called ``shifted perturbation'' $\boldsymbol{\tilde{v}} = \bs{v} - \bs{\phi}$ (see equation \ref{The background method: modified perturbation})
 and we simply refer to $\boldsymbol{\tilde{v}}$ as the perturbation from here onward. As shown in appendix \ref{The background method formulation},  a bound on the rate of energy dissipation, 
\begin{eqnarray}
\varepsilon \leq \frac{1}{(\pi r_o^2 - \pi r_i^2) \Rey L} \left[\frac{1}{a (2 -a) }  \| \bnabla \boldsymbol{U}\|_2^2 - \frac{(1-a)^2}{a(2-a)} \| \bnabla \boldsymbol{u}_{lam}\|_2^2 + 2 \upi L\right],
\label{An analytical bound: non-dimensional energy dissipation 2}
\end{eqnarray}
can be obtained for any choice of the background flow for which the functional
\begin{eqnarray}
\mathcal{H} (\boldsymbol{\tilde{v}})  = \frac{2-a}{2 \Rey} \| \bnabla \boldsymbol{\tilde{v}}\|_2^2 +  \underbrace{\int_{V} \boldsymbol{\tilde{v}} \bcdot \bnabla \boldsymbol{U}_{sym} \bcdot \boldsymbol{\tilde{v}} \; d \boldsymbol{x}}_{II},
\label{An analytical bound: the functional H}
\end{eqnarray}
(see \ref{The background method: the functional H vtilde}) is positive semi-definite. In (\ref{An analytical bound: non-dimensional energy dissipation 2}), the constant $a$ is a balance parameter (see appendix \ref{The background method formulation} for more details) that takes values between $0$ and $2$. While showing $\mathcal{H} (\boldsymbol{\tilde{v}})$ is nonnegative, we do not impose the incompressibility constraint on the perturbations $\boldsymbol{\tilde{v}}$ and only assume $\boldsymbol{\tilde{v}}$ satisfies the homogeneous boundary conditions. We make a choice of the background flow $\boldsymbol{U}$ for which $$\bnabla \boldsymbol{U}_{sym} = \frac{\bnabla  \boldsymbol{U} + \bnabla \boldsymbol{U}^{ T}}{2}$$ is non-zero only in boundary layers, which are assumed to have thicknesses $\delta_i$ and $\delta_o$ near the inner and the outer cylinder, respectively. In particular, the selected background flow $\boldsymbol{U}$ is then
\begin{eqnarray}
\boldsymbol{U}(r, \theta, z) = U(r) \boldsymbol{e}_{\theta} =
\begin{cases}
\frac{\Lambda (r_i + \delta_i) (r - r_i) - (r - r_i - \delta_i)}{\delta_i} \boldsymbol{e}_{\theta} \qquad \text{if} \quad r_i \leq r \leq r_i + \delta_i, \\
\Lambda r \boldsymbol{e}_{\theta} \qquad \text{if} \quad r_i + \delta_i < r \leq r_o - \delta_o, \\
\frac{\Lambda  (r_o - \delta_o) (r_o - r)}{\delta_o} \boldsymbol{e}_{\theta} \qquad \text{if} \quad r_o - \delta_o < r \leq r_o, \\
\end{cases}
\label{An analytical bound: definition background flow}
\end{eqnarray}
where $\Lambda$ is an $O(1)$ constant, i.e., independent of $Re$. The decision to allow for different boundary layer thicknesses is inspired from the work of \citet{kumar2020pressure}, who speculated in the context of helical pipe flows that by doing so, it is possible to capture  important geometrical aspects of problem that would otherwise not appear. As we are primarily interested in deriving bounds at asymptotically  high Reynolds numbers, for convenience, we define rescaled boundary layer thicknesses as 
\begin{eqnarray}
h_i = \frac{\delta_i}{\delta} \quad \text{and} \quad h_o = \frac{\delta_o}{\delta} \qquad \text{where} \quad \delta = \frac{1}{Re},
\end{eqnarray}
where, by construction, $h_i, h_o > 0$ and are $O(1)$. Our goal in this section is to adjust the relative size of the boundary layers ($h_i/h_o$) to optimize the bound (\ref{An analytical bound: non-dimensional energy dissipation 2}) simultaneously for different values of $\eta$ in the limit of high Reynolds number. 

We start by obtaining a simple estimate for the quantity
\begin{eqnarray}
\int_{r_i}^{r_i + \delta_i} |\tilde{v}_r| \; |\tilde{v}_\theta| dr && = \int_{r_i}^{r_i + \delta_i} \left| \int_{r_i}^{r} \frac{\partial \tilde{v}_r}{\partial r^\prime} dr^\prime \right| \; \left| \int_{r_i}^{r} \frac{\partial \tilde{v}_\theta}{\partial r^\prime} dr^\prime \right| dr \nonumber \\
&& \leq   \int_{r_i}^{r_i + \delta_i} (r - r_i)  \left[\int_{r_i}^{r_i + \delta_i} \left(\frac{\partial \tilde{v}_r}{\partial r^\prime} \right)^2 dr^\prime \right]^{1/2} \; \left[\int_{r_i}^{r_i + \delta_i} \left(\frac{\partial \tilde{v}_\theta}{\partial r^\prime} \right)^2  dr^\prime \right]^{1/2} dr \nonumber \\
&& = \frac{\delta_{i}^2}{2} \left[\int_{r_i}^{r_i + \delta_i} \left(\frac{\partial \tilde{v}_r}{\partial r^\prime} \right)^2 dr^\prime \right]^{1/2} \; \left[\int_{r_i}^{r_i + \delta_i} \left(\frac{\partial \tilde{v}_\theta}{\partial r^\prime} \right)^2  dr^\prime \right]^{1/2} \nonumber \\
&& \leq \frac{\delta_{i}^2}{4} \int_{r_i}^{r_i + \delta_i} \left(\frac{\partial \tilde{v}_r}{\partial r^\prime} \right)^2 dr^\prime + \frac{\delta_{i}^2}{4} \int_{r_i}^{r_i + \delta_i} \left(\frac{\partial \tilde{v}_\theta}{\partial r^\prime} \right)^2 dr^\prime.
\label{An analytical bound: inter II estimate 1}
\end{eqnarray}
In deriving the result, we have used the fundamental theorem of calculus in the first line, H\"{o}lder's inequality in the second line followed by an integration in $r$ to obtain the third line. Finally, we used Young's inequality to obtain the last line. In a similar manner, one can also show that
\begin{eqnarray}
\int_{r_o - \delta_o}^{r_o} |\tilde{v}_r| \; |\tilde{v}_\theta| dr \leq \frac{\delta_o^2}{4} \int_{r_o - \delta_o}^{r_o} \left(\frac{\partial \tilde{v}_r}{\partial r^\prime} \right)^2 dr^\prime + \frac{\delta_o^2}{4} \int_{r_o - \delta_o}^{r_o} \left(\frac{\partial \tilde{v}_\theta}{\partial r^\prime} \right)^2 dr^\prime.
\label{An analytical bound: inter II estimate 2}
\end{eqnarray}
Next, we note that
\begin{eqnarray}
\left| \int_{r = r_i}^{r_i + \delta_i} \tilde{v}_r \tilde{v}_{\theta} \left(\frac{d U}{dr} - \frac{U}{r}\right) rdr \right| 
\leq \max_{r_i < r < r_i + \delta_i} \left|\frac{d U}{dr} - \frac{U}{r}\right| (r_i + \delta_i) \int_{r = r_i}^{r_i + \delta_i} |\tilde{v}_r| \; |\tilde{v}_{\theta}| dr,
\label{An analytical bound: inter II estimate 3}
\end{eqnarray}
and
\begin{eqnarray}
\left| \int_{r = r_o - \delta_o}^{r_o} \tilde{v}_r \tilde{v}_{\theta} \left(\frac{d U}{dr} - \frac{U}{r}\right) rdr  \right|
\leq \max_{r_o - \delta_o < r < r_o} \left|\frac{d U}{dr} - \frac{U}{r}\right| r_o \int_{r = r_o - \delta_o}^{r_o} |\tilde{v}_r| \; |\tilde{v}_{\theta}| dr.
\label{An analytical bound: inter II estimate 4}
\end{eqnarray}
Using estimates (\ref{An analytical bound: inter II estimate 1})-(\ref{An analytical bound: inter II estimate 4}) along with the expression of the background flow (\ref{An analytical bound: definition background flow}), we finally obtain a simple bound on term $II$ in (\ref{An analytical bound: the functional H}) as
\begin{eqnarray}
|II| \leq \frac{M}{Re} \|\bnabla \boldsymbol{\tilde{v}}\|_2^2,
\end{eqnarray}
where
\begin{eqnarray}
M = \max\left\{\frac{h_i}{4} |1 - \Lambda r_i|, \; \frac{h_o}{4} |\Lambda| r_o\right\} + O(\delta).
\end{eqnarray}
This shows that the functional $\mathcal{H}$ is positive semi-definite as long as
\begin{eqnarray}
M \leq 1 - \frac{a}{2}.
\label{An analytical bound: inter optimum bound constraint}
\end{eqnarray}
Using (\ref{Problem setup: The laminar flow-field}) and (\ref{An analytical bound: definition background flow}) in (\ref{An analytical bound: non-dimensional energy dissipation 2}), we then obtain an upper bound on the dissipation as follows
\begin{eqnarray}
\varepsilon \leq  \varepsilon_{b}^a = \frac{2}{a(2-a) (r_o^2 - r_i^2)} \left(\frac{(1 - \Lambda r_i)^2 r_i}{h_i} + \frac{\Lambda^2 r_o^3}{h_o} \right) + O(\delta).
\label{An analytical bound: inter bound dissipation}
\end{eqnarray}
The upper bound obtained is called $\varepsilon_{b}^a$, and we use `$b$' in the subscript to signify that it is a bound and use `$a$' in the superscript to signify that it is obtained analytically. In final step of the procedure, we adjust the values of the unknown parameters $h_i$, $h_o$,  $\Lambda$ and $a$ to optimize the bound (\ref{An analytical bound: inter bound dissipation}) while satisfying the constraint (\ref{An analytical bound: inter optimum bound constraint}). The optimal values of the parameters, in the limit of high Reynolds number are,
\begin{eqnarray}
\Lambda = \frac{r_i}{r_i^2 + r_o^2}, \quad a = \frac{2}{3}, \quad h_o = \frac{8}{3 \Lambda r_o} \quad \text{and} \quad \frac{h_i}{h_o} = \eta.
\label{An analytical bound: optimal Lambda hi ho}  
\end{eqnarray}
The corresponding bound on the dissipation in the limit of $\Rey \to \infty$ is then given by 
\begin{eqnarray}
\varepsilon_{b, \infty}^a = \frac{27}{32} \frac{\eta}{(1 + \eta) (1 + \eta^2)^2}.
\label{An analytical bound: main term at infinity of bound on dissipation}
\end{eqnarray}
Here, we added `$\infty$' in the subscript to indicate that it is the main term of the bound in the limit $\Rey \to \infty$. Using the relationship (\ref{Bulk quantities of interest: G epsilon Nu relation}), we obtain an equivalent upper bound on the Nusselt number in the high Reynolds number limit as 
\begin{eqnarray}
Nu_{b, \infty}^a = \frac{27}{16} \frac{\eta^3}{(1 + \eta)^2 (1 + \eta^2)^2} Ta^{1/2}.
\label{An analytical bound: main term at infinity of bound on Nusselt number}
\end{eqnarray}
This expression contains a dependence on both the Taylor number (the principal flow parameter) as well as the radius ratio (the geometrical parameter). To separate out the geometrical dependence in (\ref{An analytical bound: main term at infinity of bound on Nusselt number}), we define
\begin{eqnarray}
\chi(\eta) = \frac{16 \eta^3}{(1+\eta)^2(1+\eta^2)^2},
\label{An analytical bound: the geometric scaling}
\end{eqnarray}
and call it the geometrical scaling of the bound on $Nu$. This geometrical scaling is defined in such a way that $\chi(1) = 1$ (the relevance of $\eta = 1$ being that it corresponds to the  plane Couette flow case).

Finally, by combining (\ref{An analytical bound: main term at infinity of bound on dissipation}) with the relation (\ref{Bulk quantities of interest: G epsilon relation}), we obtain an upper bound on the torque as a function of the Reynolds number
\begin{eqnarray}
G_{b, \infty}^a = \frac{27 \upi}{32} \frac{\eta^2 (1+\eta)^2}{(1 - \eta^4)^2} \Rey^2.
\label{An analytical bound: main term at infinity of bound on Reynolds number}
\end{eqnarray}
\citet{constantin1994geometric} had previously obtained a bound on the torque in Taylor--Couette flows by considering a background flow with a single boundary layer. The bound obtained by Constantin is also proportional to  $\Rey^2$, as in (\ref{An analytical bound: main term at infinity of bound on Reynolds number}). But, the coefficient in front has a different dependence on the radius ratio $\eta$. The reason for this difference is that we chose a background flow with two boundary layers and adjusted their relative thicknesses to optimize the bound. We shall see later that this optimization procedure enables us to capture the actual dependence of optimal bounds on the radius ratio.

\section{Optimal bounds}
\label{Optimal bounds: Numerical scheme}
In this section, we now proceed to obtain optimal bounds on the bulk quantities, i.e., the best possible bounds within the framework
of the background method. As described in \S\ref{Introduction}, we consider three scenarios,  `case 1', `case 2' and `case 3', in which we incrementally impose constraints on the perturbed flow field and numerically obtain the optimal bounds in each case, which allow us to systematically examine the hypothesis state in the introduction.




The general development of the background method for Taylor--Couette flow is presented in appendix \ref{The background method formulation}.  In what follows, we first describe our numerical algorithm, then proceed to present the results.

\subsection{Numerical Algorithm}
\label{Optimal bounds: Numerical scheme: Numerical Algorithm}
Here, we first describe the general numerical framework used to compute the optimal bounds, and then provide further details of the algorithm in each of the specific cases. Finding the optimal bound begins with the same background method applied to the Taylor--Couette flow as in \S \ref{An analytical bound}, which is described in appendix \ref{The background method formulation}. However, instead of using functional inequalities, we now follow the standard route toward optimal bounds, and derive a set of Euler--Lagrange equations that optimal solutions satisfy, given specific constraints in each case. The derivation is presented in appendix \ref{The background method formulation}, and the equations are given in (\ref{The background method: The main EL equations}a-d). In general, the Euler--Lagrange equations can have multiple solutions. However, we are interested in finding the unique solution that also satisfies the spectral constraint (\ref{The background method: the spectral constraint 1}).
To find this particular solution, we use the two-step algorithm first introduced by \citet{wen2013computational} in the context of porous medium convection. A remarkable property of this algorithm is that it eliminates the requirement of numerical continuation \citep{plasting2003improved}. As the two-step algorithm can be implemented at any value of the flow parameter, this flexibility has led to wider usage in several other studies of the background method to obtain the optimal bound numerically \citep{wen2015time, wen2018reduced, lee2019improved, ding2019upper, souza2020wall}. The first step of the algorithm uses a pseudo-time stepping scheme in which the Euler--Lagrange equations (\ref{The background method: The main EL equations}a-d) are converted into a time-dependent system of partial differential equations(PDEs) as follows
\begin{eqnarray}
\frac{\partial \boldsymbol{\tilde{v}}_i}{\partial t} = \frac{a}{2 (2-a)}\frac{\delta \mathcal{L}}{\delta \boldsymbol{\tilde{v}}_i}, \quad \frac{\partial U_\theta}{\partial t} = -\frac{a (2-a)}{4 \upi L} \frac{1}{r}\frac{\delta \mathcal{L}}{\delta U_\theta}, \quad \frac{\partial a}{\partial t} = - \frac{\delta \mathcal{L}}{\delta a}, \quad \bnabla \bcdot \boldsymbol{\tilde{v}} = 0,
\label{Optimal bounds: Numerical scheme: pseudo time stepping abstract}
\end{eqnarray}
where the index $i$ ranges over the $r, \theta$ and $z$ components of $\bs{\tilde{v}}$. Steady-state solutions of (\ref{Optimal bounds: Numerical scheme: pseudo time stepping abstract}) are equivalently solutions of the Euler--Lagrange equations (\ref{The background method: The main EL equations}a-d). 
Note that we multiply the Frechet derivatives with certain coefficients before introducing the time derivatives on the left-hand side. This makes the coefficient of the linear term (the Laplacian) a constant in the resultant time-dependent PDEs. Also, note that the coefficient in front of the Frechet derivative with respect to $\boldsymbol{\tilde{v}}$ is positive, while the coefficients in front of the Frechet derivatives with respect to $U_\theta$ and $a$ are negative. The reason is that we are maximizing the bound with respect to $\boldsymbol{\tilde{v}}$ while minimizing it with respect to $U_\theta$ and $a$.

\citet{ding2019upper} proved that if the pseudo-time stepping scheme leads to a steady-state solution then that solution must be the globally optimal solution of the Euler--Lagrange equations (\ref{The background method: The main EL equations}a-d), i.e., the one that leads to optimal bounds. Conveniently, the same proof extends to the case where the perturbed flow only satisfies the homogeneous boundary conditions and to the case where the perturbed flow is two-dimensional and incompressible. The proof of \citet{ding2019upper} does not guarantee the existence of a steady-state solution to (\ref{Optimal bounds: Numerical scheme: pseudo time stepping abstract}). But in all the cases that we investigated, the pseudo-time stepping scheme did relax to a steady-state solution.

The second step of the two-step algorithm is a Newton iteration which has a faster convergence rate than the pseudo-time stepping scheme but requires a good initial guess. Naturally, we use the solutions obtained at the end of the pseudo-time stepping scheme as the initial guess.

Solving the Euler--Lagrange equations in case 1 comes with two major simplifications. First, the pressure gradient term in (\ref{The background method: The main EL equations}a) disappears, as we do not impose the incompressibility constraint on the perturbation. Second, it can be shown that the optimal perturbation depends only on the radial direction (see appendix \ref{Analytical solution case 1 moma}). With these simplifications, the convergence of the pseudo-time stepping scheme is so rapid that the subsequent Newton iteration is not needed. Therefore, we only use the first step of the two-step algorithm described above. Furthermore, we found that it is also possible to solve the simplified Euler--Lagrange equations analytically in the limit $\Rey \to \infty$ using the method of matched asymptotics (solutions are presented in appendix \ref{Analytical solution case 1 moma}).

In case 2 it is also possible to make a simplification. Indeed, \citet{ding2019upper} presented numerical evidence that the optimal solution does not depend on $\theta$ when the aspect ratio $L$(i.e. the height of the cylinder) is large enough. Therefore, we choose $L = 20$, which is sufficiently large to guarantee that the optimal flow is axisymmetric. To solve the system of time-dependent PDEs (\ref{Optimal bounds: Numerical scheme: pseudo time stepping abstract}), we consider the following Fourier decomposition in the $z$ direction
\begin{eqnarray}
\boldsymbol{\tilde{v}} = \sum_{n = 1}^{N} 
\begin{bmatrix}
\tilde{v}_{r, n}(r, t) \cos (k_n z) \\
\tilde{v}_{\theta, n}(r, t) \cos (k_n z) \\
\tilde{v}_{z, n}(r, t) \sin (k_n z) 
\end{bmatrix},
\quad 
\tilde{p} = \sum_{n=1}^N \tilde{p}_n(r, t) \cos (k_n z) \quad \text{where} \quad k_n = \frac{2 \upi n}{L}.
\label{Optimal bounds: Fourier decompo}
\end{eqnarray}
The radial direction is further discretized using the Chebishev collocation method. We use a semi-implicit Crank--Nicolson scheme for the time integration where we treat the linear terms implicitly and use the second-order Adams--Bashforth extrapolation for the nonlinear terms. We use an influence matrix method to solve for the pressure at each time step \citep[see][p. 236]{peyret2013spectral}. The code is parallelized using MPI. Note that the pressure $\tilde{p}$ in (\ref{Optimal bounds: Fourier decompo}), as compared to the one in appendix \ref{The background method formulation}, has been multiplied with an  appropriate factor such that it is precisely the gradient of $\tilde{p}$ that appears in the time-evolving PDEs (\ref{Optimal bounds: Numerical scheme: pseudo time stepping abstract}). Depending on the radius ratio and Taylor number considered, we vary the number modes in the $z$ direction from $N = 200$ to $N = 6000$ and the number collocation points in the $r$ direction from $120$ to $320$. 


The numerical strategy for solving the Euler--Lagrange equations in case 3 is similar to case 2 described above. The only difference is that for the 2D incompressible perturbations, the flow quantities depend on the $\theta$ direction but are independent of $z$. Therefore, we consider the following decomposition instead
\begin{eqnarray}
\boldsymbol{\tilde{v}} = \sum_{\substack{m=-M \\ m \neq 0}}^{M} 
\begin{bmatrix}
\tilde{v}_{r, m}(r, t) e^{i m \theta} \\
\tilde{v}_{\theta, m}(r, t) e^{i m \theta} 
\end{bmatrix},
\quad 
\tilde{p} = \sum_{\substack{m=-M \\ m \neq 0}}^M \tilde{p}_m(r, t) e^{i m \theta}.
\end{eqnarray}
In this case, depending on the radius ratio and Taylor number considered, we vary the number modes in the $\theta$ direction from $M = 40$ to $M = 3000$ and the number collocation points in the $r$ direction from $120$ to $320$.

\subsection{Optimal bound results}
\label{Optimal bounds: Numerical scheme: Results}
\begin{figure}
\psfrag{A}[][][0.9]{$1.5$}
\psfrag{B}[][][0.9]{$1.7$}
\psfrag{C}[][][0.9]{$1.9$}
\psfrag{D}[][][0.9]{$2.1$}
\psfrag{E}[][][0.9]{$2.3$}
\psfrag{F}[][][0.9]{$2.5$}
\psfrag{H}[][][0.9]{$0$}
\psfrag{I}[][][0.9]{$0.2$}
\psfrag{J}[][][0.9]{$0.4$}
\psfrag{K}[][][0.9]{$0.6$}
\psfrag{L}[][][0.9]{$0.8$}
\psfrag{M}[][][0.9]{$1$}
\psfrag{1}[r][][0.9]{Analytical (\ref{An analytical bound: definition background flow})}
\psfrag{2}[r][][0.9]{case 1}
\psfrag{3}[r][][0.9]{case 2}
\psfrag{4}[r][][0.9]{case 3}
\psfrag{R}[][][1.3]{$r$}
\psfrag{U}[][][1.3][180]{$U_\theta$}
\centering
 \includegraphics[scale = 1.3]{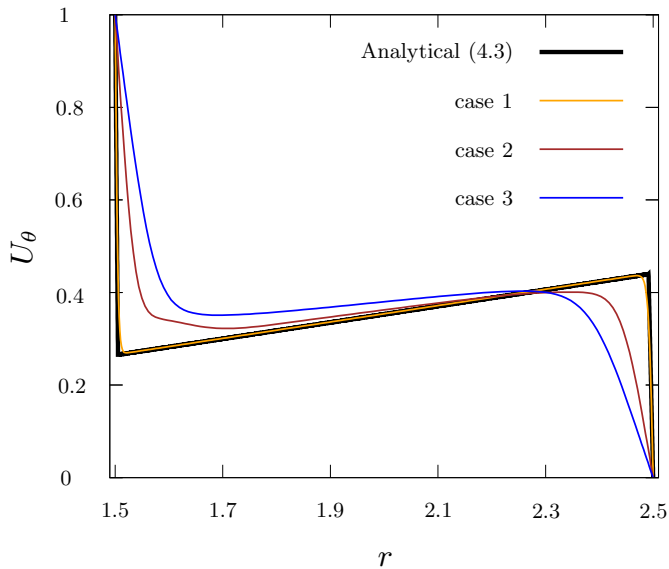}
 \caption{The optimal background flow $U_\theta(r)$ at parameter values $Ta = 10^6$ and $\eta = 0.6$. The orange color is used for case 1, brown color for case 2 and blue color for case 3. Also, shown as a black thick line is the background flow (\ref{An analytical bound: definition background flow}) used to construct the analytical bound in \S \ref{An analytical bound}, with the values of $\Lambda$, $\delta_i$ and $\delta_o$ given by (\ref{An analytical bound: optimal Lambda hi ho}) in definition (\ref{An analytical bound: definition background flow}).}
 \label{Optimal bounds: Numerical scheme: Results: BF}
\end{figure}
In this subsection, we present the optimal bounds obtained using the numerical schemes described above for each of the three different sets of constraints on the perturbations.  We begin by showing a typical optimal background flow profiles at $\eta = 0.6$ and $Ta = 10^6$ in each case in figure \ref{Optimal bounds: Numerical scheme: Results: BF}. For comparison, we have also included the background flow profile constructed in (\ref{An analytical bound: definition background flow}) to derive the original analytical bound. As can be seen in figure \ref{Optimal bounds: Numerical scheme: Results: BF}, all four background flow profiles vary as $c r$, for some constant $c$, in the bulk region. This is intuitively expected as this type of background profile makes the sign-indefinite term (which is, in a loose sense, the hardest to control in the bulk region) in the spectral constraint (\ref{The background method: the spectral constraint 1}) zero. Near the cylinders, the background flows consist of two thin boundary layers. In order to meet the prescribed boundary conditions, the gradients in these thin layers are large, which makes the sign-indefinite term nonzero. However, as the perturbation has to satisfy the homogeneous boundary conditions, the net contribution from this term will still be smaller than the positive term in (\ref{The background method: the spectral constraint 1}) as long as the boundary layer thickness is small enough. In the optimal state, the boundary layers are of just the right size so that the positive term and the sign-indefinite term balance each other out and the spectral constraint is marginally satisfied. When moving from case 1 to case 3, the restrictions on the perturbations increase, and this decreases the possibilities in which the sign-indefinite term can be negative. Therefore, the boundary layers become thicker, protruding more into the bulk region.

\begin{figure}
\centering
\psfrag{1}[r][][0.7]{$\eta = 0.1$}
\psfrag{2}[r][][0.7]{$\eta = 0.2$}
\psfrag{3}[r][][0.7]{$\eta = 0.3$}
\psfrag{4}[r][][0.7]{$\eta = 0.4$}
\psfrag{5}[r][][0.7]{$\eta = 0.5$}
\psfrag{6}[r][][0.7]{$\eta = 0.6$}
\psfrag{7}[r][][0.7]{$\eta = 0.7$}
\psfrag{8}[r][][0.7]{$\eta = 0.8$}
\psfrag{9}[r][][0.7]{$\eta = 0.9$}
\psfrag{\$}[r][][0.7]{$\eta = 0.99$}
\psfrag{A}[][][0.7]{$10^2$}
\psfrag{B}[][][0.7]{$10^3$}
\psfrag{C}[][][0.7]{$10^4$}
\psfrag{D}[][][0.7]{$10^5$}
\psfrag{E}[][][0.7]{$10^6$}
\psfrag{F}[][][0.7]{$10^7$}
\psfrag{G}[][][0.7]{$10^{8}$}
\psfrag{H}[][][0.7]{$10^{9}$}
\psfrag{I}[][][0.7]{$10^{10}$}
\psfrag{J}[][][0.7]{$10^{11}$}
\psfrag{K}[][][0.7]{$10^{12}$}
\psfrag{L}[][][0.7]{$10^{13}$}
\psfrag{M}[][][0.7]{$10^{14}$}
\psfrag{P}[][][0.7]{$0$}
\psfrag{Q}[][][0.7]{$0.02$}
\psfrag{R}[][][0.7]{$0.04$}
\psfrag{S}[][][0.7]{$0.06$}
\psfrag{T}[][][0.7]{$0.08$}
\psfrag{U}[][][0.7]{$0.1$}
\psfrag{a}[][][0.7]{$0.002$}
\psfrag{b}[][][0.7]{$0.004$}
\psfrag{c}[][][0.7]{$0.006$}
\psfrag{d}[][][0.7]{$0.008$}
\psfrag{e}[][][0.7]{$0.01$}
\psfrag{f}[][][0.7]{$0.012$}
\psfrag{g}[][][0.7]{$0.014$}
\psfrag{*}[][][0.9]{$Ta$}
\psfrag{+}[][][0.7]{$0.001$}
\psfrag{\[}[][][0.7]{$0.003$}
\psfrag{\]}[][][0.7]{$0.005$}
\psfrag{=}[][][0.7]{$0.007$}
\psfrag{!}[][][0.9][180]{$Nu^{nc}_b/Ta^{\frac{1}{2}}$}
\psfrag{\%}[][][0.9][180]{$Nu^{nc}_b/Ta^{\frac{1}{2}}/\chi(\eta)$}
\psfrag{\(}[][][0.9][180]{$Nu^{3D}_b/Ta^{\frac{1}{2}}$}
\psfrag{\)}[][][0.9][180]{$Nu^{3D}_b/Ta^{\frac{1}{2}}/\chi(\eta)$}
\psfrag{<}[][][0.9][180]{$Nu^{2D}_b/Ta^{\frac{1}{2}}$}
\psfrag{>}[][][0.9][180]{$Nu^{2D}_b/Ta^{\frac{1}{2}}/\chi(\eta)$}
\begin{tabular}{lc}
\begin{subfigure}{0.5\textwidth}
\centering
 \includegraphics[scale = 0.9]{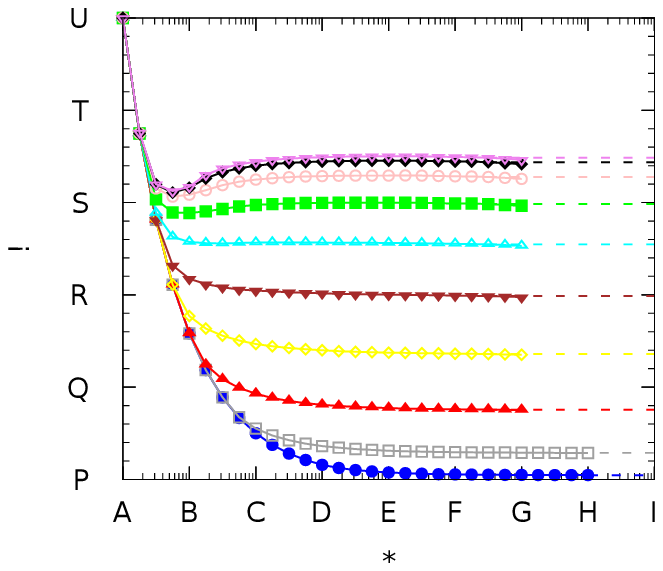}
\caption{}
\end{subfigure} &
\begin{subfigure}{0.5\textwidth}
\centering
 \includegraphics[scale = 0.9]{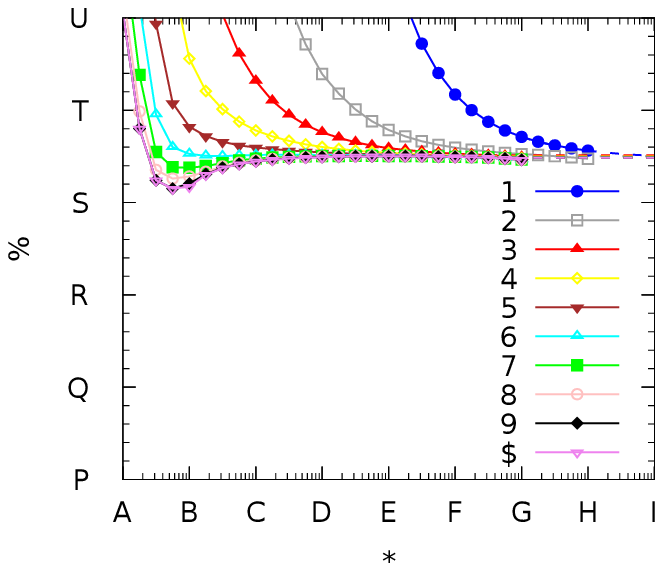}
\caption{}
\end{subfigure}
\end{tabular}
\begin{tabular}{lc}
\begin{subfigure}{0.5\textwidth}
\centering
 \includegraphics[scale = 0.9]{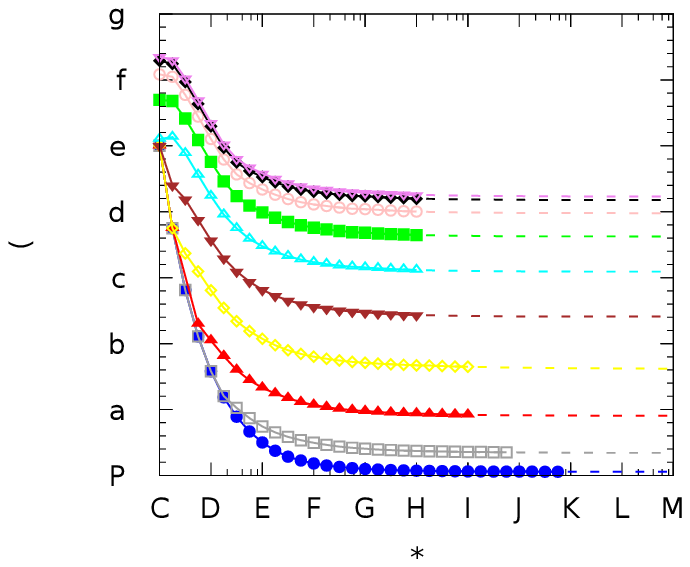}
\caption{}
\end{subfigure} &
\begin{subfigure}{0.5\textwidth}
\centering
 \includegraphics[scale = 0.9]{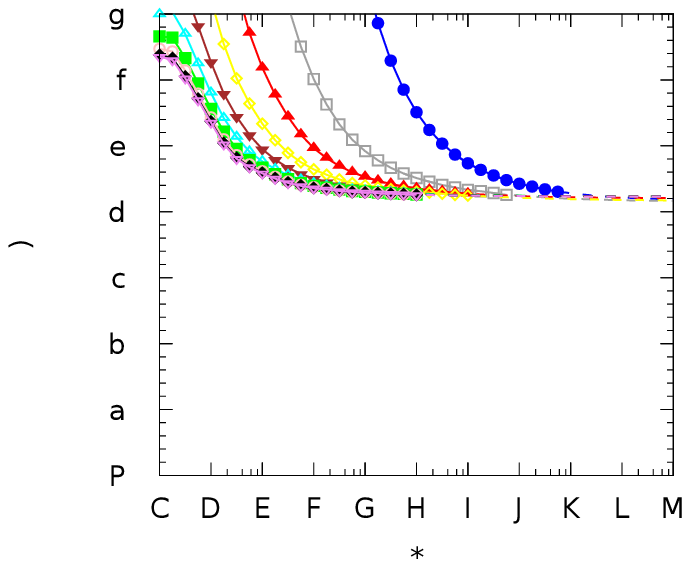}
\caption{}
\end{subfigure}
\end{tabular}
\begin{tabular}{lc}
\begin{subfigure}{0.5\textwidth}
\centering
 \includegraphics[scale = 0.9]{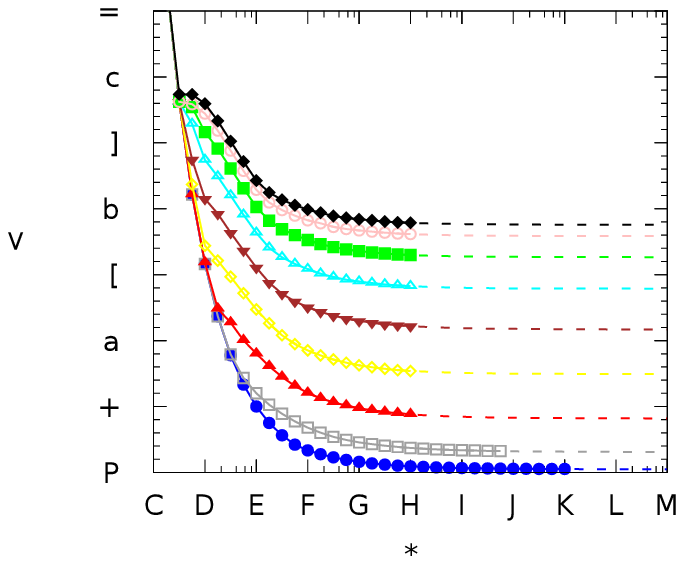}
\caption{}
\end{subfigure} &
\begin{subfigure}{0.5\textwidth}
\centering
 \includegraphics[scale = 0.9]{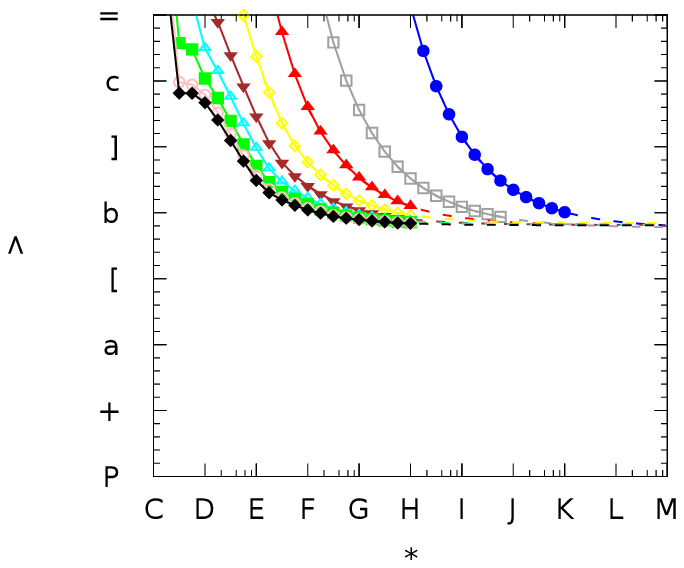}
\caption{}
\end{subfigure}
\end{tabular}
 \caption{The left column shows the optimal bound $Nu_b$ compensated with $Ta^{\frac{1}{2}}$ in case 1, case 2 and case 3 (top to bottom) as a function of the Taylor number for a wide range of radius ratios. The right column shows the same plots but further scaled with the analytical geometrical scaling $\chi(\eta)$ given by (\ref{An analytical bound: the geometric scaling}). The collapse of the curves at high Taylor numbers suggests that the bound on Nusselt number $Nu_b$ asymptotes to $c \chi(\eta) Ta^{\frac{1}{2}}$ in all three cases where the unknown constant $c$ is given in (\ref{Optimal bounds: Numerical scheme: Results: The most important conclusion of the paper}a-c)}
 \label{Optimal bounds: Numerical scheme: Results: The main plot}
\end{figure}

\begin{table}
  \begin{center}
\def~{\hphantom{0}}
  \begin{tabular}{lcccc}
      $\eta$  & $A^{nc}(\eta) / \chi(\eta)$   &  $A^{3D}(\eta) / \chi(\eta)$ &  $A^{2D}(\eta) / \chi(\eta)$ \\[3pt]
       	0.1   & $0.0694315$ & $0.00832565$ & $0.00376439$ \\[3pt]
	0.2   & $0.0700607$ & $0.00830294$ & $0.00377039$ \\[3pt]
	0.3   & $0.0702631$ & $0.00840836$ & $0.0038083$ \\[3pt]
	0.4   & $0.0699845$ & $0.00843543$ & $0.0038438$ \\[3pt]
	0.5   & $0.0698495$ & $0.00846619$ & $0.00381451$ \\[3pt]
	0.6   & $0.0697773$ & $0.00847018$ & $0.0038186$ \\[3pt]
	0.7   & $0.0697175$  & $0.00847478$ & $0.00381908$ \\[3pt]
	0.8   & $0.069738$ & $0.00846118$ & $0.00381454$ \\[3pt]
	0.9   & $0.0697065$ & $0.00846983$ & $0.00380973$ \\[3pt]
	0.99   &  $0.0697082$ & $0.0084623$ & -- \\[3pt]
  \end{tabular}
  \caption{Variation of the ratio $A(\eta)/\chi(\eta)$, where $A(\eta)$ is from the relation (\ref{Optimal bounds: Numerical scheme: Results: The best fit relation}) and $\chi(\eta)$ is given in (\ref{An analytical bound: the geometric scaling}), in case 1, case 2 and case 3 where we have respectively added `$nc$', `$3D$' and `$2D$' in the superscript to signify the case. Notice that $A(\eta)$ when scaled with $\chi(\eta)$ becomes almost invariant in $\eta$.}
  \label{Optimal bounds: Table 1}
  \end{center}
\end{table}

Figure \ref{Optimal bounds: Numerical scheme: Results: The main plot} shows the optimal bounds on the Nusselt number, $Nu_b$, as a function of the Taylor number $Ta$. We denote the bounds as $Nu_b^{nc}$ for case 1, $Nu_b^{3D}$ for case 2, and  $Nu_b^{2D}$ for case 3, and these are shown in the top, middle and bottom rows, respectively. We cover a wide range of parameters both in radius ratio (from $\eta = 0.1$ to $\eta = 0.99$) and in Taylor number. In figures \ref{Optimal bounds: Numerical scheme: Results: The main plot}a, \ref{Optimal bounds: Numerical scheme: Results: The main plot}c and \ref{Optimal bounds: Numerical scheme: Results: The main plot}e  the bound $Nu_b$ has been scaled with its expected asymptotic dependence on $Ta$, namely $Ta^{\frac{1}{2}}$.  The color and shape of the symbols each correspond to a different radius ratio, as shown in the legend.   The symbols in the plots in figure \ref{Optimal bounds: Numerical scheme: Results: The main plot} correspond to data points computed using the numerical algorithm from the previous subsection, whereas the solid lines connecting the data points are calculated using interpolation, providing a guide to the eye. For every radius ratio value, the solid line is extended up to the highest Taylor number for which the computation is performed. Beyond this point, we extrapolate using a best fit of the form
\begin{eqnarray}
f(\eta) = A(\eta) + \frac{B(\eta)}{Ta^{\alpha(\eta)}}
\label{Optimal bounds: Numerical scheme: Results: The best fit relation}
\end{eqnarray}
applied to the data of $Nu_b/Ta^{\frac{1}{2}}$ computed from the last two decades in $Ta$. For each value of $\eta$, we thus define $A(\eta)$ as the asymptotic limit of $Nu_b/Ta^{\frac{1}{2}}$ as $Ta \to \infty$. Table \ref{Optimal bounds: Table 1} summarizes the values of $A(\eta)$ obtained from this fitting procedure for different radius ratios. We have added appropriate abbreviations in the superscript of $A(\eta)$ to signify the case at hand. We remark that these extrapolations were necessary, especially for the small radius ratios, where the bound on the Nusselt number $Nu_b$ converges slowly to its asymptotic scaling in the Taylor number $Ta$. 

In figures \ref{Optimal bounds: Numerical scheme: Results: The main plot}b, \ref{Optimal bounds: Numerical scheme: Results: The main plot}d and \ref{Optimal bounds: Numerical scheme: Results: The main plot}f the bound $Nu_b$ has been scaled by $Ta^{\frac{1}{2}}$ as well as the geometrical scaling $\chi(\eta)$ obtained in (\ref{An analytical bound: the geometric scaling}). Note the striking collapse of the different radius ratio curves at high Taylor numbers in all three cases. Correspondingly, we also see from table \ref{Optimal bounds: Table 1} that the ratio $A(\eta)/\chi(\eta)$ is nearly independent of $\eta$ with less than $1.1\%$ variation in the average between the largest and smallest values. This suggests that the geometrical dependence of the bound on the Nusselt number at high Taylor number is $\chi(\eta)$ irrespective of the case considered. In case 1, the value of $A^{nc}(\eta)/\chi(\eta)$ is close to $9/128$ which is the exact asymptotic result we obtained from the method of matched asymptotics in appendix \ref{Analytical solution of the Euler--Lagrange equations in case 1 at high Reynolds number}. We also observe from table \ref{Optimal bounds: Table 1} that the value of $A/\chi(\eta)$ in case 2 and case 3 is very close to a constant for $\eta > 0.5$, but varies a little more for $\eta < 0.5$. This is likely due to the fact that the extrapolation is less accurate at small radius ratio because the computed data is further from being in the asymptotic regime compared with the case when the radius ratio is not small. For this reason, we assume that the average of $A(\eta)$ calculated for $\eta \geq 0.5$ is the correct asymptotic limit of $Nu_b/Ta^{\frac{1}{2}}$ as $Ta \to \infty$ and obtain

\begin{subequations}
\begin{eqnarray}
Nu_{b, \infty}^{nc} = \frac{9}{8}\frac{\eta^3}{(1+\eta)^2(1+\eta^2)^2} Ta^{\frac{1}{2}},  \\
Nu_{b, \infty}^{3D} = \frac{0.1354 \; \eta^3}{(1+\eta)^2(1+\eta^2)^2} Ta^{\frac{1}{2}},  \\
Nu_{b, \infty}^{2D} = \frac{0.0610 \; \eta^3}{(1+\eta)^2(1+\eta^2)^2} Ta^{\frac{1}{2}}.
\end{eqnarray}
\label{Optimal bounds: Numerical scheme: Results: The most important conclusion of the paper}
\end{subequations}
Here, we have added `$\infty$' in the subscript to point out that these are the main terms of the optimal bounds in the limit $Ta \to \infty$.

In summary, we have shown that for case 1, case 2 and case 3, the optimal bounds are respectively a factor of $1.5$, $12.46$ and $27.66$ better than the suboptimal bound (\ref{An analytical bound: main term at infinity of bound on Nusselt number}) in the high Taylor number limit. Crucially, this improvement is uniform in the radius ratio $\eta$. We had obtained the analytical expression for the geometrical scaling $\chi(\eta)$ from a fairly simple suboptimal analytical bound calculated using a choice of background flow with two boundary layers whose thicknesses were adjusted to optimize the bound. During this procedure, we had not applied any constraint on the perturbed flow $\tilde{\bs{v}}$ (other than the homogeneous boundary conditions) and further used standard calculus inequalities which are known to  overestimate the bound on $Nu_b$. Consequently, it is not at all self-evident why the optimal bounds should have the same geometrical scaling. The fact that the optimal bounds (\ref{Optimal bounds: Numerical scheme: Results: The most important conclusion of the paper}a-c), which are up to an order of magnitude better than the suboptimal bound (\ref{An analytical bound: main term at infinity of bound on Nusselt number}), preserve the same geometrical dependence on radius ratio is therefore a simple yet remarkable result. 


\subsection{Wavenumber spectrum of perturbations} 


\begin{figure}
\psfrag{A}[][][0.9]{$Ta$}
\psfrag{B}[][][0.9]{$k_{n_c}$}
\psfrag{C}[][][0.9]{$10^4$}
\psfrag{D}[][][0.9]{$10^5$}
\psfrag{E}[][][0.9]{$10^6$}
\psfrag{F}[][][0.9]{$10^7$}
\psfrag{G}[][][0.9]{$10^8$}
\psfrag{H}[][][0.9]{$10^9$}
\psfrag{I}[][][0.9]{$0$}
\psfrag{J}[][][0.9]{$0.1$}
\psfrag{K}[][][0.9]{$0.2$}
\psfrag{L}[][][0.9]{$0.3$}
\psfrag{M}[][][0.9]{$0.4$}
\psfrag{N}[][][0.9]{$0.5$}
\psfrag{O}[][][0.9]{$4$}
\psfrag{P}[][][0.9]{$4.2$}
\psfrag{Q}[][][0.9]{$4.4$}
\psfrag{R}[][][0.9]{$4.6$}
\psfrag{S}[][][0.9]{$4.8$}
\psfrag{T}[][][0.9]{$-0.2$}
\psfrag{U}[][][0.9]{$1.5$}
\psfrag{V}[][][0.9]{$1.7$}
\psfrag{W}[][][0.9]{$1.9$}
\psfrag{X}[][][0.9]{$2.1$}
\psfrag{Y}[][][0.9]{$2.3$}
\psfrag{Z}[][][0.9]{$10^{10}$}
\psfrag{a}[][][0.9]{$1$}
\psfrag{b}[][][0.9]{$2$}
\psfrag{c}[][][0.9]{$5$}
\psfrag{d}[][][0.9]{$10$}
\psfrag{e}[][][0.9]{$20$}
\psfrag{f}[][][0.9]{$50$}
\psfrag{g}[][][0.9]{$100$}
\psfrag{h}[][][0.9]{$200$}
\psfrag{i}[][][0.9]{$500$}
\psfrag{j}[][][0.9]{$1000$}
\psfrag{k}[][][0.9]{$0.6$}
\psfrag{l}[][][0.9]{$0.7$}
\psfrag{m}[][][0.9]{$0.8$}
\psfrag{n}[][][0.9]{$0.9$}
\psfrag{o}[][][0.9]{$1.1$}
\psfrag{p}[][][0.9]{$1.2$}
\psfrag{q}[][][0.9]{$2.5$}
\psfrag{r}[][][0.9]{$1.4$}
\psfrag{s}[][][0.9]{$1.6$}
\psfrag{t}[r][][0.9]{$45$}
\psfrag{u}[r][][0.9]{$46$}
\psfrag{v}[r][][0.9]{$51$}
\psfrag{w}[r][][0.8]{$122$}
\psfrag{x}[r][][0.8]{$123$}
\psfrag{y}[r][][0.8]{$158$}
\psfrag{z}[r][][0.8]{$388$}
\psfrag{0}[][0.9]{$\tilde{v}_{\theta, n_c}$}
\psfrag{1}[r][][0.8]{$15$}
\psfrag{2}[r][][0.8]{$16$}
\psfrag{3}[r][][0.8]{$42$}
\psfrag{4}[r][][0.8]{$54$}
\psfrag{5}[r][][0.8]{$55$}
\psfrag{6}[r][][0.8]{$176$}
\psfrag{7}[r][][0.8]{$177$}
\psfrag{8}[r][][0.8]{$389$}
\psfrag{9}[][][0.9][180]{$r$}
\psfrag{'}[r][][0.9]{$n_c$}
\psfrag{!}[r][][0.8]{$49$}
\psfrag{#}[r][][0.8]{$50$}
\psfrag{\$}[r][][0.8]{$53$}
\psfrag{\%}[r][][0.8]{$130$}
\psfrag{*}[r][][0.8]{$131$}
\psfrag{\(}[r][][0.8]{$156$}
\psfrag{\)}[r][][0.8]{$157$}
\psfrag{-}[r][][0.8]{$411$}
\psfrag{+}[r][][0.8]{$412$}
\psfrag{=}[r][][0.8]{$512$}
\psfrag{:}[r][][0.8]{$513$}
\psfrag{[}[r][][0.8]{$47$}
\psfrag{]}[r][][0.8]{$52$}
\psfrag{;}[r][][0.8]{$98$}
\psfrag{|}[r][][0.8]{$99$}
\psfrag{<}[r][][0.8]{$151$}
\psfrag{>}[r][][0.8]{$152$}
\psfrag{,}[r][][0.8]{$298$}
\psfrag{.}[r][][0.8]{$299$}
\psfrag{?}[r][][0.8]{$496$}
\centering
\begin{tabular}{lc}
\begin{subfigure}{0.5\textwidth}
\centering
 \includegraphics[scale = 0.64]{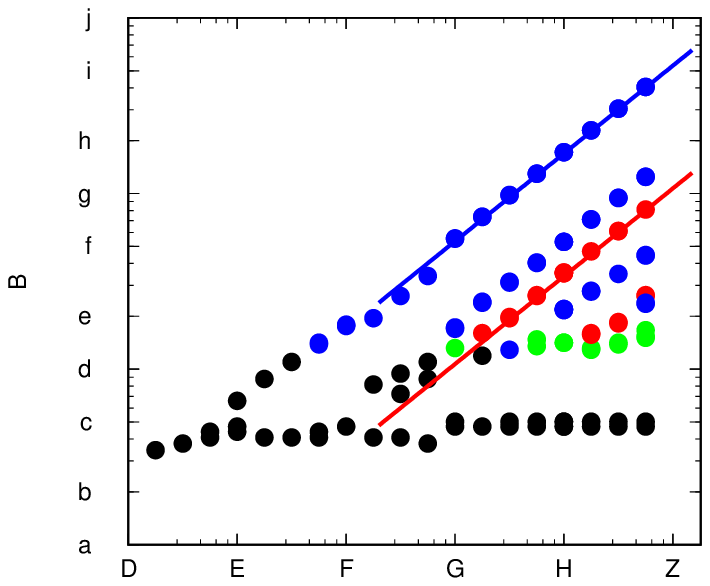}
\caption{}
\end{subfigure} &
\begin{subfigure}{0.5\textwidth}
\centering
 \includegraphics[scale = 0.64]{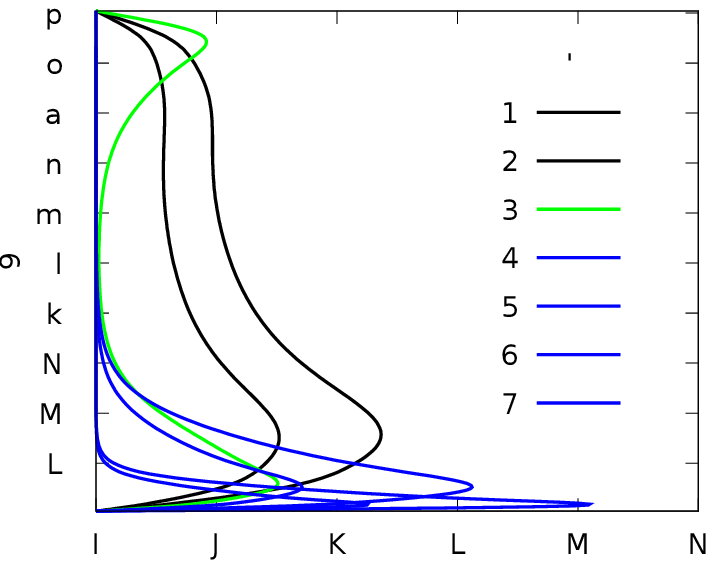}
\caption{}
\end{subfigure}
\end{tabular}
\begin{tabular}{lc}
\begin{subfigure}{0.5\textwidth}
\centering
 \includegraphics[scale = 0.64]{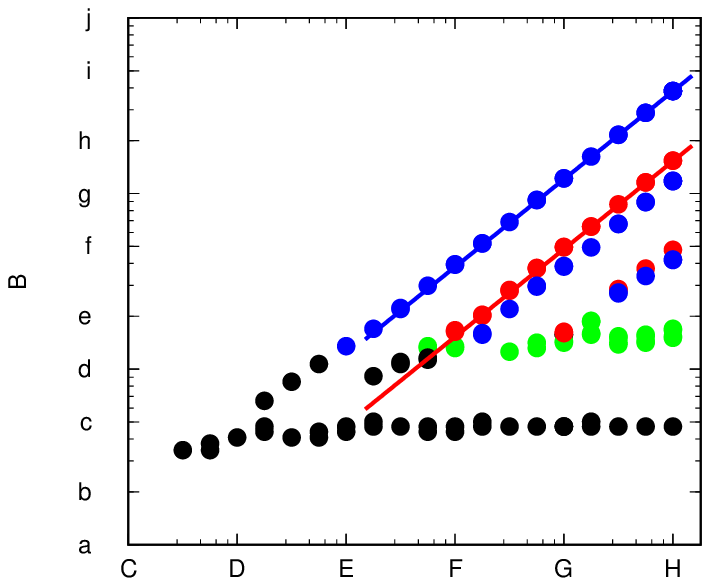}
\caption{}
\end{subfigure} &
\begin{subfigure}{0.5\textwidth}
\centering
 \includegraphics[scale = 0.64]{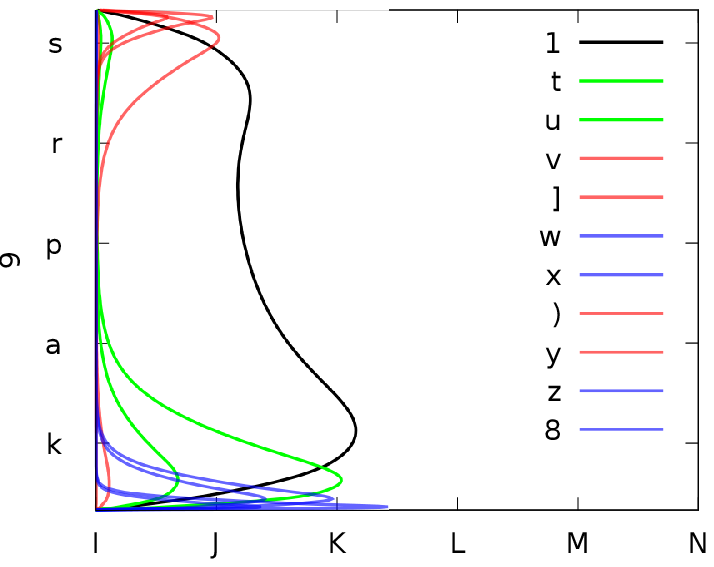}
\caption{}
\end{subfigure}
\end{tabular}
\begin{tabular}{lc}
\begin{subfigure}{0.5\textwidth}
\centering
 \includegraphics[scale = 0.64]{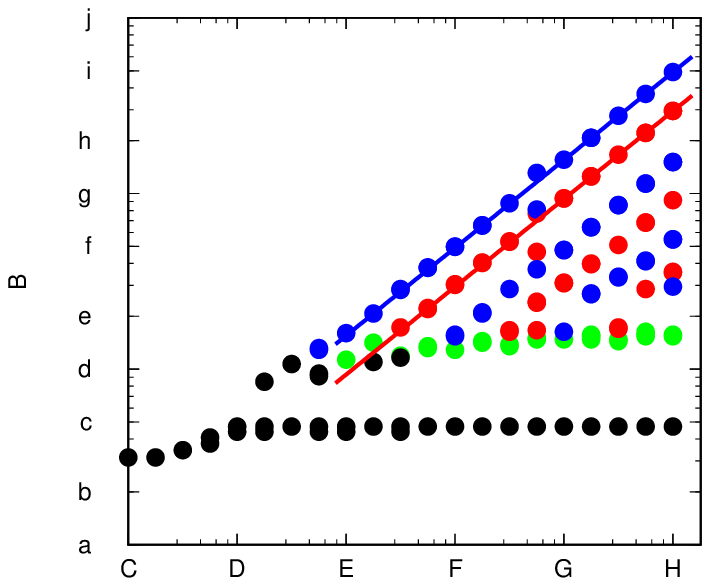}
\caption{}
\end{subfigure} &
\begin{subfigure}{0.5\textwidth}
\centering
 \includegraphics[scale = 0.64]{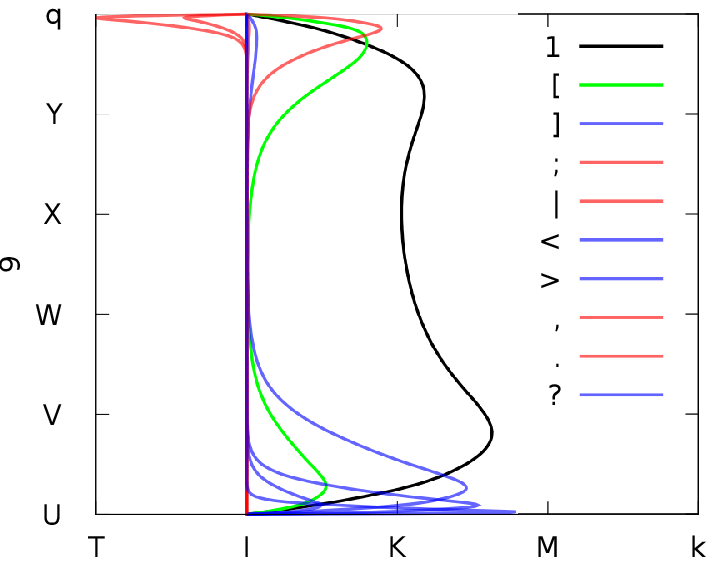}
\caption{}
\end{subfigure}
\end{tabular}
\begin{tabular}{lc}
\begin{subfigure}{0.5\textwidth}
\centering
 \includegraphics[scale = 0.64]{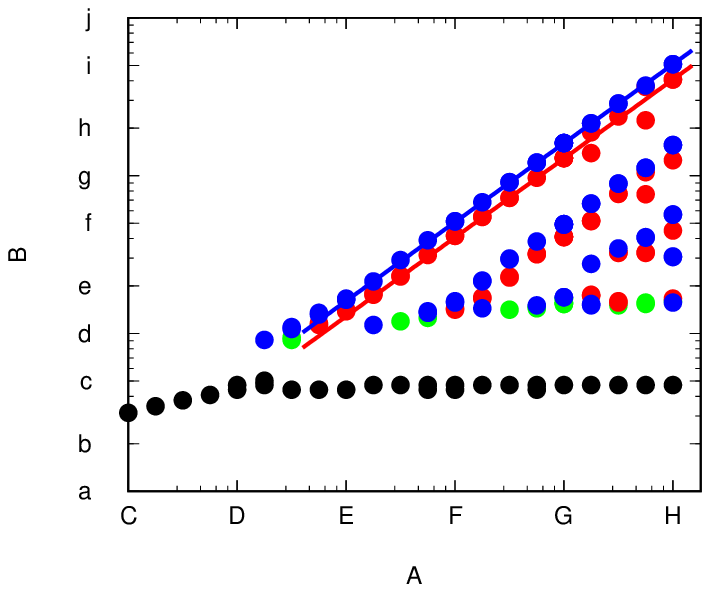}
\caption{}
\end{subfigure} &
\begin{subfigure}{0.5\textwidth}
\centering
 \includegraphics[scale = 0.64]{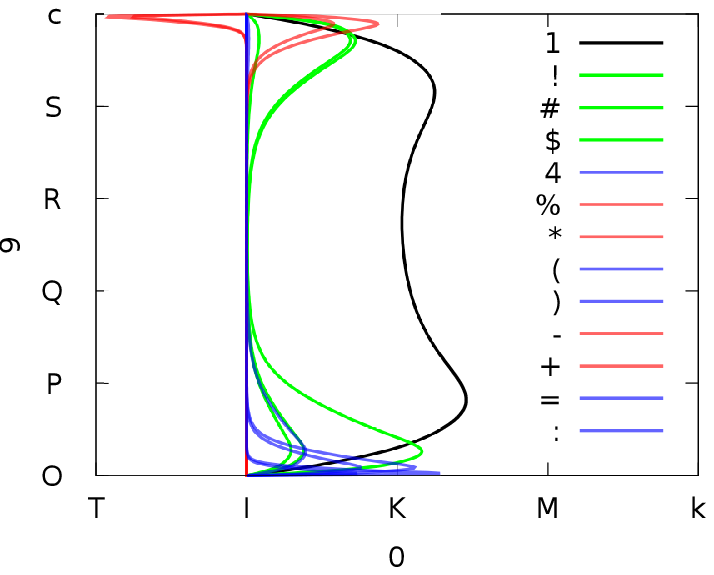}
\caption{}
\end{subfigure}
\end{tabular}
 \caption{The left column shows the wavenumbers of the critical modes (denoted using filled circles) in the optimal perturbation as a function of $Ta$ at $\eta = 0.2, 0.4, 0.6$ and $0.8$ (top to bottom). The color indicates if the critical mode is active near the inner cylinder (blue), outer cylinder (red), both cylinders (green) and in the bulk (black), according to the classification given in the main text. The blue and red solid lines are the theoretical  predictions for the critical mode with the largest wavenumber active near the inner and the outer cylinder, (see equation (\ref{Boundary layers and structure of the perturbations: highest wavenumbers with constant})), respectively. The right column shows the corresponding azimuthal component  $\tilde{v}_{\theta, n_c}(r)$ of critical modes at the same radius ratios, at $Ta = 10^8$.}
 \label{Optimal bounds: Numerical scheme: Results: 3D spectrum and mode plots}
\end{figure}

In this subsection, we investigate the wavenumber spectrum of the perturbed flow $\tilde{\bs{v}}$ with a particular focus on the small scale structures present in $\tilde{\bs{v}}$. In the optimal state, $\tilde{\bs{v}}$ contains only a finite number of modes, called the critical modes, either in the $z$ or $\theta$ direction, depending on the case considered, i.e.,
\begin{eqnarray}
\boldsymbol{\tilde{v}} = \sum_{n \in K_2} 
\begin{bmatrix}
\tilde{v}_{r, n}(r) \cos (k_n z) \\
\tilde{v}_{\theta, n}(r) \cos (k_n z) \\
\tilde{v}_{z, n}(r) \sin (k_n z) 
\end{bmatrix} \text{ in case 2} \qquad
\boldsymbol{\tilde{v}} = \sum_{m \in K_3} 
\begin{bmatrix}
\tilde{v}_{r, m}(r) e^{i m \theta} \\
\tilde{v}_{\theta, m}(r) e^{i m \theta} 
\end{bmatrix} \text{ in case 3}
\end{eqnarray}
where $K_2 \subset  \mathbb{N}$ and $K_3 \subset  \mathbb{Z} \setminus \{0\}$ are finite sets.
Moreover, as we shall demonstrate below, the smallest scales in the perturbation are present only near the boundaries.  It is thus reasonable to hypothesize that the smallest length scale in the perturbation $\tilde{\bs{v}}$  is similar to the boundary layer thickness of the background flow $\bs{U}$. To further pursue this idea, we divide the critical modes present in the perturbation into four different categories. If, for a given critical mode, more than $90\%$ of the contribution to its $L^1(dr)$ norm comes from the region $$S_{in} \coloneqq \left\{r \; | \;  r_i \leq r \leq r_i + \frac{r_o - r_i}{3}  \right\}$$ then we say that mode is active only near the inner cylinder. Similarly, if it comes from  the region $$S_{out} \coloneqq \left\{r \; | \; r_o - \frac{r_o - r_i}{3} \leq r \leq r_o  \right\}$$ then we say it is active only near the outer cylinder. Finally, if more than $90\%$ of the contribution comes from region $S_{in}$ and $S_{out}$ together then we say the mode is active near both the cylinders, otherwise we say the mode is active in the bulk.  This way of categorizing the modes may seem somewhat arbitrary at first, but looking at the shape of different critical modes, it becomes readily apparent that any other appropriate definition would have led to the same conclusion. We use the following color scheme to differentiate modes according to our classification: blue for the modes that are active near the inner cylinder, red for the modes that are active near the outer cylinder, green for the modes that are active near both the cylinders and black for the modes that are active in bulk. The right column in figure \ref{Optimal bounds: Numerical scheme: Results: 3D spectrum and mode plots} shows the plots of $\tilde{v}_{\theta, n_c}(r)$ for critical modes at $Ta = 10^8$ and for radius ratios $\eta = 0.2, 0.4, 0.6$ and $0.8$. We now see that the plots of $\tilde{v}_{\theta, n_c}(r)$ provide an unambiguous visual justification of our earlier classification of critical modes into four categories, and therefore, our classification is robust.

We first apply this categorization to the optimal perturbations found in case 2 and denote the wavenumber of the critical mode with smallest length scale that is active near the inner cylinder as $k_{in}^s$ and the one that is active near the outer cylinder as $k_{out}^s$. Assuming that our hypothesis about the similarity of the boundary layer thickness in the background flow and the smallest length scale in the perturbation is correct, then we can use the analytical expression of the boundary layer thickness from (\ref{An analytical bound: optimal Lambda hi ho}) and the relation between $\Rey$ and $Ta$ given by (\ref{Problem setup: The Taylor number}) to deduce that
\begin{eqnarray}
k_{in}^s \propto \frac{\eta^2}{(1+\eta^2)(1+\eta)^3} Ta^{\frac{1}{2}}, \qquad k_{out}^s \propto \frac{\eta^3}{(1+\eta^2)(1+\eta)^3} Ta^{\frac{1}{2}}.
\label{Boundary layers and structure of the perturbations: highest wavenumbers}
\end{eqnarray}
From these relations, we not only obtain the dependence of $k^s_{in}$ and $k^s_{out}$ on $Ta$, but also on the radius ratio $\eta$. In particular, we predict that the smallest length scales in the perturbation should become larger as $\eta \to 0$. Furthermore, at a given $\eta$, the small scale structures near the outer cylinder are predicted  to be $1/\eta$ times larger than the ones near the inner cylinder.

The left column in figure \ref{Optimal bounds: Numerical scheme: Results: 3D spectrum and mode plots} shows the wavenumbers of the critical modes in the optimal perturbations as a function of the Taylor number for four different radius ratio values $\eta = 0.2, 0.4, 0.6$ and $0.8$ (top to bottom row). By fitting these plots, we find that the constant of proportionality in (\ref{Boundary layers and structure of the perturbations: highest wavenumbers}) that best fists the data at high Taylor numbers is $C = 0.244$, therefore we expect 
\begin{eqnarray}
k_{in}^s = 0.244 \frac{\eta^2}{(1+\eta^2)(1+\eta)^3} Ta^{\frac{1}{2}}, \qquad k_{out}^s = 0.244 \frac{\eta^3}{(1+\eta^2)(1+\eta)^3} Ta^{\frac{1}{2}}.
\label{Boundary layers and structure of the perturbations: highest wavenumbers with constant}
\end{eqnarray}
These two relations are plotted in figures \ref{Optimal bounds: Numerical scheme: Results: 3D spectrum and mode plots}a, \ref{Optimal bounds: Numerical scheme: Results: 3D spectrum and mode plots}c, \ref{Optimal bounds: Numerical scheme: Results: 3D spectrum and mode plots}e and \ref{Optimal bounds: Numerical scheme: Results: 3D spectrum and mode plots}g with solid blue and red lines, respectively. We see that smallest length scales in the critical perturbation near the inner and outer boundaries, respectively, indeed follow the relations (\ref{Boundary layers and structure of the perturbations: highest wavenumbers with constant}). Furthermore, these smallest scales achieve their asymptotic scaling in Taylor number quicker than the corresponding optimal bounds on Nusselt number shown in figure \ref{Optimal bounds: Numerical scheme: Results: The main plot}, without any need for extrapolation of the data. We therefore argue that (\ref{Boundary layers and structure of the perturbations: highest wavenumbers with constant}) and figure \ref{Optimal bounds: Numerical scheme: Results: 3D spectrum and mode plots} together provide a strong validation of the analytical predictions from \S \ref{An analytical bound}.

\begin{figure}
\psfrag{A}[][][0.9]{$Ta$}
\psfrag{B}[][][0.9]{$m_c$}
\psfrag{C}[][][0.9]{$10^4$}
\psfrag{D}[][][0.9]{$10^5$}
\psfrag{E}[][][0.9]{$10^6$}
\psfrag{F}[][][0.9]{$10^7$}
\psfrag{G}[][][0.9]{$10^8$}
\psfrag{H}[][][0.9]{$10^9$}
\psfrag{I}[][][0.9]{$0$}
\psfrag{J}[][][0.9]{$0.1$}
\psfrag{K}[][][0.9]{$0.2$}
\psfrag{L}[][][0.9]{$0.3$}
\psfrag{M}[][][0.9]{$0.4$}
\psfrag{N}[][][0.9]{$0.5$}
\psfrag{O}[][][0.9]{$4$}
\psfrag{P}[][][0.9]{$4.2$}
\psfrag{Q}[][][0.9]{$4.4$}
\psfrag{R}[][][0.9]{$4.6$}
\psfrag{S}[][][0.9]{$4.8$}
\psfrag{T}[][][0.9]{$-0.2$}
\psfrag{U}[][][0.9]{$1.5$}
\psfrag{V}[][][0.9]{$1.7$}
\psfrag{W}[][][0.9]{$1.9$}
\psfrag{X}[][][0.9]{$2.1$}
\psfrag{Y}[][][0.9]{$2.3$}
\psfrag{Z}[][][0.9]{$10^{10}$}
\psfrag{a}[][][0.9]{$1$}
\psfrag{b}[][][0.9]{$2$}
\psfrag{c}[][][0.9]{$5$}
\psfrag{d}[][][0.9]{$10$}
\psfrag{e}[][][0.9]{$20$}
\psfrag{f}[][][0.9]{$50$}
\psfrag{g}[][][0.9]{$100$}
\psfrag{h}[][][0.9]{$200$}
\psfrag{i}[][][0.9]{$500$}
\psfrag{j}[][][0.9]{$1000$}
\psfrag{k}[][][0.9]{$0.6$}
\psfrag{l}[][][0.9]{$0.7$}
\psfrag{m}[][][0.9]{$0.8$}
\psfrag{n}[][][0.9]{$0.9$}
\psfrag{o}[][][0.9]{$1.1$}
\psfrag{p}[][][0.9]{$1.2$}
\psfrag{q}[][][0.9]{$2.5$}
\psfrag{r}[][][0.9]{$1.4$}
\psfrag{s}[][][0.9]{$1.6$}
\psfrag{t}[r][][0.9]{$15$}
\psfrag{u}[r][][0.9]{$16$}
\psfrag{v}[r][][0.9]{$19$}
\psfrag{w}[r][][0.9]{$44$}
\psfrag{x}[r][][0.9]{$45$}
\psfrag{y}[r][][0.9]{$158$}
\psfrag{z}[r][][0.9]{$388$}
\psfrag{0}[][0.9]{$\tilde{v}_{\theta, m_c}^c$}
\psfrag{1}[][][0.9]{$-0.1$}
\psfrag{2}[][][0.9]{$-0.2$}
\psfrag{3}[][][0.9]{$-0.3$}
\psfrag{4}[r][][0.9]{$3$}
\psfrag{5}[r][][0.9]{$4$}
\psfrag{6}[r][][0.9]{$9$}
\psfrag{7}[r][][0.9]{$10$}
\psfrag{8}[r][][0.9]{$389$}
\psfrag{9}[b][][0.9][180]{$r$}
\psfrag{!}[r][][0.9]{$94$}
\psfrag{#}[r][][0.9]{$95$}
\psfrag{\$}[r][][0.9]{$96$}
\psfrag{\%}[r][][0.9]{$337$}
\psfrag{*}[r][][0.9]{$338$}
\psfrag{\(}[r][][0.9]{$156$}
\psfrag{\)}[r][][0.9]{$157$}
\psfrag{-}[r][][0.9]{$411$}
\psfrag{+}[r][][0.9]{$25$}
\psfrag{=}[r][][0.9]{$97$}
\psfrag{:}[][][0.9]{$2000$}
\psfrag{[}[r][][0.9]{$11$}
\psfrag{]}[r][][0.9]{$36$}
\psfrag{;}[r][][0.9]{$37$}
\psfrag{|}[r][][0.9]{$39$}
\psfrag{<}[r][][0.9]{$124$}
\psfrag{>}[r][][0.9]{$125$}
\psfrag{,}[r][][0.9]{$6$}
\psfrag{.}[r][][0.9]{$299$}
\psfrag{?}[r][][0.9]{$43$}
\centering
\begin{tabular}{lc}
\begin{subfigure}{0.5\textwidth}
\centering
 \includegraphics[scale = 0.68]{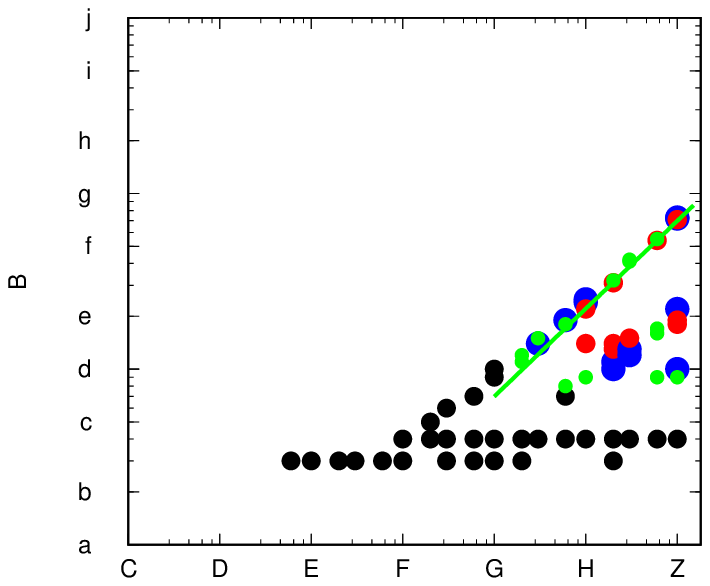}
\caption{}
\end{subfigure} &
\begin{subfigure}{0.5\textwidth}
\centering
 \includegraphics[scale = 0.68]{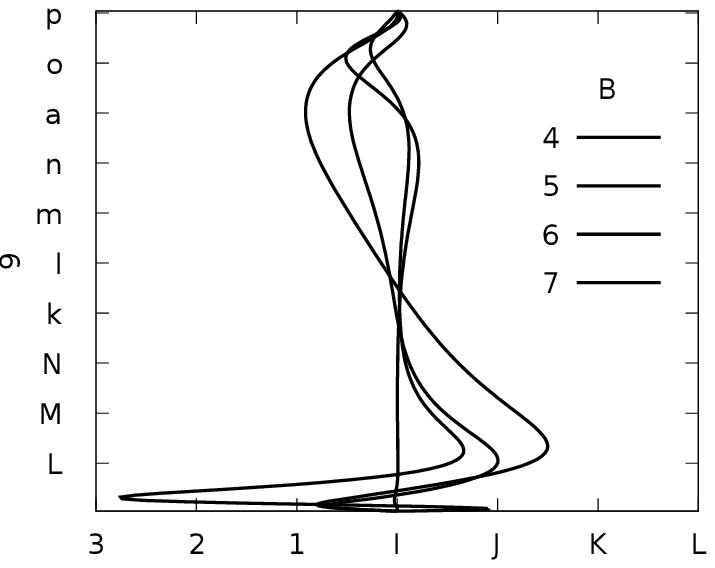}
\caption{}
\end{subfigure}
\end{tabular}
\begin{tabular}{lc}
\begin{subfigure}{0.5\textwidth}
\centering
 \includegraphics[scale = 0.68]{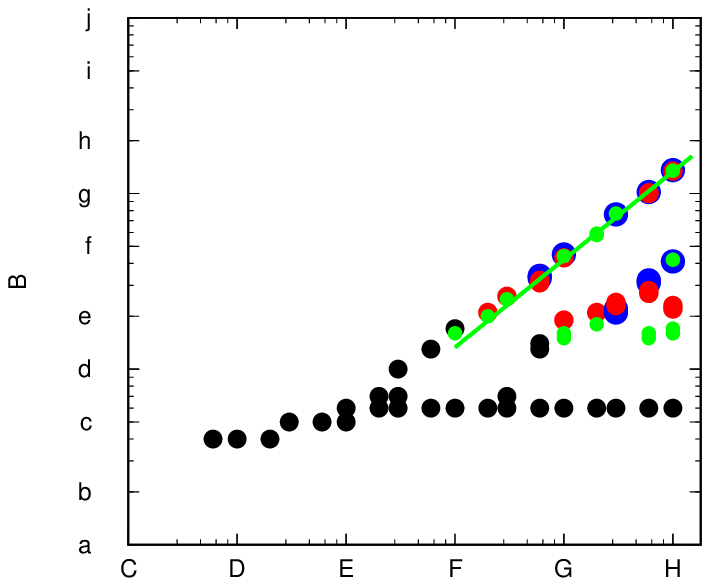}
\caption{}
\end{subfigure} &
\begin{subfigure}{0.5\textwidth}
\centering
 \includegraphics[scale = 0.68]{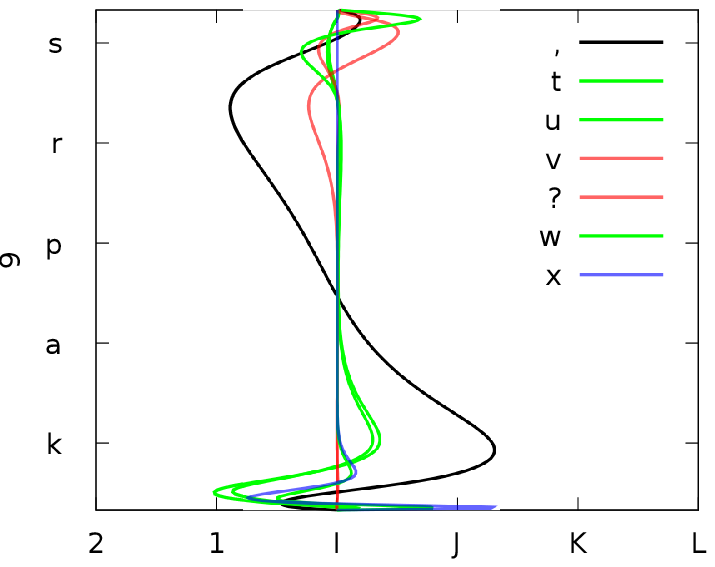}
\caption{}
\end{subfigure}
\end{tabular}
\begin{tabular}{lc}
\begin{subfigure}{0.5\textwidth}
\centering
 \includegraphics[scale = 0.68]{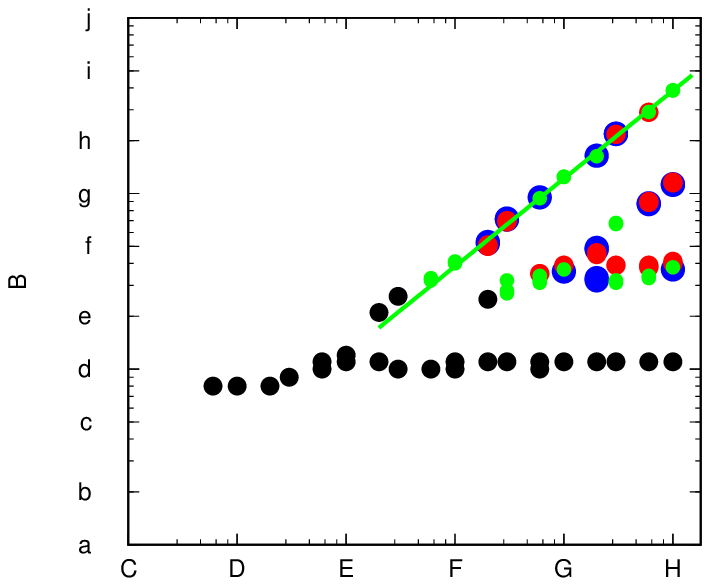}
\caption{}
\end{subfigure} &
\begin{subfigure}{0.5\textwidth}
\centering
 \includegraphics[scale = 0.68]{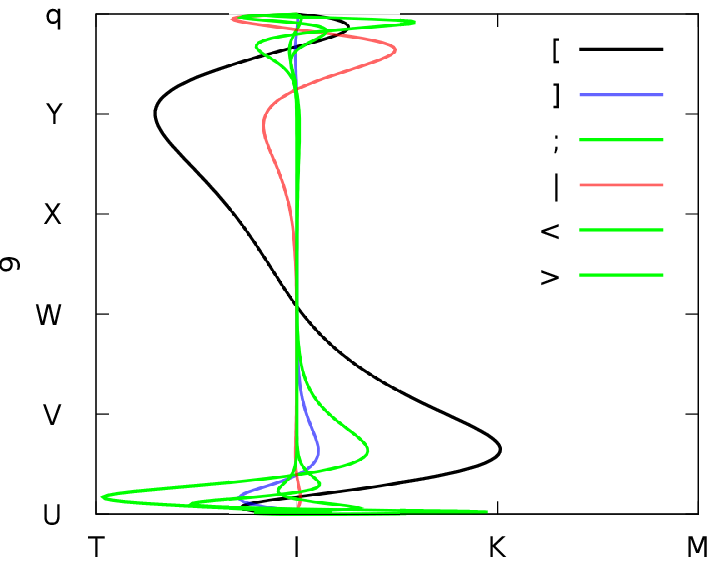}
\caption{}
\end{subfigure}
\end{tabular}
\begin{tabular}{lc}
\begin{subfigure}{0.5\textwidth}
\centering
 \includegraphics[scale = 0.68]{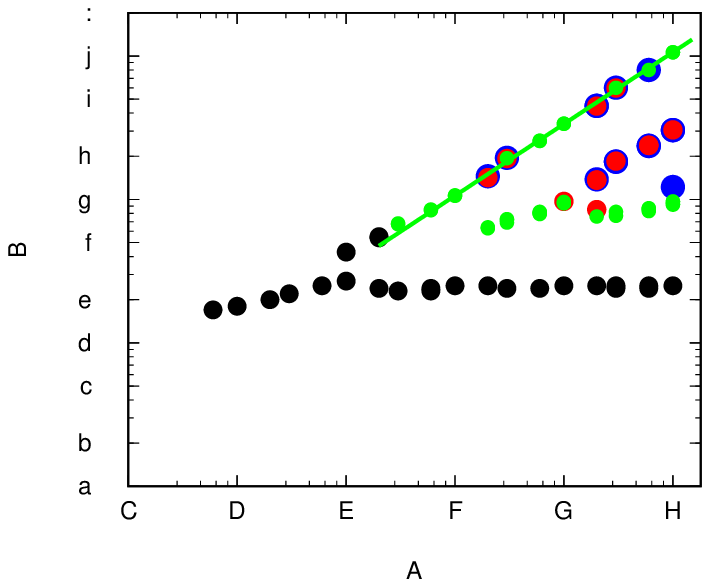}
\caption{}
\end{subfigure} &
\begin{subfigure}{0.5\textwidth}
\centering
 \includegraphics[scale = 0.68]{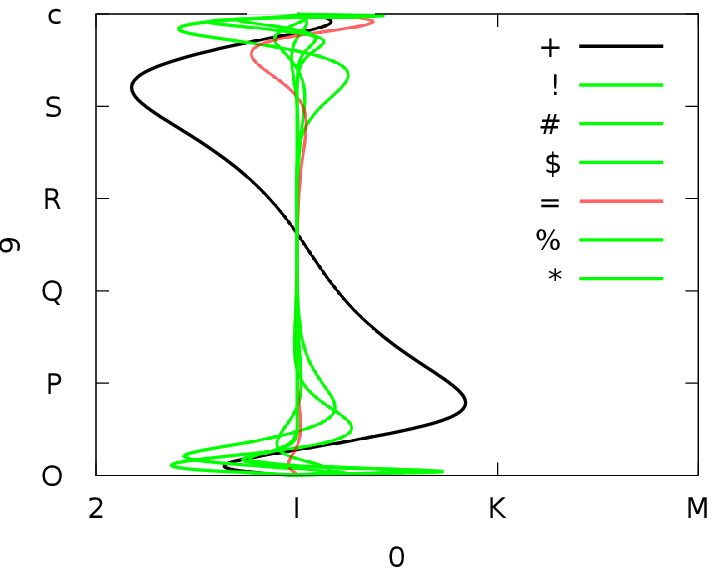}
\caption{}
\end{subfigure}
\end{tabular}
 \caption{The plots in the left column shows the wavenumber $m_c$ of the critical modes that constitutes the two dimensional optimal perturbation as a function of the Taylor number for radius ratios $\eta = 0.2, 0.4, 0.6$ and $0.8$ (top to bottom). We use the same color scheme as in figure \ref{Optimal bounds: Numerical scheme: Results: 3D spectrum and mode plots} to distinguish different critical modes. The solid green line is the relation (\ref{Boundary layers and structure of the perturbations: wavenumber prediction 2D}) which predicts the largest critical wavenumber. The right column shows the plots of $\tilde{v}^c_{\theta, m_c}$, the coefficient of $\cos m_c \theta$ in the azimuthal component of the velocity, at $Ta = 10^8$ and the same radius ratios as the left-hand side plots.}
 \label{Optimal bounds: Numerical scheme: Results: 2D spectrum and mode plots}
\end{figure}

We can use similar ideas to predict the scaling of the smallest length scales in optimal perturbations in case 3. Using (\ref{An analytical bound: optimal Lambda hi ho}), one would anticipate $m_i^s \propto r_i/\delta_i$ and $m_o^s \propto r_o/\delta_o$, where $m_i^s$ is the largest wavenumber of a critical mode active near the inner cylinder and $m_o^s$ is the largest wavenumber of a critical mode active near the outer cylinder. The plots on the left-hand side column in figure \ref{Optimal bounds: Numerical scheme: Results: 2D spectrum and mode plots} shows the wavenumbers of critical modes in the 2D optimal perturbations as a function of the Taylor number at radius ratios $0.2, 0.4, 0.6$ and $0.8$.  We apply the same mode identification method, and use the same color scheme to differentiate the critical modes as before. From these plots, we can fit the data at high Taylor numbers, to measure the constant of proportionality in the expressions for  $m_i^s$ and $m_o^s$, leading to
\begin{eqnarray}
m_i^s = m_o^s = \frac{0.126\eta^3}{(1-\eta^4)(1+\eta)^2} Ta^{\frac{1}{2}}.
\label{Boundary layers and structure of the perturbations: wavenumber prediction 2D}
\end{eqnarray} 
We see that the wavenumber of the critical mode with smallest length scale that is active near the inner cylinder and outer cylinder are equal.
The relation (\ref{Boundary layers and structure of the perturbations: wavenumber prediction 2D}), shown as a solid green line on the left-hand side of figure \ref{Optimal bounds: Numerical scheme: Results: 2D spectrum and mode plots}, does seem to predict the largest wavenumbers at high Taylor numbers correctly.

As in figure \ref{Optimal bounds: Numerical scheme: Results: 3D spectrum and mode plots}, the right-hand side column of figure \ref{Optimal bounds: Numerical scheme: Results: 2D spectrum and mode plots} shows the function $v^c_{\theta, m_c}(r)$, defined as the coefficient of $\cos m_c \theta$ in the expression $$\tilde{v}_{\theta, m_c}(r) e^{i m_c \theta} + \tilde{v}_{\theta, -m_c}(r) e^{- i m_c \theta},$$ where $m_c$ refers to a critical mode. The main difference between the shape of modes $\tilde{v}_{\theta, n_c}(r)$ in figure \ref{Optimal bounds: Numerical scheme: Results: 3D spectrum and mode plots} compared with $v^c_{\theta, m_c}(r)$ in figure \ref{Optimal bounds: Numerical scheme: Results: 2D spectrum and mode plots}, is that the mean of $\tilde{v}_{\theta, m_c}^{c}(r)$ is zero, i.e.,
\begin{eqnarray}
\int_{r_i}^{r_o} \tilde{v}_{\theta, m_c}^{c}(r) dr = 0.
\label{Boundary layers and structure of the perturbations: impact of incompressibility}
\end{eqnarray}
This condition comes from incompressibility, which leads to (\ref{Boundary layers and structure of the perturbations: impact of incompressibility}) in 2D, but does not in 3D because the $z$-component, $\tilde{v}_z$, is nonzero. As a result, in 2D, modes which are active solely near the cylinders oscillates in the boundary layer to ensure that (\ref{Boundary layers and structure of the perturbations: impact of incompressibility}) is satisfied.

\section{A note on the applicability of the background method}
\label{A note on the applicability of the background method}
In one of our previous studies \citep{kumar2020pressure}, we presented a sufficient criterion to determine when the background method can be applied, for a given flow geometry and boundary conditions.  We demonstrated that it can be used with any flow problem (tangential-velocity-driven or pressure-driven) with impermeable boundaries, provided the boundaries have the shape of streamtubes of the following flow
\begin{eqnarray}
\bs{V} = \mathsfbi{A} \bs{x} + \bs{V}_0.
\label{A note on the applicability of the background method: criterion flow}
\end{eqnarray} 
Here, $\mathsfbi{A}$ is a constant skew-symmetric tensor, $\bs{V}_0$ is a constant vector and $\bs{x}$ is the position vector. For these types of problems, one can further show that the upper bound on the dissipation becomes independent of viscosity at high Reynolds numbers.  In this section, we explore the complementary question of whether there exist  flow configurations for which the background method {\it cannot} be applied.  

Indeed, the applicability of the background method depends on the existence of an incompressible background flow (which also satisfies the inhomogeneous boundary conditions) such that the following functional is positive semi-definite
\begin{eqnarray}
\mathcal{H}(\bs{v}) = \left[\frac{1}{2 \Rey} \| \bnabla \boldsymbol{\tilde{v}}\|_2^2 +  \int_{V} \boldsymbol{\tilde{v}} \bcdot \bnabla \boldsymbol{U}_{sym} \bcdot \boldsymbol{\tilde{v}} \; d \boldsymbol{x} + \int_{V} \boldsymbol{U} \bcdot \bnabla \boldsymbol{U} \bcdot \boldsymbol{\tilde{v}} \; d \boldsymbol{x}\right],
\label{A note on the applicability of the background method: the functional H}
\end{eqnarray}
for any perturbations $\bs{\tilde{v}}$ that satisfies the homogeneous boundary conditions. Consequently, proving that the background method {\it cannot} be applied reduces to the problem of finding  a perturbation or a family of perturbations such that there is no background flow $\bs{U}$ for which $\mathcal{H}$ is positive semi-definite.

We start by giving a few examples where the applicability of the background flow can be rigorously dismissed. We first consider the case of Taylor--Couette flow with suction at the inner cylinder. The energy stability analysis of this problem was considered by \citet{gallet2010destabilizing}. The  boundary conditions for this problem are:
\begin{eqnarray}
\bs{u} = - \bs{e}_r  + \omega_i r_i \bs{e}_\theta \quad \text{at} \quad r = r_i, \qquad \bs{u} = - \frac{r_i}{r_o} \bs{e}_r + \omega_o r_o \bs{e}_\theta \quad \text{at} \quad r = r_o,
\end{eqnarray}
where the Reynolds number $\Rey$ is defined such that $\bs{u} \bcdot \bs{e}_r = -1$ at the inner cylinder. The non-dimensional angular velocities of the inner and outer cylinder are $\omega_i$ and $\omega_o$, respectively. In this problem, the flow is constricted to a narrow area as it moves from the outer cylinder (inlet) to the inner cylinder (outlet). We restrict ourselves to two dimensions but the arguments given below are valid in three dimensions as well. The domain of interest is $V = [r_i, r_o] \times [0, 2\upi]$. 

We consider a perturbation $\bs{\tilde{v}}$ of the form 
\begin{eqnarray}
\bs{\tilde{v}} = (\tilde{v}_r, \tilde{v}_\theta) = \left(0, \; v_{0} r (r - r_i) (r - r_o)\right),
\label{A note on the applicability of the background method: dangerous perturbed flow TC with suction}
\end{eqnarray}
whose amplitude is $v_{0}$. Note that $\tilde{\bs{v}}$ satisfies the homogeneous boundary conditions and is incompressible.  We now demonstrate that for this perturbation, the spectral constraint can never be satisfied above a certain Reynolds number regardless of the choice of  background flow $\bs{U}$.

To show that the spectral constraint (\ref{A note on the applicability of the background method: the functional H}) is not satisfied, we have to show that the second term is negative and that its absolute value is larger than the first term. Being linear in $\bs{\tilde{v}}$, the last term can be made arbitrarily small compared with the first two terms by choosing $v_{0}$ in (\ref{A note on the applicability of the background method: dangerous perturbed flow TC with suction}) to be large enough for any given background flow $\bs{U}$. As such, it does not play any role in the following argument. 

The calculation of the first term is straightforward:
\begin{eqnarray}
\frac{1}{2 \Rey} \| \bnabla \boldsymbol{\tilde{v}}\|_2^2 = \frac{\upi}{60 \Rey} (r_i + r_0) (5r_i^2 + 5r_o^2 + 2) v_{0}^2.
\label{A note on the applicability of the background method: first term final calc}
\end{eqnarray}
In the calculation of the second term, we take advantage of the fact that the chosen perturbation (\ref{A note on the applicability of the background method: dangerous perturbed flow TC with suction})  is independent of $\theta$, so
\begin{eqnarray}
\int_{V} \boldsymbol{\tilde{v}} \bcdot \bnabla \boldsymbol{U} \bcdot \boldsymbol{\tilde{v}} \; d \boldsymbol{x} = \int_{r_i}^{r_o} \tilde{v}_\theta^2 \left[\int_{0}^{2 \upi}\left(\frac{1}{r} \frac{\partial U_\theta}{\partial \theta} + \frac{U_r}{r}\right) d\theta \right] r dr.
\label{A note on the applicability of the background method: second term inter calc}
\end{eqnarray}
Now, using periodicity as well as the incompressibility condition satisfied by the background flow $\bs{U}$, the following holds
\begin{eqnarray}
\int_{0}^{2 \upi} \frac{\partial U_\theta}{\partial \theta} d \theta = 0, \qquad \int_{0}^{2 \upi} U_r d \theta = - \frac{2 \upi r_i}{r}.
\label{A note on the applicability of the background method: cond. for any bf}
\end{eqnarray}
Using (\ref{A note on the applicability of the background method: dangerous perturbed flow TC with suction}) and (\ref{A note on the applicability of the background method: cond. for any bf}) in (\ref{A note on the applicability of the background method: second term inter calc}) gives
\begin{eqnarray}
\int_{V} \boldsymbol{\tilde{v}} \bcdot \bnabla \boldsymbol{U} \bcdot \boldsymbol{\tilde{v}} \; d \boldsymbol{x} = -  \frac{\upi}{30} (r_i^2 + r_i r_o) v_0^2.
\label{A note on the applicability of the background method: second term final calc}
\end{eqnarray}
From (\ref{A note on the applicability of the background method: first term final calc}) and (\ref{A note on the applicability of the background method: second term final calc}), we deduce that the spectral constraint (\ref{A note on the applicability of the background method: the functional H}) will not be satisfied if 
\begin{eqnarray}
\Rey > \frac{5r_i^2 + 5r_o^2 + 2}{2 r_i},
\label{A note on the applicability of the background method: Re fail for TC with suction}
\end{eqnarray}
a condition that is, remarkably,  independent of the choice of $\bs{U}$. Note that in the limit of $r_i / r_o \to 1$, the Reynolds number beyond which the method fails goes to infinity. This limit recovers the case of a plane Couette flow with suction and injection at the walls \citep{doering2000energy}, where the background method can indeed be applied, so (\ref{A note on the applicability of the background method: Re fail for TC with suction}) is consistent with these results. 

\begin{figure}
\centering
 \includegraphics[scale = 0.08]{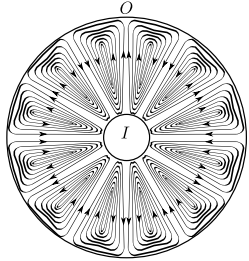}
 \caption{A cartoon of the streamlines of the flow field given by (\ref{A note on the applicability of the background method: dangerous perturbed flow TC with injection incompressible}).}
 \label{A note on the applicability of the background method: injection convering diverging}
\end{figure}

A similar type of condition on the Reynolds number can be derived in the problem of the Taylor--Couette flow with injection at the inner cylinder, i.e.,  
\begin{eqnarray}
\bs{u} = \bs{e}_r  + \omega_i r_i \bs{e}_\theta \quad \text{at} \quad r = r_i, \qquad \bs{u} = \frac{r_i}{r_o} \bs{e}_r + \omega_o r_o \bs{e}_\theta \quad \text{at} \quad r = r_o.
\end{eqnarray}
In this problem, the flow overall expands into a larger area as it moves from the inner cylinder (inlet) to the outer cylinder (outlet). For this case, we can use similar arguments but with the new perturbed flow 
\begin{eqnarray}
\bs{\tilde{v}} = (\tilde{v}_r, \tilde{v}_\theta) = \left(v_0 r (r - r_i) (r - r_o), \; 0\right).
\label{A note on the applicability of the background method: dangerous perturbed flow TC with injection}
\end{eqnarray}
The perturbation this time is not incompressible but can be shown to yield a negative $\mathcal{H}(\tilde{\bs{v}})$ regardless of the background flow $\bs{U}$, for sufficiently large Reynolds number. However, noting that the perturbation (\ref{A note on the applicability of the background method: dangerous perturbed flow TC with injection}) is radial, one may then expect that an incompressible perturbation, which is composed of vortices stretched in the radial direction will also yield a negative $\mathcal{H}(\tilde{\bs{v}})$. This observation led us to consider consider the following streamfunction:
\begin{eqnarray}
\psi_{v} = v_0 r^2 (r-r_i)^2(r-r_o)^2 \sin m\theta \quad \text{where } m \in \mathbb{N}.
\end{eqnarray}
We define the corresponding velocity field $\bs{\tilde{v}} = (\tilde{v}_r, \tilde{v}_\theta)$ as
\begin{eqnarray}
\tilde{v}_r = \frac{1}{r} \frac{\partial \psi_v}{\partial \theta}, \quad \tilde{v}_\theta = -\frac{\partial \psi_v}{\partial r}. 
\label{A note on the applicability of the background method: dangerous perturbed flow TC with injection incompressible}
\end{eqnarray}
This velocity field is divergence free and satisfies the homogeneous boundary conditions at the surface of the cylinders. The streamlines of $\tilde{\bs{v}}$ are depicted in figure \ref{A note on the applicability of the background method: injection convering diverging}.
Next define a family of rotation operators $\mathcal{Q}_{\varphi}:\mathcal{D_\sigma}(V) \to \mathcal{D_\sigma}(V)$, indexed with $\varphi$, on the space of divergence-free vector fields that satisfies the homogeneous boundary conditions at $\partial V$ as
\begin{eqnarray}
\mathcal{Q}_{\varphi}(\bs{\tilde{v}})(r, \theta) = \bs{\tilde{v}}(r, \theta + \varphi) \quad \forall (r, \theta) \in V.
\end{eqnarray} 
A tedious calculation, first involving an integration in $\varphi$ and then using the arguments similar to the suction problem above, shows
\begin{eqnarray}
\int_{0}^{2 \upi} \mathcal{H}(\mathcal{Q}_{\varphi}(\bs{\tilde{v}})) d \varphi = \qquad \qquad \qquad \qquad \qquad \qquad \qquad \qquad \qquad \qquad \qquad \qquad  && \nonumber \\
 \frac{\upi(r_i + r_o)}{1260}\bigg[\frac{246 r_i^4 + 138 r_o^4 + 108 r_o^3 + 120 r_i^2 r_o^2 + 108 r_i^2 r_o + 8m^2(r_o^3 - r_i^3) + m^4}{2 \Rey}  && \nonumber \\
 - r_i (m^2 - 6(r_i^2 + r_o^2))\bigg]. &&
\label{A note on the applicability of the background method: integral on family}
\end{eqnarray}
This calculation implies that if 
\begin{eqnarray}
m > \left\lceil \sqrt{6r_i^2 + 6r_o^2} \right\rceil, 
\label{A note on the applicability of the background method: dangerous perturbed flow TC with injection cond. on wavenumber m}
\end{eqnarray}
where $\left\lceil \; \cdot \; \right\rceil$ is the ceiling function, then for 
\begin{eqnarray}
\Rey > \frac{246 r_i^4 + 138 r_o^4 + 108 r_o^3 + 120 r_i^2 r_o^2 + 108 r_i^2 r_o + 8m^2(r_o^3 - r_i^3) + m^4}{2 r_i \left(m^2 - 6r_i^2 - 6r_o^2\right)}
\end{eqnarray}
the integral (\ref{A note on the applicability of the background method: integral on family}) is negative, which implies
there is at least one $\varphi \in [0, 2 \upi]$ such that $ \mathcal{H}(\mathcal{Q}_{\varphi}(\bs{\tilde{v}})) < 0$. More generally,
there exist a set $S \subset [0, 2 \upi]$, depending on the background flow $\bs{U}$, of positive measure ($\mu(S) > 0$) such that $\mathcal{Q}_{\varphi}(\bs{\tilde{v}}) < 0$ for any $\varphi \in S$, i.e., the spectral constraint is not satisfied. Note that the condition (\ref{A note on the applicability of the background method: dangerous perturbed flow TC with injection cond. on wavenumber m}) is basically saying that the vortices in the incompressible perturbed flow field (\ref{A note on the applicability of the background method: dangerous perturbed flow TC with injection incompressible}) should be stretched in the radial direction, which we expected from the example of the compressible perturbed flow (\ref{A note on the applicability of the background method: dangerous perturbed flow TC with injection}). 


The key message from these two problems is that if there is a converging flow, then one can rule out the applicability of the background method by creating  a perturbation whose streamlines are perpendicular to the direction of the mean flow, while in the case of a  diverging flow, one can use a perturbation whose streamlines are parallel to the direction of the mean flow instead. Of course, in both the cases, we need to make sure that the perturbation satisfies the homogeneous boundary conditions.


Combining these ideas  suggests that one cannot apply the background method to flows in a converging-diverging nozzle, either because one can choose the perturbation to be composed of vortices that stretch in the perpendicular direction to the flow in the converging section or parallel to the flow in the diverging section. Using the same arguments, one would then also conjecture that the background method can in general not  be applied to flows between rough walls. Indeed, in this case, one can always find vertical sections where the flow expands or compresses and then one could use the same strategy to choose perturbations for which $\mathcal{H}(\tilde{\bs{v}}) < 0$. However, note that, in this case, the compression or expansion is small, i.e., the gap width on averages decreases only by a factor of $(1 - \epsilon)$ in the converging part or increases by a factor of $(1 + \epsilon)$ in the diverging part, where $\epsilon$ is the non-dimensional roughness scale. This problem is analogous to the converging-diverging nozzle if the Reynolds number is based on the surface roughness $\epsilon$. Therefore, for the Reynolds number based on the average gap width, we expect that the spectral constraint (\ref{A note on the applicability of the background method: the functional H}) will not be satisfied if $\Rey \gtrsim \epsilon^{-1}$.

This still leaves the problem open for the flow systems which do not have a converging or diverging section, for example, flow in tortuous channels. We believe that even for these problems the spectral constraint will fail to hold past a certain Reynolds number for any background flow. Therefore, we conjecture that the sufficient condition for the applicability of the background method mentioned in the beginning of this section is also a necessary condition.

\section{Discussion and conclusion}
\label{Discussion and conclusion}
\subsection{Summary and implications}
In this paper, we computed optimal bounds on mean quantities in the Taylor--Couette flow problem with a stationary outer cylinder, with particular focus on the dependence of these bounds on the system geometry. Along the way, we studied the energy stability of the laminar flow in \S \ref{Energy stability analysis}. The main finding of this section was that for a value of radius ratio $r_i/r_o$ below $0.0556$, the marginally stable flow at the energy stability threshold is not composed of the  well-known axisymmetric Taylor vortices but is instead a fully three-dimensional flow field.

To uncover the functional dependence of the optimal bounds on the radius ratio at large Taylor number, we began by deriving a suboptimal but analytical bound with the use of standard inequalities and a choice of background flow with two boundary layers (one near the inner cylinder and one near the outer cylinder) whose thicknesses were then adjusted to optimize the bound. We then argued that the dependence on the radius ratio captured by this analytical bound should also be the same for the optimal bounds at large Taylor numbers. We systematically verified this statement by obtaining distinct optimal bounds under three circumstances. In the first case, we imposed no constraints on the perturbation other than the homogeneous boundary conditions (case 1). Next, we allowed for three-dimensional incompressible perturbations (case 2), and finally, we considered two-dimensional incompressible perturbations (case 3). In the high Taylor number limit, we see an improvement of $1.5$, $12.46$ and $27.66$, respectively, over the analytical bound as we move from case 1 to case 3, and that improvement is the  same for all radius ratios. This result is  striking and non-trivial because there is no known transformation of variables which makes the Euler--Lagrange equations (\ref{The background method: The main EL equations}a-d) of the optimal bounds independent of the radius ratio. 



In \S \ref{A note on the applicability of the background method}, we rigorously dismissed the applicability of the background method for two flow problems. The limitation of the background method is previously known in the context of Rayleigh--B\'enard convection at infinite Prandtl number \citep{nobili2017limitations}, where it was shown that using a different method a tighter bound can be obtained as compared to the background method. Here, we have shown that past a certain Reynolds number, no bound can be obtained using the background method applied to Taylor--Couette flow with suction or injection at the inner cylinder, i.e., there is no background flow that satisfies the spectral constraint even when the incompressibility condition on the perturbation is imposed. Generalizing these results then suggests that the spectral condition may not be satisfied for flow problems that contain converging or diverging sections, such as flow in a converging-diverging nozzle or flow between the rough walls.


Our study brings into light the significance (or lack of significance, to be more precise) of the incompressibility constraint on the perturbation while calculating optimal bounds, especially in the  limit of high Reynolds number, which is generally of interest in turbulent flows. As we showed in the present study, dropping the incompressibility constraint on the perturbations altogether still recovers the correct dependence of the bounds on both the principal flow parameter ($Ta$, or equivalently $\Rey$) and on the domain geometry (through the radius ratio). One cannot help but wonder whether the same holds true for other flow problems, including for instance the case of convection.  It is a fundamental question of concern, as not imposing the incompressibility constraint tremendously decreases the computational cost of the optimal bound calculation.  In the particular example studied here, in fact, not imposing the incompressibility condition allowed us to solve the Euler--Lagrange equations analytically using the method of matched asymptotics. This could also be potentially helpful in other studies involving the background method where it is relatively difficult to establish the scaling of the optimal bound even numerically, perhaps because the bounds involve logarithms \citep{fantuzzi2018boundsA, fantuzzi2020new} or a scaling other than a simple power-law \citep{kumar2021IH}. These ideas can also be of relevance  to other variational approaches such as the  wall-to-wall transport problem \citep{hassanzadeh2014wall, tobasco2017optimal, motoki2018optimal, motoki2018maximal, doering2019optimal, souza2020wall, tobasco2021optimal} which asks the question of what is the maximum heat transfer for a fixed energy  or enstrophy budget. 


\subsection{Comparison with the DNS}

\begin{figure}
\centering
\psfrag{a}[r][][0.7]{$\eta = 0.909$}
\psfrag{b}[r][][0.7]{$\eta = 0.714$}
\psfrag{c}[r][][0.7]{$\eta = 0.5 \; \; \; \;$}
\psfrag{d}[r][][0.7]{$\eta = 0.357$}
\psfrag{A}[][][0.9]{$1$}
\psfrag{B}[][][0.9]{$10$}
\psfrag{C}[][][0.9]{$10^2$}
\psfrag{D}[][][0.9]{$10^3$}
\psfrag{Z}[][][0.9]{$10^7$}
\psfrag{Y}[][][0.9]{$10^8$}
\psfrag{X}[][][0.9]{$10^9$}
\psfrag{W}[][][0.9]{$10^{10}$}
\psfrag{V}[][][0.9]{$10^{11}$}
\psfrag{P}[][][0.9][180]{$Nu/\chi(\eta)$}
\psfrag{R}[][][0.9][180]{$Nu$}
\psfrag{T}[l][][0.9]{$Ta$}
\begin{tabular}{lc}
\begin{subfigure}{0.5\textwidth}
\centering
 \includegraphics[scale = 0.9]{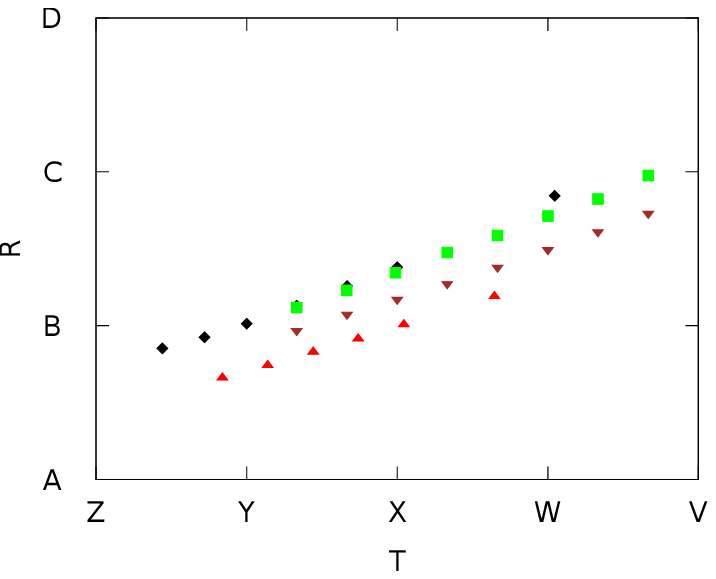}
\caption{}
\end{subfigure} &
\begin{subfigure}{0.5\textwidth}
\centering
 \includegraphics[scale = 0.9]{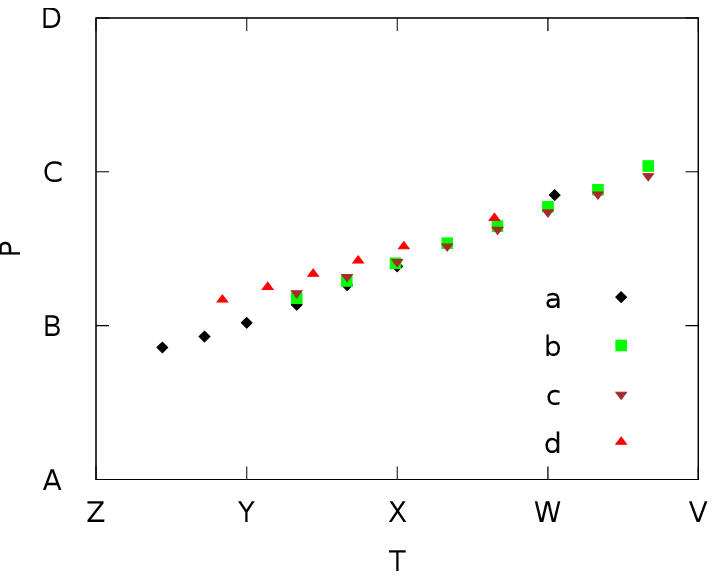}
\caption{}
\end{subfigure}
\end{tabular}
 \caption{(a) The Nusselt number ($Nu$) from DNS as a function of the Taylor number ($Ta$). (b) The Nusselt number ($Nu$) scaled with the geometrical scaling $X(\eta)$ given by (\ref{An analytical bound: the geometric scaling}) as a function of Taylor number ($Ta$). In these figures the DNS results are taken from \citep{ostilla2014exploring} for $\eta = 0.5, 0.714$ and $0.909$ and from \citep{froitzheim2019angular} for $\eta = 0.357$.}
 \label{Discussion, conclusion and summary: Comparison experiment}
\end{figure}

We now briefly analyze our results from a more practical point of view and ask the question of whether the dependence of the Nusselt number on the radius ratio obtained in this paper bears any relationship with that of the actual turbulent flow. Note that the asymptotic dependence of the optimal bound on the Taylor number is known to overestimate the actual Nusselt number in turbulent Taylor Couette flows by a logarithmic factor in $Ta$ \citep{grossmann2016high}.  As such, we cannot directly compare our results to the data, but instead merely ask the question of whether the geometric prefactor $g(\eta)$ in the expression $Nu(\eta,Ta) = g(\eta) f(Ta)$ measured in turbulent Taylor--Couette flows bears any resemblance with the prefactor $\chi(\eta)$ obtained in our optimal bound calculation, see equation (\ref{An analytical bound: the geometric scaling}).


We first test this idea on the direct numerical simulations (DNS) data from \citet{ostilla2014exploring} and \citet{froitzheim2019angular}. On the left-hand panel in figure \ref{Discussion, conclusion and summary: Comparison experiment}, we have plotted $Nu$ vs $Ta$ from these DNS, and on the right-hand panel, we show the same data divided by $\chi(\eta)$. We see that the rescaled data does become more compact and appears to fall on a single curve. This observation gives us confidence that the geometrical dependence of the bound $\chi(\eta)$ obtained in this paper is a good approximation to that of the actual Nusselt number $Nu$ measured in turbulent Taylor Couette flows. However, we note that the data has not yet reached the asymptotic scaling corresponding to the high $Ta$ regime, so the comparison at this point remains tentative. We also note that a different prediction for $Nu(\eta, Ta)$ has recently been obtained by  \citet{berghout2020calculation}  using the idea of Monin--Obukhov theory for thermally stratified turbulent boundary layers. Their scaling the asymptotic limit of high $Ta$ number is given as
\begin{eqnarray}
Nu \sim 4 \kappa^2 \frac{\eta^3}{(1+\eta)^2} \frac{Ta^{\frac{1}{2}}}{\log^2 Ta} = 0.6084 \frac{\eta^3}{(1+\eta)^2} \frac{Ta^{\frac{1}{2}}}{\log^2 Ta},
\label{Comparison with the DNS: MO scaling}
\end{eqnarray}
where $\kappa = 0.39$ is the von K\'arm\'an constant. The geometrical dependence in (\ref{Comparison with the DNS: MO scaling}) differs from $\chi(\eta)$ by a factor of $(1+\eta^2)^2$. However, it is reassuring to see that both expressions are proportional to  $\eta^3$ in the limit of small radius ratio. A definitive answer to the question of whether the geometrical scaling $\chi(\eta)$ given by our bound is exact or just an approximation would require a precise comparison with the turbulent data at very high Taylor numbers collected for a  range of radius ratios spanning the entire interval $(0, 1)$, which is at present a challenge for the numerical computations.

\subsection{Further generalizations}

We end this paper by discussing a few important consequences and generalizations of our study as well as  future outlooks. The first one of these consequences concerns  the bound on dissipation. The optimal bound on the Nusselt number for case 2 (3D incompressible perturbations) combined with the relations (\ref{Bulk quantities of interest: G epsilon Nu relation}) and (\ref{Problem setup: The Taylor number}) gives us the optimal bound on the dissipation
\begin{eqnarray}
\varepsilon_{b, \infty}^{3D} = 0.0677 \frac{\eta}{(1 + \eta) (1 + \eta^2)^2}.
\label{Discussion, conclusion and summary: main term at infinity of bound on dissipation 3D}
\end{eqnarray}
This bound tends to $0.00846$ in the limit $\eta \to 1$, which is within $1\%$ of the optimal bound obtained by \citet{plasting2003improved} for the plane Couette flow, namely $0.008553$. The consistency between the two results shows that our work can, in retrospect, be viewed as a generalization of the result of \citet{plasting2003improved} to Taylor--Couette flow for an arbitrary radius ratio. 


The second item is related to our previous work \citep{kumar2020pressure} on the dependence of the bound on the friction factor $\lambda$ on the radius of curvature $\kappa$ and torsion $\tau$ for a pressure-driven flow in a helical pipe. We were able to employ a similar boundary layer optimization technique together with standard inequalities as we did here to obtain the following analytical bound on the friction factor in high $\Rey$ limit
\begin{eqnarray}
\lambda_{b, \infty}^a = \frac{27}{8} I(\kappa, \tau),
\end{eqnarray}
where
\begin{eqnarray}
I(\kappa, \tau) = \frac{1}{2 \upi} \int_{0}^{2 \upi} ((1 - \kappa \cos \alpha)^2 + \tau^2)^{3/2} (1 - \kappa \cos \alpha) \; d \alpha.
\end{eqnarray}
However, the complexity of the helical pipe geometry makes it impossible in practice to compute the corresponding optimal bound. Nevertheless, in the light of results from the present study, and assuming that we captured the geometrical dependence correctly, one can in principle compute the prefactor in a limit where the optimal bound {\it can} be computed, namely the case of a straight pipe, for which $\kappa = \tau = 0$. This bound was computed by \citet{plasting2005friction} to be $\lambda_{b, \infty}^{3D}(0,0) = 0.27$,  and using this result we then expect that the optimal bound for helical pipes in the limit of high Reynolds number is 
\begin{eqnarray}
\lambda_{b,\infty}^{3D} = 0.27 I(\kappa, \tau).
\end{eqnarray}

Finally, the results presented in this paper potentially open the door to solving many important outstanding problems in engineering. Indeed, within that context we are often interested in finding the optimal geometry of the system or the object involved that minimizes or maximizes a certain flow quantity subject to some physical constraint. These types of problems therefore, demand a careful study of the effect of the domain shape on a flow quantity. From this perspective,  our study has broader implications. Even though we ruled out the applicability of the background method to a large class of problems (see \S \ref{A note on the applicability of the background method}), this still leaves a number of interesting problems open for analysis. For example, two problems which have been investigated using direct numerical simulations before but where an application of the background method can provide further insights are the Taylor--Couette flow with axisymmetric grooved walls \citep{zhu2016direct} and pressure-driven flow in a pipe with an elliptic cross-section \citep{nikitin2005direct}. Another problem where the background method has previously been used but capturing the exact domain shape dependence in the bounds were not the primary focus are the flow of fluid in an arbitrary domain driven by moving boundaries \citep{wang1997time}. Our study suggests an interesting avenue towards solving these problems, by using the background method, together with perturbations that are not assumed to be incompressible, which, as we demonstrated here, can greatly simplify the calculation.

\section*{Acknowledgement}
This paper is dedicated to Charlie Doering, whose work has been instrumental in motivating the author's research. A.K. thanks P. Garaud for a careful read of the paper and for providing comments that improved the quality of the paper.

\section*{Declaration of interests}
The author reports no conflict of interest.

\appendix
\section{The background method}
\label{The background method formulation}
In this section, we formulate the background method to obtain an upper bound on the quantity 
\begin{eqnarray}
\frac{1}{\Rey} \overline{ \| \bnabla \boldsymbol{u}\|_2^2},
\label{The background method: scaled L2 norm of the velocity}
\end{eqnarray}
in Taylor--Couette flow. It is clear from (\ref{Bulk quantities of interest: non-dimensional energy dissipation 2}) that an upper bound on this quantity immediately provides an upper bound on the dissipation $\varepsilon$. 

We begin by writing the total flow field $\boldsymbol{u}$ as a sum of two divergence-free flow fields,
\begin{eqnarray}
\boldsymbol{u} = \boldsymbol{U} + \boldsymbol{v}.
\label{The background method: the background decomposition}
\end{eqnarray}
We call $\boldsymbol{U}$ the background flow, and require that it satisfies same boundary conditions as $\boldsymbol{u}$ and is only a function of space.
We call $\boldsymbol{v}$, the perturbation, or perturbed flow, which satisfies homogeneous boundary conditions. The governing equation for the perturbation, obtained by substituting (\ref{The background method: the background decomposition}) in (\ref{Problem setup: momentum equation}), is given by
\begin{eqnarray}
\frac{\partial \boldsymbol{v}}{\partial t} + \boldsymbol{U} \bcdot \bnabla \boldsymbol{U} + \boldsymbol{U} \bcdot \bnabla \boldsymbol{v} + \boldsymbol{v} \bcdot \bnabla \boldsymbol{U} + \boldsymbol{v} \bcdot \bnabla \boldsymbol{v} = - \bnabla p + \frac{1}{\Rey} \nabla^2 \boldsymbol{U} + \frac{1}{\Rey} \nabla^2 \boldsymbol{v}.
\label{The background method: Governing equation perturbation}
\end{eqnarray}
We then obtain the evolution equation of the energy in the perturbed flow by taking the dot product of (\ref{The background method: Governing equation perturbation}) with $\boldsymbol{v}$ and integrating over the volume,
\begin{eqnarray}
\frac{d \|\boldsymbol{v}\|_2^2}{d t} = \frac{1}{\Rey} \int_{V} \nabla^2 \boldsymbol{U} \bcdot \boldsymbol{v} \; d \boldsymbol{x} - \frac{1}{\Rey} \|\bnabla  \boldsymbol{v}\|_2^2 - \int_{V} \boldsymbol{v} \bcdot \bnabla \boldsymbol{U} \bcdot \boldsymbol{v} \; d \boldsymbol{x} - \int_{V} \boldsymbol{U} \bcdot \bnabla \boldsymbol{U} \bcdot \boldsymbol{v} \; d \boldsymbol{x}, \nonumber \\
\label{The background method: energy perturbation}
\end{eqnarray}
Now using integration by parts, we can write
\begin{eqnarray}
\int_{V} \nabla^2 \boldsymbol{U} \bcdot \boldsymbol{v} \; d \boldsymbol{x} = - a \int_{V} \bnabla \boldsymbol{U} \boldsymbol{\colon} \bnabla \boldsymbol{v} \; d \boldsymbol{x} + (1-a) \int_{V} \nabla^2 \boldsymbol{U} \bcdot \boldsymbol{v} \; d \boldsymbol{x}.
\label{The background method: simple integration by parts}
\end{eqnarray}
where, in the index notation,
\begin{eqnarray}
\bnabla \boldsymbol{U} \boldsymbol{\colon} \bnabla \boldsymbol{v} = \partial_{i} \bs{v}_j \partial_{i} \bs{U}_j.
\end{eqnarray}
At the same time, one also has the following identity
\begin{eqnarray}
\bnabla \boldsymbol{U} \boldsymbol{:} \bnabla \boldsymbol{v} = \frac{|\bnabla \boldsymbol{u}|^2 - |\bnabla \boldsymbol{U}|^2 - |\bnabla \boldsymbol{v}|^2}{2}.
\label{The background method: the identity}
\end{eqnarray}
Using (\ref{The background method: simple integration by parts}) and (\ref{The background method: the identity}) in (\ref{The background method: energy perturbation}) leads to
\begin{eqnarray}
\frac{d \|\boldsymbol{v}\|_2^2}{d t} + \frac{a \| \bnabla \boldsymbol{u}\|_2^2}{2 \Rey} = \frac{a \| \bnabla \boldsymbol{U}\|_2^2}{2 \Rey} - \frac{(2 - a)}{2 \Rey} \| \bnabla \boldsymbol{v}\|_2^2  + \frac{(1 - a)}{\Rey} \int_{V} \nabla^2 \boldsymbol{U} \bcdot \boldsymbol{v} \; d \boldsymbol{x} \nonumber \\
- \int_{V} \boldsymbol{v} \bcdot \bnabla \boldsymbol{U} \bcdot \boldsymbol{v} \; d \boldsymbol{x} - \int_{V} \boldsymbol{U} \bcdot \bnabla \boldsymbol{U} \bcdot \boldsymbol{v} \; d \boldsymbol{x}.
\label{The background method: an inter modified ev. eqn.}
\end{eqnarray}
The introduction of a balance parameter `$a$' in the background formulation goes back to \citet{nicodemus1997improved}. Now it can be shown within the framework of the background method that the quantity $\|\boldsymbol{v}\|_2^2$ is uniformly bounded in time \citep[see][for example]{PhysRevLett.69.1648}. As a result, the long-time average of the time derivative of $\|\boldsymbol{v}\|_2^2$ vanishes. Therefore, taking the long-time average of the equation (\ref{The background method: an inter modified ev. eqn.}) leads to the following bound
\begin{eqnarray}
\frac{1}{\Rey} \overline{ \| \bnabla \boldsymbol{u}\|_2^2} \leq \frac{1}{\Rey} \| \bnabla \boldsymbol{U}\|_2^2 - \frac{2}{a}\overline{\mathcal{F} (\boldsymbol{v})}, 
\label{The background method: general bound}
\end{eqnarray}
where
\begin{eqnarray}
\mathcal{F} (\boldsymbol{v}) = && \left[\frac{2-a}{2 \Rey} \| \bnabla \boldsymbol{v}\|_2^2  - \frac{(1 - a)}{\Rey} \int_{V} \nabla^2 \boldsymbol{U} \bcdot \boldsymbol{v} \; d \boldsymbol{x} \right. \nonumber \\
&& \qquad \qquad \left. + \int_{V} \boldsymbol{v} \bcdot \bnabla \boldsymbol{U} \bcdot \boldsymbol{v} \; d \boldsymbol{x} +  \int_{V} \boldsymbol{U} \bcdot \bnabla \boldsymbol{U} \bcdot \boldsymbol{v} \; d \boldsymbol{x} \right].
\end{eqnarray}
This formulation of the background method is general, until this point. From here onward, we restrict the background flow $\boldsymbol{U}$ to be unidirectional, of the form
\begin{eqnarray}
\boldsymbol{U} = U_\theta(r) \boldsymbol{e}_{\theta}.
\label{The background method: unidirectional assumption}
\end{eqnarray}
At this point, we give proof of a straightforward but important lemma. 
\begin{lemma}
Let the domain $V$ be given by (\ref{Problem setup: The domain of interest}). Then for a continuous function $f:[r_i, r_o] \to \mathbb{R}$ and a divergence-free vector field $\boldsymbol{w}:V \to \mathbb{R}^3$ such that $\left. \boldsymbol{w} \right|_{r = r_i} = \left. \boldsymbol{w} \right|_{r = r_o} = \boldsymbol{0}$ and periodic in the $z$ direction, the following holds:
\begin{eqnarray}
\int_{V} f(r) w_r \; d \boldsymbol{x} = 0,
\end{eqnarray}
where, $w_r$ is the radial component of $\boldsymbol{w}$.
\label{The background method: easy lemma}
\end{lemma}
\proof
Let
\begin{eqnarray}
F(r, \theta, z) = \int_{r^\prime = r_i}^{r} f(r^\prime) \; d r^\prime.
\end{eqnarray}
Then we can write
\begin{eqnarray}
\int_{V} f(r) w_r d \boldsymbol{x} = \int_{V} \bnabla F \bcdot \boldsymbol{w} \; d \boldsymbol{x} = \int_{V} \bnabla \boldsymbol \bcdot (F \boldsymbol{w}) d \boldsymbol{x} = 0
\end{eqnarray}
where we used the divergence theorem and the boundary conditions on $\boldsymbol{w}$ to obtain the last equality.
\QEDB
\\
The assumption (\ref{The background method: unidirectional assumption}) combined with lemma \ref{The background method: easy lemma} implies
\begin{eqnarray}
\int_{V} \boldsymbol{U} \bcdot \bnabla \boldsymbol{U} \bcdot \boldsymbol{v} \; d \boldsymbol{x} = 0.
\end{eqnarray}
The functional $\mathcal{F}$ therefore takes the following form
\begin{eqnarray}
\mathcal{F} (\boldsymbol{v}) = \left[\frac{2-a}{2 \Rey} \| \bnabla \boldsymbol{v}\|_2^2 + \int_{V} \boldsymbol{v} \bcdot \bnabla \boldsymbol{U} \bcdot \boldsymbol{v} \; d \boldsymbol{x} - \frac{(1 - a)}{\Rey} \int_{V} \nabla^2 \boldsymbol{U} \bcdot \boldsymbol{v} \; d \boldsymbol{x} \right].
\end{eqnarray}
If the infimum of this functional $\mathcal{F}$ over all the divergence-free vector fields $\boldsymbol{v}$ is finite then it may not be zero as $\mathcal{F}$ is not homogeneous due the presence of a linear term. Therefore, similar to \citet{doering1998bounds} and \citet{plasting2003improved}, we define a shifted perturbation as
\begin{eqnarray}
\tilde{\boldsymbol{v}} = \boldsymbol{v} - \boldsymbol{\phi},
\label{The background method: modified perturbation}
\end{eqnarray}
where both $\tilde{\boldsymbol{v}}$ and $\boldsymbol{\phi}$ are divergence-free and satisfy homogeneous boundary conditions at the surface of the cylinders, and select $\boldsymbol{\phi}$ to eliminate the linear term when the bound (\ref{The background method: general bound}) is written in terms of $\tilde{\boldsymbol{v}}$.

We substitute (\ref{The background method: modified perturbation}) in (\ref{The background method: general bound}) and use (\ref{The background method: unidirectional assumption}) and lemma \ref{The background method: easy lemma} whenever required. We obtain the following linear term in $\boldsymbol{\tilde{v}}$:
\begin{eqnarray}
\frac{2}{a \Rey} \int_{V} \left[(2-a) \nabla^2 \boldsymbol{\phi} + (1-a) \nabla^2 \boldsymbol{U}\right] \bcdot \tilde{\boldsymbol{v}} \; d \boldsymbol{x}.
\label{The background method: inter specific bound}
\end{eqnarray}
Therefore, for this linear term to be zero, we require
\begin{eqnarray}
(2-a) \nabla^2 \boldsymbol{\phi} + (1-a) \nabla^2 \boldsymbol{U} = 0.
\end{eqnarray}
Without loss of generality, we can select the unidirectional solution 
\begin{eqnarray}
\boldsymbol{\phi} = - \frac{1-a}{2-a} \left[U_\theta - u_{lam, \theta}\right] \boldsymbol{e}_{\theta}.
\end{eqnarray}
Using this expression for $\boldsymbol{\phi}$, the bound in terms of $\boldsymbol{\tilde{v}}$ now reads
\begin{eqnarray}
\frac{1}{\Rey} \overline{ \| \bnabla \boldsymbol{u}\|_2^2} \leq \frac{1}{a (2 -a) \Rey}  \| \bnabla \boldsymbol{U}\|_2^2 - \frac{(1-a)^2}{a(2-a)\Rey} \| \bnabla \boldsymbol{u}_{lam}\|_2^2 - \frac{2}{a}\mathcal{H} (\boldsymbol{\tilde{v}}), 
\label{The background method: bound inter modified pert}
\end{eqnarray}
where
\begin{eqnarray}
\mathcal{H} (\boldsymbol{\tilde{v}})  = \left[\frac{2-a}{2 \Rey} \| \bnabla \boldsymbol{\tilde{v}}\|_2^2 +  \int_{V} \boldsymbol{\tilde{v}} \bcdot \bnabla \boldsymbol{U} \bcdot \boldsymbol{\tilde{v}} \; d \boldsymbol{x}\right].
\label{The background method: the functional H vtilde}
\end{eqnarray}
If we choose a background flow $\boldsymbol{U}$ such that the functional $\mathcal{H}$ is positive semi-definite on the space of divergence-free vector field $\boldsymbol{\tilde{v}}$, i.e.
\begin{eqnarray}
\inf_{\substack{ \boldsymbol{\tilde{v}} \\ \bnabla \bcdot \boldsymbol{\tilde{v}} = 0}} \mathcal{H} (\boldsymbol{\tilde{v}}) \geq 0,
\label{The background method: the spectral constraint 1}
\end{eqnarray}
then the bound (\ref{The background method: bound inter modified pert}) simply is
\begin{eqnarray}
\frac{1}{\Rey} \overline{ \| \bnabla \boldsymbol{u}\|_2^2} \leq \frac{1}{a (2 -a) \Rey}  \| \bnabla \boldsymbol{U}\|_2^2 - \frac{(1-a)^2}{a(2-a)\Rey} \| \bnabla \boldsymbol{u}_{lam}\|_2^2.
\label{The background method: final bound}
\end{eqnarray}
The positive semi-definite condition (\ref{The background method: the spectral constraint 1}) on $\mathcal{H}$ is referred to as the spectral constraint. Since the functional $\mathcal{H}$ is quadratic and homogeneous, we can rewrite the spectral constraint as
\begin{eqnarray}
\mathcal{H} (\boldsymbol{\tilde{v}}) \geq 0 \quad \forall \boldsymbol{\tilde{v}} \quad \text{such that} \quad \bnabla \bcdot \boldsymbol{\tilde{v}} = 0 \quad \text{and} \quad \|\boldsymbol{\tilde{v}}\|_2 = 1.
\label{The background method: the spectral constraint 2}
\end{eqnarray}
Using the Euler--Lagrange equations, the spectral constraint (\ref{The background method: the spectral constraint 2}) is equivalent to the non-negativity of the smallest eigenvalue $\lambda$ of the following self-adjoint spectral problem
\begin{subequations}
\begin{eqnarray}
\bnabla \bcdot \boldsymbol{\tilde{v}} = 0, \\
2 \lambda \boldsymbol{\tilde{v}} = \frac{(2-a)}{\Rey}  \nabla^2 \boldsymbol{\tilde{v}} - 2 \boldsymbol{\tilde{v}} \bcdot \bnabla \boldsymbol{U}_{sym} - \bnabla \tilde{p}.
\end{eqnarray}
\label{The background method: the spectral constraint 3}
\end{subequations}
Here, $\tilde{p}$ and $\lambda$ are the Lagrange multipliers for the constraints $\bnabla \bcdot \boldsymbol{\tilde{v}} = 0$ and $1 - \|\boldsymbol{\tilde{v}}\|_2^2 = 0$.

Now, to optimize the bound (\ref{The background method: bound inter modified pert}) under the incompressibility constraint on $\boldsymbol{\tilde{v}}$, we write the following Lagrangian
\begin{eqnarray}
\mathcal{L} = \frac{1}{a (2 -a) \Rey}  \| \bnabla \boldsymbol{U}\|_2^2 - \frac{(1-a)^2}{a(2-a)\Rey} \| \bnabla \boldsymbol{u}_{lam}\|_2^2 - \frac{2}{a}\mathcal{H} (\boldsymbol{\tilde{v}}) + \int_{V} \tilde{p} \; \bnabla \bcdot \boldsymbol{\tilde{v}} \; d \boldsymbol{x}.
\end{eqnarray}
Letting the first variation (the Frechet derivative) of this functional with respective to $\boldsymbol{\tilde{v}}$, $\tilde{p}$, $\boldsymbol{U}$ and $a$ to zero, leads to 
\begin{subequations}
\begin{eqnarray}
&& \frac{\delta \mathcal{L}}{\delta \boldsymbol{\tilde{v}}} = \frac{2 (2-a)}{a \Rey}  \nabla^2 \boldsymbol{\tilde{v}} - \frac{4}{a}\boldsymbol{\tilde{v}} \bcdot \bnabla \boldsymbol{U}_{sym} - \bnabla \tilde{p} = 0, \\
&& \frac{\delta \mathcal{L}}{\delta \tilde{p}} = \bnabla \bcdot \boldsymbol{\tilde{v}} = 0, \\
&& \frac{\delta \mathcal{L}}{\delta U_\theta} = -\frac{4 \upi L}{a (2-a) \Rey} \left(r \frac{d^2 U_\theta}{d r^2} + \frac{d U_\theta}{d r} - \frac{U_\theta}{r}\right) + \frac{1}{ r} \frac{d}{dr} \left(\frac{2 r^2}{a} \int_{\theta = 0}^{2 \upi} \int_{z = 0}^{L} \tilde{v}_r \tilde{v}_\theta d \theta d z\right) = 0, \nonumber\\ \\
&& \frac{\delta \mathcal{L}}{\delta a} = \frac{2(a-1)}{a^2(2-a)^2 \Rey} \left(\|\bnabla \boldsymbol{U}\|_2^2 - \|\bnabla \boldsymbol{u}_{lam}\|_2^2\right) + \frac{2}{a^2} \left(\frac{1}{\Rey}\|\bnabla \boldsymbol{\tilde{v}}\|_2^2 + \int_{V} \boldsymbol{\tilde{v}} \bcdot \bnabla \boldsymbol{U} \bcdot \boldsymbol{\tilde{v}} \; d \boldsymbol{x} \right) = 0. \nonumber \\
\end{eqnarray}
\label{The background method: The main EL equations}
\end{subequations}
In general, these equations do not have a unique solution. However, the solution to these equations for which the background flow also satisfies the spectral constraint (\ref{The background method: the spectral constraint 1}), or equivalently, all the eigenvalues of the eigenvalue problem (\ref{The background method: the spectral constraint 3} a,b) are non-negative, is unique.

\section{A useful lemma}
\label{Analytical solution case 1 moma}
Here we prove that the marginally stable perturbations in the energy stability analysis \S \ref{Energy stability analysis} or optimal perturbations in \S \ref{Optimal bounds: Numerical scheme} only depend on radius when they are not required to be incompressible.
\begin{lemma}
\label{A useful lemma: A useful lemma}
Let $\mathcal{D}(V)$ be the set of smooth velocity fields (not necessarily incompressible) that satisfy the homogeneous boundary conditions. For a given choice of the balance parameter $0 < a < 2$ and of the unidirectional background flow $\boldsymbol{U} = U_\theta(r) \boldsymbol{e}_\theta$, the functional $\mathcal{H}(\boldsymbol{\tilde{v}})$ (given by (\ref{The background method: the functional H vtilde})) achieves a minimum when $\boldsymbol{\tilde{v}}$ is a function of the radial direction only. Furthermore, if the background flow satisfies $(dU_\theta / dr - U_\theta/r) \leq 0$ then the optimal perturbed flow corresponds to $\tilde{v}_r = \tilde{v}_\theta$.
\label{lemma of only radial functions}
\end{lemma}
\begin{remark}
Although we do not prove that the optimal background flow satisfies $(dU_\theta / dr - U_\theta/r) \leq 0$, this condition was found to hold in every numerical computations of optimal bounds in all the three cases considered in our paper as well as for the choice of the background flow in analytical construction presented in \S \ref{An analytical bound}. Therefore, it is natural to make the assumption that $(dU_\theta / dr - U_\theta/r) \leq 0$.
\end{remark}
\proof
In the first part of the lemma, it is sufficient to show that for every $\boldsymbol{\tilde{v}} \in \mathcal{D}(V)$ there exist $\boldsymbol{\tilde{v}}_0 \in \mathcal{D}(V)$ with $\boldsymbol{\tilde{v}}_0(\boldsymbol{x}) = \boldsymbol{\tilde{v}}_0(r)$ such that $\mathcal{H}(\boldsymbol{\tilde{v}}_0) \leq \mathcal{H}(\boldsymbol{\tilde{v}})$.
\begin{eqnarray}
\mathcal{H} (\boldsymbol{\tilde{v}})  && = \left[\frac{2-a}{2 \Rey} \| \bnabla \boldsymbol{\tilde{v}}\|_2^2 +  \int_{V} \boldsymbol{\tilde{v}} \bcdot \bnabla \boldsymbol{U} \bcdot \boldsymbol{\tilde{v}} \; d \boldsymbol{x}\right] \nonumber \\
&& = \left[\int_{z = 0}^{L} \int_{\theta = 0}^{2 \pi}\left( \int_{r = r_i}^{r_o} \frac{2-a}{2 \Rey} | \bnabla \boldsymbol{\tilde{v}}|^2 +  \boldsymbol{\tilde{v}} \bcdot \bnabla \boldsymbol{U} \bcdot \boldsymbol{\tilde{v}} \; \; r d r \right) \; \;  d \theta d z \right] \nonumber \\
&& \geq \left[\int_{z = 0}^{L} \int_{\theta = 0}^{2 \pi} \inf_{\substack{ 0 \leq \theta \leq 2 \upi \\ 0 \leq z \leq L}} \left( \int_{r = r_i}^{r_o} \frac{2-a}{2 \Rey} | \bnabla \boldsymbol{\tilde{v}}|^2 +  \boldsymbol{\tilde{v}} \bcdot \bnabla \boldsymbol{U} \bcdot \boldsymbol{\tilde{v}} \; \; r d r \right) \; \;  d \theta d z \right] \nonumber \\
&& = \mathcal{H} (\boldsymbol{\tilde{v}}_0), 
\end{eqnarray}
where $\boldsymbol{\tilde{v}}_0 (\boldsymbol{x}) = \boldsymbol{\tilde{v}}(r, \theta_0, z_0)$ and $\theta_0$, $z_0$ corresponds to the values for which the infimum in third line is achieved. 

In the second part, for every perturbation $\bs{\tilde{v}} = (\tilde{v}_r, \tilde{v}_\theta)$, we define a modified perturbation
\begin{eqnarray}
\bs{\hat{v}} = \left(\frac{\sqrt{\tilde{v}_r^2 + \tilde{v}_\theta^2}}{\sqrt{2}}, \frac{\sqrt{\tilde{v}_r^2 + \tilde{v}_\theta^2}}{\sqrt{2}}\right).
\end{eqnarray}
So, if the initial perturbations $\bs{\tilde{v}}$ are weakly differentiable in space then so is the modified perturbation $\bs{\hat{v}}$. Therefore, all the operations below apply. For this modified perturbation, we have
\begin{eqnarray}
\| \bnabla \boldsymbol{\hat{v}}\|_2^2 && = \frac{\tilde{v}_r^2}{\tilde{v}_r^2 + \tilde{v}_\theta^2} \left(\frac{\partial \tilde{v}_r}{\partial r}\right)^2 + \frac{\tilde{v}_\theta^2}{\tilde{v}_r^2 + \tilde{v}_\theta^2} \left(\frac{\partial \tilde{v}_\theta}{\partial r}\right)^2 + \frac{\tilde{v}_r \tilde{v}_\theta}{\tilde{v}_r^2 + \tilde{v}_\theta^2} \frac{\partial \tilde{v}_r}{\partial r} \frac{\partial \tilde{v}_\theta}{\partial r} + \frac{\tilde{v}_r^2 + \tilde{v}_\theta^2}{r} \nonumber \\
&& \leq \left(\frac{\partial \tilde{v}_r}{\partial r}\right)^2 + \left(\frac{\partial \tilde{v}_\theta}{\partial r}\right)^2 + \frac{\tilde{v}_r^2 + \tilde{v}_\theta^2}{r} = \| \bnabla \boldsymbol{\tilde{v}}\|_2^2,
\label{A useful lemma: dissipation relation modified pert}
\end{eqnarray}
where we used  Young's inequality on the third term on the right-hand side in the first line to obtain the second line. Now the assumption on $U_\theta$ implies
\begin{eqnarray}
\frac{\tilde{v}_r^2 + \tilde{v}_\theta^2}{2} \left(\frac{dU_\theta}{dr} - \frac{U_\theta}{r} \right) \leq \tilde{v}_r \tilde{v}_\theta \left(\frac{dU_\theta}{dr} - \frac{U_\theta}{r} \right),
\label{A useful lemma: indefinite term relation modified pert}
\end{eqnarray}
again through the use of Young's inequality. Combining (\ref{A useful lemma: dissipation relation modified pert}) and (\ref{A useful lemma: indefinite term relation modified pert}) with the definition of $\mathcal{H}(\bs{v})$, leads to 
\begin{eqnarray}
\mathcal{H}(\bs{\hat{v}}) \leq \mathcal{H}(\bs{\tilde{v}}).
\end{eqnarray}
Finally, noting that $\hat{v}_r = \hat{v}_\theta$ proves the lemma.
\QEDB
\\

\section{Analytical solution of the Euler--Lagrange equations in case 1 at high Reynolds number}
\label{Analytical solution of the Euler--Lagrange equations in case 1 at high Reynolds number}
Before writing the Euler--Lagrange equations, we recall the simplifications pertaining to case 1. From lemma \ref{lemma of only radial functions}, we note that the optimal perturbations depends only on the radial direction and that $\tilde{v}_r = \tilde{v}_\theta$. Finally, noting that the Lagrangian $\mathcal{L}$ in case 1 does not involve the pressure term, as we do not impose the incompressibility condition, therefore the simplified Euler--Lagrange equations (\ref{The background method: The main EL equations}a-d) in case 1 are given by
\begin{subequations}
\begin{eqnarray}
&& \frac{(2-a)}{\Rey} \left(\frac{d^2 \tilde{v}_r}{d r^2} + \frac{1}{r}\frac{d \tilde{v}_r}{d r} - \frac{\tilde{v}_r}{r^2}\right)  - \tilde{v}_{r} \left(\frac{d U_\theta}{d r} - \frac{U_\theta}{r}\right) = 0, \\
&& -\frac{1}{(2-a) \Rey} \left(r \frac{d^2 U_\theta}{d r^2} + \frac{d U_\theta}{d r} - \frac{U_\theta}{r}\right) + \frac{1}{r} \frac{d (r^2\tilde{v}_r^2)}{dr} = 0,  \\
&& \frac{(a-1)}{(2-a)^2 \Rey} \left(\|\bnabla \boldsymbol{U}\|_2^2 - \|\bnabla \boldsymbol{u}_{lam}\|_2^2\right) + \frac{1}{\Rey}\|\bnabla \boldsymbol{\tilde{v}}\|_2^2 + \int_{V} \boldsymbol{\tilde{v}} \bcdot \bnabla \boldsymbol{U} \bcdot \boldsymbol{\tilde{v}} \; d \boldsymbol{x}  = 0. 
\end{eqnarray}
\label{EL equations case 1}
\end{subequations}
These equations need to be solved with boundary conditions
\begin{subequations}
\begin{eqnarray}
U_\theta  = 1, \quad \tilde{v}_r = 0 \qquad \text{at} \quad r = r_i, \\
U_\theta  = 0, \quad \tilde{v}_r = 0 \qquad \text{at} \quad r = r_o.
\end{eqnarray}
\label{EL equations case 1: boundary conditions}
\end{subequations}
As $\tilde{v}_z$ does not enter into the computations, it can be taken to be zero; as such $\bs{\tilde{v}}$ here should be understood as $(\tilde{v}_r, \tilde{v}_r, 0)$.

These equations can be solved using the method of matched asymptotics as described below. We consider three different regions: the inner boundary layer, the bulk and the outer boundary layer. We use the following scaled coordinates for the inner and outer boundary layer, respectively:
\begin{eqnarray}
s_i = \frac{r - r_i}{\delta}, \qquad s_o = \frac{r_o - r}{\delta},
\end{eqnarray}
where 
\begin{eqnarray}
\delta = \frac{1}{\Rey}.
\end{eqnarray}
We will use $in$, $bulk$ and $out$ in the superscript of the variables to indicate in which region the variable is being considered. Before proceeding further, we make the following change of variables
\begin{eqnarray}
U = \frac{U_\theta}{r}, \quad \tilde{v} = r \tilde{v}_r.
\label{EL equations case 1: change of variables}
\end{eqnarray}
Next, we write separate expansions for the variables in each of the three different regions as
\begin{subequations}
\begin{eqnarray}
&& \tilde{v}^{in}(s_i) = \tilde{v}_{0}^{in}(s_i) + \delta \; \tilde{v}_{1}^{in}(s_i) + \delta^2 \; \tilde{v}_{2}^{in}(s_i) + \dots, \\
&& \tilde{v}^{bulk}(r) = \tilde{v}_{0}^{bulk}(r) + \delta \; \tilde{v}_{1}^{bulk}(r) + \delta^2 \; \tilde{v}_{2}^{bulk}(r) + \dots, \\
&& \tilde{v}^{in}(s_o) = \tilde{v}_{0}^{out}(s_o) + \delta \; \tilde{v}_{1}^{out}(s_o) + \delta^2 \; \tilde{v}_{2}^{out}(s_o) + \dots.
\end{eqnarray}
\end{subequations}
A similar expansion can be written for $U$. Finally, we also use a simple expansion for the balance parameter 
\begin{eqnarray}
a = a_0 + \delta a_1 + \delta^2 a_2 + \dots.
\end{eqnarray}
Substituting the change of variables (\ref{EL equations case 1: change of variables}) and the series expansions of these new variables in (\ref{EL equations case 1}a-c), one can find out the leading order equations in different regions which then need to be solved with the boundary conditions (\ref{EL equations case 1: boundary conditions}a-b) and the following matching conditions
\begin{subequations}
\begin{eqnarray}
&& U_{0}^{in}(s_i \to \infty) = U_{0}^{bulk}(r = r_i), \quad U_{0}^{bulk}(r = r_o) = U_{0}^{out}(s_o \to \infty), \\
&& \tilde{v}_{0}^{in}(s_i \to \infty) = \tilde{v}_{0}^{bulk}(r = r_i), \quad \tilde{v}_{0}^{bulk}(r = r_o) = \tilde{v}_{0}^{out}(s_o \to \infty).
\end{eqnarray}
\end{subequations}
Upon solving the resultant set of equations, we find that the leading order term in  the background flow in the three different regions is given by 
\begin{subequations}
\begin{eqnarray}
&& U_{\theta}^{in} = \frac{r}{r_i}\left(1 - \frac{4 \sqrt{2}}{3} \alpha \tanh \left(\frac{\alpha s_i}{\sqrt{2}}\right) \right) + O\left(\delta\right), \\
&& U_{\theta}^{bulk} = \frac{r_i r}{r_i^2 + r_o^2} + O\left(\delta\right), \\
&& U_{\theta}^{out} = \frac{r}{r_o}\left(\frac{4 \sqrt{2}}{3} \beta \tanh \left(\frac{\beta s_o}{\sqrt{2}}\right) \right) + O\left(\delta\right),
\end{eqnarray}
\label{Case 1: Perturbations satisfying the homogeneous boundary conditions: background flow}
\end{subequations}
whereas the perturbed flow field is given by
\begin{subequations}
\begin{eqnarray}
&& \tilde{v}_{r}^{in} = \tilde{v}_{\theta}^{in} = \frac{\alpha r_i}{r} \tanh \left(\frac{\alpha s_i}{\sqrt{2}}\right) + O\left(\delta\right), \\
&& \tilde{v}_{r}^{out} = \tilde{v}_{\theta}^{out} = \frac{\beta r_o}{r} \tanh \left(\frac{\beta s_o}{\sqrt{2}}\right) + O\left(\delta\right), \\
&& \tilde{v}_{r}^{bulk} = \tilde{v}_{\theta}^{bulk} = \left(\frac{3 r_i r_o^2}{4 \sqrt{2} (r_i^2 + r_o^2)} \right) \frac{1}{r} + O\left(\delta\right),
\end{eqnarray}
\label{MOMA: pert sol}
\end{subequations}
where $\alpha$ and $\beta$ depend on $\eta$ and are given by 
\begin{eqnarray}
\alpha = \frac{3}{4 \sqrt{2}} \frac{1}{1 + \eta^2}, \quad \beta = \frac{3}{4 \sqrt{2}} \frac{\eta}{1 + \eta^2}.
\end{eqnarray}
The balance parameter takes the value $a = 2/3 + O(\delta)$. Using the expression of the background flow (\ref{Case 1: Perturbations satisfying the homogeneous boundary conditions: background flow}a-c) in (\ref{The background method: final bound}) and the relationships between different mean quantities (\ref{Bulk quantities of interest: non-dimensional energy dissipation 2}) and (\ref{Bulk quantities of interest: G epsilon Nu relation}), the leading order term in the  bound on the Nusselt number in the limit of high Reynolds number (or equivalently high Taylor number), is given by
\begin{eqnarray}
Nu_{b}^{nc} = \frac{9}{8} \frac{\eta^3}{(1 + \eta)^2 (1 + \eta^2)^2} Ta^{1/2}.
\end{eqnarray}
This bound is $2/3$ of the bound (\ref{An analytical bound: main term at infinity of bound on Nusselt number}) obtained using standard inequalities. This improvement has also been confirmed from the numerical results.

\bibliographystyle{jfm}
\bibliography{Reference}
\end{document}